\documentclass[longauth]{aa}
\pdfoutput=1
\usepackage{euclid}
\usepackage[T1]{fontenc}
\usepackage{geometry} 
\usepackage[section]{placeins}
\usepackage{amsmath}
\usepackage[colorlinks=true, allcolors=blue]{hyperref}
\usepackage{mathtools}
\usepackage[switch, modulo]{lineno}
\usepackage[usenames,dvipsnames]{xcolor} 
\usepackage[normalem]{ulem}
\usepackage[capitalise,nameinlink]{cleveref}
\usepackage{enumitem}
\usepackage{csquotes}
\usepackage{xspace}
\usepackage{orcidlink}

\usepackage{listings}
\usepackage{color}
\definecolor{dkgreen}{rgb}{0,0.6,0}
\definecolor{grey}{rgb}{0.5,0.5,0.5}
\definecolor{mauve}{rgb}{0.58,0,0.82}
\bibpunct{(}{)}{;}{a}{}{,}

\newcommand{\montepython}{\texttt{MontePython}}
\newcommand{\class}{\texttt{CLASS}}

\newcommand{\bcemu}{\texttt{BCemu}}

\newcommand{\loga}{\log_{10}(a_{\rm dark} / {\rm Mpc}^{-1})}
\newcommand{\logaxi}{\log_{10}(a_{\rm dark} \xi_{\rm idr}^4 / {\rm Mpc}^{-1})}

\defcitealias{Blanchard:2019oqi}{EP:VII}%{EC20}

\mathchardef\mhyphen="2D

\begin{document}

\title{\Euclid preparation}
\subtitle{Sensitivity to non-standard particle dark matter model}    

\newcommand{\orcid}[1]{\orcidlink{#1}}
%%%% please do not edit the author list -- contact ECEB Bureau for changes
%\newcommand{\orcid}[1]{} %% if already defined in aa.cls: comment, or use renewcommand			   
\author{Euclid Collaboration: J.~Lesgourgues\orcid{0000-0001-7627-353X}\thanks{\email{Julien.Lesgourgues@physik.rwth-aachen.de}}\inst{\ref{aff1}}
\and J.~Schwagereit\orcid{0009-0001-9987-1433}\inst{\ref{aff1}}
\and J.~Bucko\orcid{0000-0002-1662-1042}\inst{\ref{aff2}}
\and G.~Parimbelli\orcid{0000-0002-2539-2472}\inst{\ref{aff3},\ref{aff4},\ref{aff5}}
\and S.~K.~Giri\orcid{0000-0002-2560-536X}\inst{\ref{aff2},\ref{aff6}}
\and F.~Hervas-Peters\inst{\ref{aff7},\ref{aff2}}
\and A.~Schneider\orcid{0000-0001-7055-8104}\inst{\ref{aff2}}
\and M.~Archidiacono\orcid{0000-0003-4952-9012}\inst{\ref{aff8},\ref{aff9}}
\and F.~Pace\orcid{0000-0001-8039-0480}\inst{\ref{aff10},\ref{aff11},\ref{aff12}}
\and Z.~Sakr\orcid{0000-0002-4823-3757}\inst{\ref{aff13},\ref{aff14},\ref{aff15}}
\and A.~Amara\inst{\ref{aff16}}
\and L.~Amendola\orcid{0000-0002-0835-233X}\inst{\ref{aff13}}
\and S.~Andreon\orcid{0000-0002-2041-8784}\inst{\ref{aff17}}
\and N.~Auricchio\orcid{0000-0003-4444-8651}\inst{\ref{aff18}}
\and H.~Aussel\orcid{0000-0002-1371-5705}\inst{\ref{aff7}}
\and M.~Baldi\orcid{0000-0003-4145-1943}\inst{\ref{aff19},\ref{aff18},\ref{aff20}}
\and S.~Bardelli\orcid{0000-0002-8900-0298}\inst{\ref{aff18}}
\and R.~Bender\orcid{0000-0001-7179-0626}\inst{\ref{aff21},\ref{aff22}}
\and C.~Bodendorf\inst{\ref{aff21}}
\and D.~Bonino\orcid{0000-0002-3336-9977}\inst{\ref{aff12}}
\and E.~Branchini\orcid{0000-0002-0808-6908}\inst{\ref{aff23},\ref{aff24},\ref{aff17}}
\and M.~Brescia\orcid{0000-0001-9506-5680}\inst{\ref{aff25},\ref{aff26},\ref{aff27}}
\and J.~Brinchmann\orcid{0000-0003-4359-8797}\inst{\ref{aff28}}
\and S.~Camera\orcid{0000-0003-3399-3574}\inst{\ref{aff10},\ref{aff11},\ref{aff12}}
\and V.~Capobianco\orcid{0000-0002-3309-7692}\inst{\ref{aff12}}
\and C.~Carbone\orcid{0000-0003-0125-3563}\inst{\ref{aff29}}
\and V.~F.~Cardone\inst{\ref{aff30},\ref{aff31}}
\and J.~Carretero\orcid{0000-0002-3130-0204}\inst{\ref{aff32},\ref{aff33}}
\and S.~Casas\orcid{0000-0002-4751-5138}\inst{\ref{aff1}}
\and M.~Castellano\orcid{0000-0001-9875-8263}\inst{\ref{aff30}}
\and S.~Cavuoti\orcid{0000-0002-3787-4196}\inst{\ref{aff26},\ref{aff27}}
\and A.~Cimatti\inst{\ref{aff34}}
\and G.~Congedo\orcid{0000-0003-2508-0046}\inst{\ref{aff35}}
\and C.~J.~Conselice\orcid{0000-0003-1949-7638}\inst{\ref{aff36}}
\and L.~Conversi\orcid{0000-0002-6710-8476}\inst{\ref{aff37},\ref{aff38}}
\and Y.~Copin\orcid{0000-0002-5317-7518}\inst{\ref{aff39}}
\and F.~Courbin\orcid{0000-0003-0758-6510}\inst{\ref{aff40}}
\and H.~M.~Courtois\orcid{0000-0003-0509-1776}\inst{\ref{aff41}}
\and A.~Da~Silva\orcid{0000-0002-6385-1609}\inst{\ref{aff42},\ref{aff43}}
\and H.~Degaudenzi\orcid{0000-0002-5887-6799}\inst{\ref{aff44}}
\and A.~M.~Di~Giorgio\orcid{0000-0002-4767-2360}\inst{\ref{aff45}}
\and M.~Douspis\inst{\ref{aff46}}
\and F.~Dubath\orcid{0000-0002-6533-2810}\inst{\ref{aff44}}
\and X.~Dupac\inst{\ref{aff38}}
\and S.~Dusini\orcid{0000-0002-1128-0664}\inst{\ref{aff47}}
\and M.~Farina\orcid{0000-0002-3089-7846}\inst{\ref{aff45}}
\and S.~Farrens\orcid{0000-0002-9594-9387}\inst{\ref{aff7}}
\and S.~Ferriol\inst{\ref{aff39}}
\and P.~Fosalba\orcid{0000-0002-1510-5214}\inst{\ref{aff48},\ref{aff49}}
\and M.~Frailis\orcid{0000-0002-7400-2135}\inst{\ref{aff50}}
\and E.~Franceschi\orcid{0000-0002-0585-6591}\inst{\ref{aff18}}
\and M.~Fumana\orcid{0000-0001-6787-5950}\inst{\ref{aff29}}
\and S.~Galeotta\orcid{0000-0002-3748-5115}\inst{\ref{aff50}}
\and B.~Gillis\orcid{0000-0002-4478-1270}\inst{\ref{aff35}}
\and C.~Giocoli\orcid{0000-0002-9590-7961}\inst{\ref{aff18},\ref{aff51}}
\and A.~Grazian\orcid{0000-0002-5688-0663}\inst{\ref{aff52}}
\and F.~Grupp\inst{\ref{aff21},\ref{aff22}}
\and L.~Guzzo\orcid{0000-0001-8264-5192}\inst{\ref{aff8},\ref{aff17},\ref{aff9}}
\and S.~V.~H.~Haugan\orcid{0000-0001-9648-7260}\inst{\ref{aff53}}
\and H.~Hoekstra\orcid{0000-0002-0641-3231}\inst{\ref{aff54}}
\and W.~Holmes\inst{\ref{aff55}}
\and I.~Hook\orcid{0000-0002-2960-978X}\inst{\ref{aff56}}
\and F.~Hormuth\inst{\ref{aff57}}
\and A.~Hornstrup\orcid{0000-0002-3363-0936}\inst{\ref{aff58},\ref{aff59}}
\and K.~Jahnke\orcid{0000-0003-3804-2137}\inst{\ref{aff60}}
\and B.~Joachimi\orcid{0000-0001-7494-1303}\inst{\ref{aff61}}
\and E.~Keih\"anen\orcid{0000-0003-1804-7715}\inst{\ref{aff62}}
\and S.~Kermiche\orcid{0000-0002-0302-5735}\inst{\ref{aff63}}
\and A.~Kiessling\orcid{0000-0002-2590-1273}\inst{\ref{aff55}}
\and B.~Kubik\orcid{0009-0006-5823-4880}\inst{\ref{aff39}}
\and M.~Kunz\orcid{0000-0002-3052-7394}\inst{\ref{aff64}}
\and H.~Kurki-Suonio\orcid{0000-0002-4618-3063}\inst{\ref{aff65},\ref{aff66}}
\and R.~Laureijs\inst{\ref{aff67}}
\and S.~Ligori\orcid{0000-0003-4172-4606}\inst{\ref{aff12}}
\and P.~B.~Lilje\orcid{0000-0003-4324-7794}\inst{\ref{aff53}}
\and V.~Lindholm\orcid{0000-0003-2317-5471}\inst{\ref{aff65},\ref{aff66}}
\and I.~Lloro\inst{\ref{aff68}}
\and D.~Maino\inst{\ref{aff8},\ref{aff29},\ref{aff9}}
\and E.~Maiorano\orcid{0000-0003-2593-4355}\inst{\ref{aff18}}
\and O.~Mansutti\orcid{0000-0001-5758-4658}\inst{\ref{aff50}}
\and O.~Marggraf\orcid{0000-0001-7242-3852}\inst{\ref{aff69}}
\and K.~Markovic\orcid{0000-0001-6764-073X}\inst{\ref{aff55}}
\and N.~Martinet\orcid{0000-0003-2786-7790}\inst{\ref{aff70}}
\and F.~Marulli\orcid{0000-0002-8850-0303}\inst{\ref{aff71},\ref{aff18},\ref{aff20}}
\and R.~Massey\orcid{0000-0002-6085-3780}\inst{\ref{aff72}}
\and E.~Medinaceli\orcid{0000-0002-4040-7783}\inst{\ref{aff18}}
\and S.~Mei\orcid{0000-0002-2849-559X}\inst{\ref{aff73}}
\and Y.~Mellier\inst{\ref{aff74},\ref{aff75}}
\and M.~Meneghetti\orcid{0000-0003-1225-7084}\inst{\ref{aff18},\ref{aff20}}
\and E.~Merlin\orcid{0000-0001-6870-8900}\inst{\ref{aff30}}
\and G.~Meylan\inst{\ref{aff40}}
\and M.~Moresco\orcid{0000-0002-7616-7136}\inst{\ref{aff71},\ref{aff18}}
\and L.~Moscardini\orcid{0000-0002-3473-6716}\inst{\ref{aff71},\ref{aff18},\ref{aff20}}
\and E.~Munari\orcid{0000-0002-1751-5946}\inst{\ref{aff50},\ref{aff76}}
\and R.~Nakajima\inst{\ref{aff69}}
\and S.-M.~Niemi\inst{\ref{aff67}}
\and J.~W.~Nightingale\orcid{0000-0002-8987-7401}\inst{\ref{aff77},\ref{aff78}}
\and C.~Padilla\orcid{0000-0001-7951-0166}\inst{\ref{aff79}}
\and S.~Paltani\orcid{0000-0002-8108-9179}\inst{\ref{aff44}}
\and F.~Pasian\orcid{0000-0002-4869-3227}\inst{\ref{aff50}}
\and K.~Pedersen\inst{\ref{aff80}}
\and W.~J.~Percival\orcid{0000-0002-0644-5727}\inst{\ref{aff81},\ref{aff82},\ref{aff83}}
\and V.~Pettorino\inst{\ref{aff67}}
\and G.~Polenta\orcid{0000-0003-4067-9196}\inst{\ref{aff84}}
\and M.~Poncet\inst{\ref{aff85}}
\and L.~A.~Popa\inst{\ref{aff86}}
\and F.~Raison\orcid{0000-0002-7819-6918}\inst{\ref{aff21}}
\and R.~Rebolo\inst{\ref{aff87},\ref{aff88}}
\and A.~Renzi\orcid{0000-0001-9856-1970}\inst{\ref{aff89},\ref{aff47}}
\and J.~Rhodes\inst{\ref{aff55}}
\and G.~Riccio\inst{\ref{aff26}}
\and E.~Romelli\orcid{0000-0003-3069-9222}\inst{\ref{aff50}}
\and M.~Roncarelli\orcid{0000-0001-9587-7822}\inst{\ref{aff18}}
\and R.~Saglia\orcid{0000-0003-0378-7032}\inst{\ref{aff22},\ref{aff21}}
\and D.~Sapone\orcid{0000-0001-7089-4503}\inst{\ref{aff90}}
\and B.~Sartoris\orcid{0000-0003-1337-5269}\inst{\ref{aff22},\ref{aff50}}
\and R.~Scaramella\orcid{0000-0003-2229-193X}\inst{\ref{aff30},\ref{aff31}}
\and P.~Schneider\orcid{0000-0001-8561-2679}\inst{\ref{aff69}}
\and T.~Schrabback\orcid{0000-0002-6987-7834}\inst{\ref{aff91}}
\and A.~Secroun\orcid{0000-0003-0505-3710}\inst{\ref{aff63}}
\and G.~Seidel\orcid{0000-0003-2907-353X}\inst{\ref{aff60}}
\and S.~Serrano\orcid{0000-0002-0211-2861}\inst{\ref{aff48},\ref{aff3},\ref{aff92}}
\and C.~Sirignano\orcid{0000-0002-0995-7146}\inst{\ref{aff89},\ref{aff47}}
\and G.~Sirri\orcid{0000-0003-2626-2853}\inst{\ref{aff20}}
\and L.~Stanco\orcid{0000-0002-9706-5104}\inst{\ref{aff47}}
\and P.~Tallada-Cresp\'{i}\orcid{0000-0002-1336-8328}\inst{\ref{aff32},\ref{aff33}}
\and I.~Tereno\inst{\ref{aff42},\ref{aff93}}
\and R.~Toledo-Moreo\orcid{0000-0002-2997-4859}\inst{\ref{aff94}}
\and F.~Torradeflot\orcid{0000-0003-1160-1517}\inst{\ref{aff33},\ref{aff32}}
\and I.~Tutusaus\orcid{0000-0002-3199-0399}\inst{\ref{aff14}}
\and L.~Valenziano\orcid{0000-0002-1170-0104}\inst{\ref{aff18},\ref{aff95}}
\and T.~Vassallo\orcid{0000-0001-6512-6358}\inst{\ref{aff22},\ref{aff50}}
\and A.~Veropalumbo\orcid{0000-0003-2387-1194}\inst{\ref{aff17},\ref{aff24}}
\and Y.~Wang\orcid{0000-0002-4749-2984}\inst{\ref{aff96}}
\and J.~Weller\orcid{0000-0002-8282-2010}\inst{\ref{aff22},\ref{aff21}}
\and G.~Zamorani\orcid{0000-0002-2318-301X}\inst{\ref{aff18}}
\and E.~Zucca\orcid{0000-0002-5845-8132}\inst{\ref{aff18}}
\and A.~Biviano\orcid{0000-0002-0857-0732}\inst{\ref{aff50},\ref{aff76}}
\and A.~Boucaud\orcid{0000-0001-7387-2633}\inst{\ref{aff73}}
\and E.~Bozzo\orcid{0000-0002-8201-1525}\inst{\ref{aff44}}
\and C.~Burigana\orcid{0000-0002-3005-5796}\inst{\ref{aff97},\ref{aff95}}
\and M.~Calabrese\orcid{0000-0002-2637-2422}\inst{\ref{aff98},\ref{aff29}}
\and C.~Colodro-Conde\inst{\ref{aff87}}
\and G.~De~Lucia\orcid{0000-0002-6220-9104}\inst{\ref{aff50}}
\and D.~Di~Ferdinando\inst{\ref{aff20}}
\and J.~A.~Escartin~Vigo\inst{\ref{aff21}}
\and G.~Fabbian\orcid{0000-0002-3255-4695}\inst{\ref{aff99},\ref{aff100}}
\and R.~Farinelli\inst{\ref{aff18}}
\and J.~Gracia-Carpio\inst{\ref{aff21}}
\and S.~Ili\'c\orcid{0000-0003-4285-9086}\inst{\ref{aff101},\ref{aff14}}
\and G.~Mainetti\inst{\ref{aff102}}
\and M.~Martinelli\orcid{0000-0002-6943-7732}\inst{\ref{aff30},\ref{aff31}}
\and N.~Mauri\orcid{0000-0001-8196-1548}\inst{\ref{aff34},\ref{aff20}}
\and C.~Neissner\orcid{0000-0001-8524-4968}\inst{\ref{aff79},\ref{aff33}}
\and A.~A.~Nucita\inst{\ref{aff103},\ref{aff104},\ref{aff105}}
\and V.~Scottez\inst{\ref{aff74},\ref{aff106}}
\and M.~Tenti\orcid{0000-0002-4254-5901}\inst{\ref{aff20}}
\and M.~Viel\orcid{0000-0002-2642-5707}\inst{\ref{aff76},\ref{aff50},\ref{aff5},\ref{aff107},\ref{aff108}}
\and M.~Wiesmann\orcid{0009-0000-8199-5860}\inst{\ref{aff53}}
\and Y.~Akrami\orcid{0000-0002-2407-7956}\inst{\ref{aff109},\ref{aff110}}
\and S.~Anselmi\orcid{0000-0002-3579-9583}\inst{\ref{aff47},\ref{aff89},\ref{aff111}}
\and C.~Baccigalupi\orcid{0000-0002-8211-1630}\inst{\ref{aff5},\ref{aff50},\ref{aff107},\ref{aff76}}
\and M.~Ballardini\orcid{0000-0003-4481-3559}\inst{\ref{aff112},\ref{aff18},\ref{aff113}}
\and D.~Bertacca\orcid{0000-0002-2490-7139}\inst{\ref{aff89},\ref{aff52},\ref{aff47}}
\and L.~Blot\orcid{0000-0002-9622-7167}\inst{\ref{aff114},\ref{aff111}}
\and H.~B\"ohringer\orcid{0000-0001-8241-4204}\inst{\ref{aff21},\ref{aff115}}
\and S.~Borgani\orcid{0000-0001-6151-6439}\inst{\ref{aff116},\ref{aff76},\ref{aff50},\ref{aff107}}
\and S.~Bruton\orcid{0000-0002-6503-5218}\inst{\ref{aff117}}
\and R.~Cabanac\orcid{0000-0001-6679-2600}\inst{\ref{aff14}}
\and A.~Calabro\orcid{0000-0003-2536-1614}\inst{\ref{aff30}}
\and A.~Cappi\inst{\ref{aff18},\ref{aff118}}
\and C.~S.~Carvalho\inst{\ref{aff93}}
\and G.~Castignani\orcid{0000-0001-6831-0687}\inst{\ref{aff18}}
\and T.~Castro\orcid{0000-0002-6292-3228}\inst{\ref{aff50},\ref{aff107},\ref{aff76},\ref{aff108}}
\and K.~C.~Chambers\orcid{0000-0001-6965-7789}\inst{\ref{aff119}}
\and S.~Contarini\orcid{0000-0002-9843-723X}\inst{\ref{aff21},\ref{aff71}}
\and A.~R.~Cooray\orcid{0000-0002-3892-0190}\inst{\ref{aff120}}
\and S.~Davini\orcid{0000-0003-3269-1718}\inst{\ref{aff24}}
\and B.~De~Caro\inst{\ref{aff47},\ref{aff89}}
\and S.~de~la~Torre\inst{\ref{aff70}}
\and G.~Desprez\inst{\ref{aff121}}
\and A.~D\'iaz-S\'anchez\orcid{0000-0003-0748-4768}\inst{\ref{aff122}}
\and S.~Di~Domizio\orcid{0000-0003-2863-5895}\inst{\ref{aff23},\ref{aff24}}
\and H.~Dole\orcid{0000-0002-9767-3839}\inst{\ref{aff46}}
\and S.~Escoffier\orcid{0000-0002-2847-7498}\inst{\ref{aff63}}
\and A.~G.~Ferrari\orcid{0009-0005-5266-4110}\inst{\ref{aff34},\ref{aff20}}
\and P.~G.~Ferreira\orcid{0000-0002-3021-2851}\inst{\ref{aff123}}
\and I.~Ferrero\orcid{0000-0002-1295-1132}\inst{\ref{aff53}}
\and F.~Finelli\orcid{0000-0002-6694-3269}\inst{\ref{aff18},\ref{aff95}}
\and F.~Fornari\orcid{0000-0003-2979-6738}\inst{\ref{aff95}}
\and L.~Gabarra\orcid{0000-0002-8486-8856}\inst{\ref{aff123}}
\and K.~Ganga\orcid{0000-0001-8159-8208}\inst{\ref{aff73}}
\and J.~Garc\'ia-Bellido\orcid{0000-0002-9370-8360}\inst{\ref{aff109}}
\and E.~Gaztanaga\orcid{0000-0001-9632-0815}\inst{\ref{aff3},\ref{aff48},\ref{aff124}}
\and F.~Giacomini\orcid{0000-0002-3129-2814}\inst{\ref{aff20}}
\and G.~Gozaliasl\orcid{0000-0002-0236-919X}\inst{\ref{aff125},\ref{aff65}}
\and H.~Hildebrandt\orcid{0000-0002-9814-3338}\inst{\ref{aff126}}
\and J.~Hjorth\orcid{0000-0002-4571-2306}\inst{\ref{aff127}}
\and A.~Jimenez~Mun\~noz\orcid{0009-0004-5252-185X}\inst{\ref{aff128}}
\and S.~Joudaki\orcid{0000-0001-8820-673X}\inst{\ref{aff124}}
\and J.~J.~E.~Kajava\orcid{0000-0002-3010-8333}\inst{\ref{aff129},\ref{aff130}}
\and V.~Kansal\orcid{0000-0002-4008-6078}\inst{\ref{aff131},\ref{aff132}}
\and D.~Karagiannis\orcid{0000-0002-4927-0816}\inst{\ref{aff133},\ref{aff134}}
\and C.~C.~Kirkpatrick\inst{\ref{aff62}}
\and L.~Legrand\orcid{0000-0003-0610-5252}\inst{\ref{aff135}}
\and G.~Libet\inst{\ref{aff85}}
\and A.~Loureiro\orcid{0000-0002-4371-0876}\inst{\ref{aff136},\ref{aff137}}
\and J.~Macias-Perez\orcid{0000-0002-5385-2763}\inst{\ref{aff128}}
\and G.~Maggio\orcid{0000-0003-4020-4836}\inst{\ref{aff50}}
\and M.~Magliocchetti\orcid{0000-0001-9158-4838}\inst{\ref{aff45}}
\and F.~Mannucci\orcid{0000-0002-4803-2381}\inst{\ref{aff138}}
\and R.~Maoli\orcid{0000-0002-6065-3025}\inst{\ref{aff139},\ref{aff30}}
\and C.~J.~A.~P.~Martins\orcid{0000-0002-4886-9261}\inst{\ref{aff140},\ref{aff28}}
\and S.~Matthew\inst{\ref{aff35}}
\and L.~Maurin\orcid{0000-0002-8406-0857}\inst{\ref{aff46}}
\and R.~B.~Metcalf\orcid{0000-0003-3167-2574}\inst{\ref{aff71},\ref{aff18}}
\and M.~Migliaccio\inst{\ref{aff141},\ref{aff142}}
\and P.~Monaco\orcid{0000-0003-2083-7564}\inst{\ref{aff116},\ref{aff50},\ref{aff107},\ref{aff76}}
\and C.~Moretti\orcid{0000-0003-3314-8936}\inst{\ref{aff5},\ref{aff108},\ref{aff50},\ref{aff76},\ref{aff107}}
\and G.~Morgante\inst{\ref{aff18}}
\and S.~Nadathur\orcid{0000-0001-9070-3102}\inst{\ref{aff124}}
\and Nicholas~A.~Walton\orcid{0000-0003-3983-8778}\inst{\ref{aff143}}
\and L.~Patrizii\inst{\ref{aff20}}
\and A.~Pezzotta\orcid{0000-0003-0726-2268}\inst{\ref{aff21}}
\and M.~P\"ontinen\orcid{0000-0001-5442-2530}\inst{\ref{aff65}}
\and V.~Popa\inst{\ref{aff86}}
\and C.~Porciani\orcid{0000-0002-7797-2508}\inst{\ref{aff69}}
\and D.~Potter\orcid{0000-0002-0757-5195}\inst{\ref{aff2}}
\and P.~Reimberg\orcid{0000-0003-3410-0280}\inst{\ref{aff74}}
\and I.~Risso\orcid{0000-0003-2525-7761}\inst{\ref{aff4}}
\and P.-F.~Rocci\inst{\ref{aff46}}
\and M.~Sahl\'en\orcid{0000-0003-0973-4804}\inst{\ref{aff144}}
\and A.~G.~S\'anchez\orcid{0000-0003-1198-831X}\inst{\ref{aff21}}
\and J.~A.~Schewtschenko\inst{\ref{aff35}}
\and E.~Sefusatti\orcid{0000-0003-0473-1567}\inst{\ref{aff50},\ref{aff76},\ref{aff107}}
\and M.~Sereno\orcid{0000-0003-0302-0325}\inst{\ref{aff18},\ref{aff20}}
\and P.~Simon\inst{\ref{aff69}}
\and A.~Spurio~Mancini\orcid{0000-0001-5698-0990}\inst{\ref{aff145},\ref{aff146}}
\and J.~Steinwagner\inst{\ref{aff21}}
\and C.~Tao\orcid{0000-0001-7961-8177}\inst{\ref{aff63}}
\and N.~Tessore\orcid{0000-0002-9696-7931}\inst{\ref{aff61}}
\and G.~Testera\inst{\ref{aff24}}
\and R.~Teyssier\orcid{0000-0001-7689-0933}\inst{\ref{aff147}}
\and S.~Toft\orcid{0000-0003-3631-7176}\inst{\ref{aff59},\ref{aff148},\ref{aff149}}
\and S.~Tosi\orcid{0000-0002-7275-9193}\inst{\ref{aff23},\ref{aff17},\ref{aff24}}
\and A.~Troja\orcid{0000-0003-0239-4595}\inst{\ref{aff89},\ref{aff47}}
\and M.~Tucci\inst{\ref{aff44}}
\and C.~Valieri\inst{\ref{aff20}}
\and J.~Valiviita\orcid{0000-0001-6225-3693}\inst{\ref{aff65},\ref{aff66}}
\and D.~Vergani\orcid{0000-0003-0898-2216}\inst{\ref{aff18}}
\and G.~Verza\orcid{0000-0002-1886-8348}\inst{\ref{aff150},\ref{aff100}}}
										   
%%%% please do not edit the affiliation list -- contact ECEB Bureau for changes
\institute{Institute for Theoretical Particle Physics and Cosmology (TTK), RWTH Aachen University, 52056 Aachen, Germany\label{aff1}
\and
Department of Astrophysics, University of Zurich, Winterthurerstrasse 190, 8057 Zurich, Switzerland\label{aff2}
\and
Institute of Space Sciences (ICE, CSIC), Campus UAB, Carrer de Can Magrans, s/n, 08193 Barcelona, Spain\label{aff3}
\and
Dipartimento di Fisica, Universit\`a degli studi di Genova, and INFN-Sezione di Genova, via Dodecaneso 33, 16146, Genova, Italy\label{aff4}
\and
SISSA, International School for Advanced Studies, Via Bonomea 265, 34136 Trieste TS, Italy\label{aff5}
\and
Nordita, KTH Royal Institute of Technology and Stockholm 1859 University, Hannes Alfv\'ens v\"{a}g 12, Stockholm, SE-106 91, Sweden\label{aff6}
\and
Universit\'e Paris-Saclay, Universit\'e Paris Cit\'e, CEA, CNRS, AIM, 91191, Gif-sur-Yvette, France\label{aff7}
\and
Dipartimento di Fisica "Aldo Pontremoli", Universit\`a degli Studi di Milano, Via Celoria 16, 20133 Milano, Italy\label{aff8}
\and
INFN-Sezione di Milano, Via Celoria 16, 20133 Milano, Italy\label{aff9}
\and
Dipartimento di Fisica, Universit\`a degli Studi di Torino, Via P. Giuria 1, 10125 Torino, Italy\label{aff10}
\and
INFN-Sezione di Torino, Via P. Giuria 1, 10125 Torino, Italy\label{aff11}
\and
INAF-Osservatorio Astrofisico di Torino, Via Osservatorio 20, 10025 Pino Torinese (TO), Italy\label{aff12}
\and
Institut f\"ur Theoretische Physik, University of Heidelberg, Philosophenweg 16, 69120 Heidelberg, Germany\label{aff13}
\and
Institut de Recherche en Astrophysique et Plan\'etologie (IRAP), Universit\'e de Toulouse, CNRS, UPS, CNES, 14 Av. Edouard Belin, 31400 Toulouse, France\label{aff14}
\and
Universit\'e St Joseph; Faculty of Sciences, Beirut, Lebanon\label{aff15}
\and
School of Mathematics and Physics, University of Surrey, Guildford, Surrey, GU2 7XH, UK\label{aff16}
\and
INAF-Osservatorio Astronomico di Brera, Via Brera 28, 20122 Milano, Italy\label{aff17}
\and
INAF-Osservatorio di Astrofisica e Scienza dello Spazio di Bologna, Via Piero Gobetti 93/3, 40129 Bologna, Italy\label{aff18}
\and
Dipartimento di Fisica e Astronomia, Universit\`a di Bologna, Via Gobetti 93/2, 40129 Bologna, Italy\label{aff19}
\and
INFN-Sezione di Bologna, Viale Berti Pichat 6/2, 40127 Bologna, Italy\label{aff20}
\and
Max Planck Institute for Extraterrestrial Physics, Giessenbachstr. 1, 85748 Garching, Germany\label{aff21}
\and
Universit\"ats-Sternwarte M\"unchen, Fakult\"at f\"ur Physik, Ludwig-Maximilians-Universit\"at M\"unchen, Scheinerstrasse 1, 81679 M\"unchen, Germany\label{aff22}
\and
Dipartimento di Fisica, Universit\`a di Genova, Via Dodecaneso 33, 16146, Genova, Italy\label{aff23}
\and
INFN-Sezione di Genova, Via Dodecaneso 33, 16146, Genova, Italy\label{aff24}
\and
Department of Physics "E. Pancini", University Federico II, Via Cinthia 6, 80126, Napoli, Italy\label{aff25}
\and
INAF-Osservatorio Astronomico di Capodimonte, Via Moiariello 16, 80131 Napoli, Italy\label{aff26}
\and
INFN section of Naples, Via Cinthia 6, 80126, Napoli, Italy\label{aff27}
\and
Instituto de Astrof\'isica e Ci\^encias do Espa\c{c}o, Universidade do Porto, CAUP, Rua das Estrelas, PT4150-762 Porto, Portugal\label{aff28}
\and
INAF-IASF Milano, Via Alfonso Corti 12, 20133 Milano, Italy\label{aff29}
\and
INAF-Osservatorio Astronomico di Roma, Via Frascati 33, 00078 Monteporzio Catone, Italy\label{aff30}
\and
INFN-Sezione di Roma, Piazzale Aldo Moro, 2 - c/o Dipartimento di Fisica, Edificio G. Marconi, 00185 Roma, Italy\label{aff31}
\and
Centro de Investigaciones Energ\'eticas, Medioambientales y Tecnol\'ogicas (CIEMAT), Avenida Complutense 40, 28040 Madrid, Spain\label{aff32}
\and
Port d'Informaci\'{o} Cient\'{i}fica, Campus UAB, C. Albareda s/n, 08193 Bellaterra (Barcelona), Spain\label{aff33}
\and
Dipartimento di Fisica e Astronomia "Augusto Righi" - Alma Mater Studiorum Universit\`a di Bologna, Viale Berti Pichat 6/2, 40127 Bologna, Italy\label{aff34}
\and
Institute for Astronomy, University of Edinburgh, Royal Observatory, Blackford Hill, Edinburgh EH9 3HJ, UK\label{aff35}
\and
Jodrell Bank Centre for Astrophysics, Department of Physics and Astronomy, University of Manchester, Oxford Road, Manchester M13 9PL, UK\label{aff36}
\and
European Space Agency/ESRIN, Largo Galileo Galilei 1, 00044 Frascati, Roma, Italy\label{aff37}
\and
ESAC/ESA, Camino Bajo del Castillo, s/n., Urb. Villafranca del Castillo, 28692 Villanueva de la Ca\~nada, Madrid, Spain\label{aff38}
\and
Universit\'e Claude Bernard Lyon 1, CNRS/IN2P3, IP2I Lyon, UMR 5822, Villeurbanne, F-69100, France\label{aff39}
\and
Institute of Physics, Laboratory of Astrophysics, Ecole Polytechnique F\'ed\'erale de Lausanne (EPFL), Observatoire de Sauverny, 1290 Versoix, Switzerland\label{aff40}
\and
UCB Lyon 1, CNRS/IN2P3, IUF, IP2I Lyon, 4 rue Enrico Fermi, 69622 Villeurbanne, France\label{aff41}
\and
Departamento de F\'isica, Faculdade de Ci\^encias, Universidade de Lisboa, Edif\'icio C8, Campo Grande, PT1749-016 Lisboa, Portugal\label{aff42}
\and
Instituto de Astrof\'isica e Ci\^encias do Espa\c{c}o, Faculdade de Ci\^encias, Universidade de Lisboa, Campo Grande, 1749-016 Lisboa, Portugal\label{aff43}
\and
Department of Astronomy, University of Geneva, ch. d'Ecogia 16, 1290 Versoix, Switzerland\label{aff44}
\and
INAF-Istituto di Astrofisica e Planetologia Spaziali, via del Fosso del Cavaliere, 100, 00100 Roma, Italy\label{aff45}
\and
Universit\'e Paris-Saclay, CNRS, Institut d'astrophysique spatiale, 91405, Orsay, France\label{aff46}
\and
INFN-Padova, Via Marzolo 8, 35131 Padova, Italy\label{aff47}
\and
Institut d'Estudis Espacials de Catalunya (IEEC),  Edifici RDIT, Campus UPC, 08860 Castelldefels, Barcelona, Spain\label{aff48}
\and
Institut de Ciencies de l'Espai (IEEC-CSIC), Campus UAB, Carrer de Can Magrans, s/n Cerdanyola del Vall\'es, 08193 Barcelona, Spain\label{aff49}
\and
INAF-Osservatorio Astronomico di Trieste, Via G. B. Tiepolo 11, 34143 Trieste, Italy\label{aff50}
\and
Istituto Nazionale di Fisica Nucleare, Sezione di Bologna, Via Irnerio 46, 40126 Bologna, Italy\label{aff51}
\and
INAF-Osservatorio Astronomico di Padova, Via dell'Osservatorio 5, 35122 Padova, Italy\label{aff52}
\and
Institute of Theoretical Astrophysics, University of Oslo, P.O. Box 1029 Blindern, 0315 Oslo, Norway\label{aff53}
\and
Leiden Observatory, Leiden University, Einsteinweg 55, 2333 CC Leiden, The Netherlands\label{aff54}
\and
Jet Propulsion Laboratory, California Institute of Technology, 4800 Oak Grove Drive, Pasadena, CA, 91109, USA\label{aff55}
\and
Department of Physics, Lancaster University, Lancaster, LA1 4YB, UK\label{aff56}
\and
Felix Hormuth Engineering, Goethestr. 17, 69181 Leimen, Germany\label{aff57}
\and
Technical University of Denmark, Elektrovej 327, 2800 Kgs. Lyngby, Denmark\label{aff58}
\and
Cosmic Dawn Center (DAWN), Denmark\label{aff59}
\and
Max-Planck-Institut f\"ur Astronomie, K\"onigstuhl 17, 69117 Heidelberg, Germany\label{aff60}
\and
Department of Physics and Astronomy, University College London, Gower Street, London WC1E 6BT, UK\label{aff61}
\and
Department of Physics and Helsinki Institute of Physics, Gustaf H\"allstr\"omin katu 2, 00014 University of Helsinki, Finland\label{aff62}
\and
Aix-Marseille Universit\'e, CNRS/IN2P3, CPPM, Marseille, France\label{aff63}
\and
Universit\'e de Gen\`eve, D\'epartement de Physique Th\'eorique and Centre for Astroparticle Physics, 24 quai Ernest-Ansermet, CH-1211 Gen\`eve 4, Switzerland\label{aff64}
\and
Department of Physics, P.O. Box 64, 00014 University of Helsinki, Finland\label{aff65}
\and
Helsinki Institute of Physics, Gustaf H{\"a}llstr{\"o}min katu 2, University of Helsinki, Helsinki, Finland\label{aff66}
\and
European Space Agency/ESTEC, Keplerlaan 1, 2201 AZ Noordwijk, The Netherlands\label{aff67}
\and
NOVA optical infrared instrumentation group at ASTRON, Oude Hoogeveensedijk 4, 7991PD, Dwingeloo, The Netherlands\label{aff68}
\and
Universit\"at Bonn, Argelander-Institut f\"ur Astronomie, Auf dem H\"ugel 71, 53121 Bonn, Germany\label{aff69}
\and
Aix-Marseille Universit\'e, CNRS, CNES, LAM, Marseille, France\label{aff70}
\and
Dipartimento di Fisica e Astronomia "Augusto Righi" - Alma Mater Studiorum Universit\`a di Bologna, via Piero Gobetti 93/2, 40129 Bologna, Italy\label{aff71}
\and
Department of Physics, Centre for Extragalactic Astronomy, Durham University, South Road, DH1 3LE, UK\label{aff72}
\and
Universit\'e Paris Cit\'e, CNRS, Astroparticule et Cosmologie, 75013 Paris, France\label{aff73}
\and
Institut d'Astrophysique de Paris, 98bis Boulevard Arago, 75014, Paris, France\label{aff74}
\and
Institut d'Astrophysique de Paris, UMR 7095, CNRS, and Sorbonne Universit\'e, 98 bis boulevard Arago, 75014 Paris, France\label{aff75}
\and
IFPU, Institute for Fundamental Physics of the Universe, via Beirut 2, 34151 Trieste, Italy\label{aff76}
\and
School of Mathematics, Statistics and Physics, Newcastle University, Herschel Building, Newcastle-upon-Tyne, NE1 7RU, UK\label{aff77}
\and
Department of Physics, Institute for Computational Cosmology, Durham University, South Road, DH1 3LE, UK\label{aff78}
\and
Institut de F\'{i}sica d'Altes Energies (IFAE), The Barcelona Institute of Science and Technology, Campus UAB, 08193 Bellaterra (Barcelona), Spain\label{aff79}
\and
Department of Physics and Astronomy, University of Aarhus, Ny Munkegade 120, DK-8000 Aarhus C, Denmark\label{aff80}
\and
Waterloo Centre for Astrophysics, University of Waterloo, Waterloo, Ontario N2L 3G1, Canada\label{aff81}
\and
Department of Physics and Astronomy, University of Waterloo, Waterloo, Ontario N2L 3G1, Canada\label{aff82}
\and
Perimeter Institute for Theoretical Physics, Waterloo, Ontario N2L 2Y5, Canada\label{aff83}
\and
Space Science Data Center, Italian Space Agency, via del Politecnico snc, 00133 Roma, Italy\label{aff84}
\and
Centre National d'Etudes Spatiales -- Centre spatial de Toulouse, 18 avenue Edouard Belin, 31401 Toulouse Cedex 9, France\label{aff85}
\and
Institute of Space Science, Str. Atomistilor, nr. 409 M\u{a}gurele, Ilfov, 077125, Romania\label{aff86}
\and
Instituto de Astrof\'isica de Canarias, Calle V\'ia L\'actea s/n, 38204, San Crist\'obal de La Laguna, Tenerife, Spain\label{aff87}
\and
Departamento de Astrof\'isica, Universidad de La Laguna, 38206, La Laguna, Tenerife, Spain\label{aff88}
\and
Dipartimento di Fisica e Astronomia "G. Galilei", Universit\`a di Padova, Via Marzolo 8, 35131 Padova, Italy\label{aff89}
\and
Departamento de F\'isica, FCFM, Universidad de Chile, Blanco Encalada 2008, Santiago, Chile\label{aff90}
\and
Universit\"at Innsbruck, Institut f\"ur Astro- und Teilchenphysik, Technikerstr. 25/8, 6020 Innsbruck, Austria\label{aff91}
\and
Satlantis, University Science Park, Sede Bld 48940, Leioa-Bilbao, Spain\label{aff92}
\and
Instituto de Astrof\'isica e Ci\^encias do Espa\c{c}o, Faculdade de Ci\^encias, Universidade de Lisboa, Tapada da Ajuda, 1349-018 Lisboa, Portugal\label{aff93}
\and
Universidad Polit\'ecnica de Cartagena, Departamento de Electr\'onica y Tecnolog\'ia de Computadoras,  Plaza del Hospital 1, 30202 Cartagena, Spain\label{aff94}
\and
INFN-Bologna, Via Irnerio 46, 40126 Bologna, Italy\label{aff95}
\and
Infrared Processing and Analysis Center, California Institute of Technology, Pasadena, CA 91125, USA\label{aff96}
\and
INAF, Istituto di Radioastronomia, Via Piero Gobetti 101, 40129 Bologna, Italy\label{aff97}
\and
Astronomical Observatory of the Autonomous Region of the Aosta Valley (OAVdA), Loc. Lignan 39, I-11020, Nus (Aosta Valley), Italy\label{aff98}
\and
School of Physics and Astronomy, Cardiff University, The Parade, Cardiff, CF24 3AA, UK\label{aff99}
\and
Center for Computational Astrophysics, Flatiron Institute, 162 5th Avenue, 10010, New York, NY, USA\label{aff100}
\and
Universit\'e Paris-Saclay, CNRS/IN2P3, IJCLab, 91405 Orsay, France\label{aff101}
\and
Centre de Calcul de l'IN2P3/CNRS, 21 avenue Pierre de Coubertin 69627 Villeurbanne Cedex, France\label{aff102}
\and
Department of Mathematics and Physics E. De Giorgi, University of Salento, Via per Arnesano, CP-I93, 73100, Lecce, Italy\label{aff103}
\and
INAF-Sezione di Lecce, c/o Dipartimento Matematica e Fisica, Via per Arnesano, 73100, Lecce, Italy\label{aff104}
\and
INFN, Sezione di Lecce, Via per Arnesano, CP-193, 73100, Lecce, Italy\label{aff105}
\and
Junia, EPA department, 41 Bd Vauban, 59800 Lille, France\label{aff106}
\and
INFN, Sezione di Trieste, Via Valerio 2, 34127 Trieste TS, Italy\label{aff107}
\and
ICSC - Centro Nazionale di Ricerca in High Performance Computing, Big Data e Quantum Computing, Via Magnanelli 2, Bologna, Italy\label{aff108}
\and
Instituto de F\'isica Te\'orica UAM-CSIC, Campus de Cantoblanco, 28049 Madrid, Spain\label{aff109}
\and
CERCA/ISO, Department of Physics, Case Western Reserve University, 10900 Euclid Avenue, Cleveland, OH 44106, USA\label{aff110}
\and
Laboratoire Univers et Th\'eorie, Observatoire de Paris, Universit\'e PSL, Universit\'e Paris Cit\'e, CNRS, 92190 Meudon, France\label{aff111}
\and
Dipartimento di Fisica e Scienze della Terra, Universit\`a degli Studi di Ferrara, Via Giuseppe Saragat 1, 44122 Ferrara, Italy\label{aff112}
\and
Istituto Nazionale di Fisica Nucleare, Sezione di Ferrara, Via Giuseppe Saragat 1, 44122 Ferrara, Italy\label{aff113}
\and
Kavli Institute for the Physics and Mathematics of the Universe (WPI), University of Tokyo, Kashiwa, Chiba 277-8583, Japan\label{aff114}
\and
Ludwig-Maximilians-University, Schellingstrasse 4, 80799 Munich, Germany\label{aff115}
\and
Dipartimento di Fisica - Sezione di Astronomia, Universit\`a di Trieste, Via Tiepolo 11, 34131 Trieste, Italy\label{aff116}
\and
Minnesota Institute for Astrophysics, University of Minnesota, 116 Church St SE, Minneapolis, MN 55455, USA\label{aff117}
\and
Universit\'e C\^{o}te d'Azur, Observatoire de la C\^{o}te d'Azur, CNRS, Laboratoire Lagrange, Bd de l'Observatoire, CS 34229, 06304 Nice cedex 4, France\label{aff118}
\and
Institute for Astronomy, University of Hawaii, 2680 Woodlawn Drive, Honolulu, HI 96822, USA\label{aff119}
\and
Department of Physics \& Astronomy, University of California Irvine, Irvine CA 92697, USA\label{aff120}
\and
Department of Astronomy \& Physics and Institute for Computational Astrophysics, Saint Mary's University, 923 Robie Street, Halifax, Nova Scotia, B3H 3C3, Canada\label{aff121}
\and
Departamento F\'isica Aplicada, Universidad Polit\'ecnica de Cartagena, Campus Muralla del Mar, 30202 Cartagena, Murcia, Spain\label{aff122}
\and
Department of Physics, Oxford University, Keble Road, Oxford OX1 3RH, UK\label{aff123}
\and
Institute of Cosmology and Gravitation, University of Portsmouth, Portsmouth PO1 3FX, UK\label{aff124}
\and
Department of Computer Science, Aalto University, PO Box 15400, Espoo, FI-00 076, Finland\label{aff125}
\and
Ruhr University Bochum, Faculty of Physics and Astronomy, Astronomical Institute (AIRUB), German Centre for Cosmological Lensing (GCCL), 44780 Bochum, Germany\label{aff126}
\and
DARK, Niels Bohr Institute, University of Copenhagen, Jagtvej 155, 2200 Copenhagen, Denmark\label{aff127}
\and
Univ. Grenoble Alpes, CNRS, Grenoble INP, LPSC-IN2P3, 53, Avenue des Martyrs, 38000, Grenoble, France\label{aff128}
\and
Department of Physics and Astronomy, Vesilinnantie 5, 20014 University of Turku, Finland\label{aff129}
\and
Serco for European Space Agency (ESA), Camino bajo del Castillo, s/n, Urbanizacion Villafranca del Castillo, Villanueva de la Ca\~nada, 28692 Madrid, Spain\label{aff130}
\and
ARC Centre of Excellence for Dark Matter Particle Physics, Melbourne, Australia\label{aff131}
\and
Centre for Astrophysics \& Supercomputing, Swinburne University of Technology,  Hawthorn, Victoria 3122, Australia\label{aff132}
\and
School of Physics and Astronomy, Queen Mary University of London, Mile End Road, London E1 4NS, UK\label{aff133}
\and
Department of Physics and Astronomy, University of the Western Cape, Bellville, Cape Town, 7535, South Africa\label{aff134}
\and
ICTP South American Institute for Fundamental Research, Instituto de F\'{\i}sica Te\'orica, Universidade Estadual Paulista, S\~ao Paulo, Brazil\label{aff135}
\and
Oskar Klein Centre for Cosmoparticle Physics, Department of Physics, Stockholm University, Stockholm, SE-106 91, Sweden\label{aff136}
\and
Astrophysics Group, Blackett Laboratory, Imperial College London, London SW7 2AZ, UK\label{aff137}
\and
INAF-Osservatorio Astrofisico di Arcetri, Largo E. Fermi 5, 50125, Firenze, Italy\label{aff138}
\and
Dipartimento di Fisica, Sapienza Universit\`a di Roma, Piazzale Aldo Moro 2, 00185 Roma, Italy\label{aff139}
\and
Centro de Astrof\'{\i}sica da Universidade do Porto, Rua das Estrelas, 4150-762 Porto, Portugal\label{aff140}
\and
Dipartimento di Fisica, Universit\`a di Roma Tor Vergata, Via della Ricerca Scientifica 1, Roma, Italy\label{aff141}
\and
INFN, Sezione di Roma 2, Via della Ricerca Scientifica 1, Roma, Italy\label{aff142}
\and
Institute of Astronomy, University of Cambridge, Madingley Road, Cambridge CB3 0HA, UK\label{aff143}
\and
Theoretical astrophysics, Department of Physics and Astronomy, Uppsala University, Box 515, 751 20 Uppsala, Sweden\label{aff144}
\and
Department of Physics, Royal Holloway, University of London, TW20 0EX, UK\label{aff145}
\and
Mullard Space Science Laboratory, University College London, Holmbury St Mary, Dorking, Surrey RH5 6NT, UK\label{aff146}
\and
Department of Astrophysical Sciences, Peyton Hall, Princeton University, Princeton, NJ 08544, USA\label{aff147}
\and
Cosmic Dawn Center (DAWN)\label{aff148}
\and
Niels Bohr Institute, University of Copenhagen, Jagtvej 128, 2200 Copenhagen, Denmark\label{aff149}
\and
Center for Cosmology and Particle Physics, Department of Physics, New York University, New York, NY 10003, USA\label{aff150}}

%\emailAdd{casas@physik.rwth-aachen.de}
%\emailAdd{lesgourg@physik.rwth-aachen.de}
%\emailAdd{nils.science@gmail.com}
\date{}
\authorrunning{J. Lesgourgues et al.}

\date{\today}

\abstract
{
% Context
The \Euclid mission of the European Space Agency will provide weak gravitational lensing and galaxy clustering surveys that can be used to constrain the standard cosmological model and its extensions, with an opportunity to test the properties of dark matter beyond the minimal cold dark matter paradigm.
% Aims
We present forecasts from the combination of the \Euclid\ weak lensing and photometric galaxy clustering data on the parameters describing four interesting and representative non-minimal dark matter models: a mixture of cold and warm dark matter relics; unstable dark matter decaying either into massless or massive relics; and dark matter experiencing feeble interactions with relativistic relics.
% Methods
We model these scenarios at the level of the non-linear matter power spectrum using emulators trained on dedicated $N$-body simulations. We use a mock \Euclid{} likelihood and Monte Carlo Markov Chains to fit mock data and infer error bars on dark matter parameters marginalised over other parameters.
% Results
We find that the \Euclid{} photometric probe (alone or in combination with cosmic microwave background data from the \Planck{} satellite) will be sensitive to the effect of each of the four dark matter models considered here. The improvement will be particularly spectacular for decaying and interacting dark matter models. With \Euclid{}, the bounds on some dark matter parameters can improve by up to two orders of magnitude compared to current limits. We discuss the dependence of predicted uncertainties on different assumptions: inclusion of photometric galaxy clustering data, minimum angular scale taken into account, modelling of baryonic feedback effects.
% Conclusions
We conclude that the \Euclid{} mission will be able to measure quantities related to the dark sector of particle physics with unprecedented sensitivity. This will provide important information for model building in high-energy physics. Any hint of a deviation from the minimal cold dark matter paradigm would have profound implications for cosmology and particle physics.}


%\runningpagewiselinenumbers
%\linenumbers
\maketitle
\tableofcontents
%\flushbottom
%\input{intro.tex}
\section{Introduction}

Understanding the nature of dark matter (DM) is one of the priority targets within the communities of cosmology, astroparticle, and high-energy physics. Over the past decade, the Large Hadron Collider (LHC) results and the absence of direct or indirect DM detection have shown that the situation concerning the nature of DM is wide open. Weakly Interacting Massive Particles (WIMPs) are only one candidate among many possibilities \citep{Bertone:2004pz,Feng:2010gw}. Particle-like DM could have a large range of plausible mass, lifetime, annihilation cross-section, and scattering cross-sections.

The standard cosmological model makes the working assumption of a purely stable, decoupled, and cold dark matter (CDM) species, which can be modelled as dust in simulations of the evolution of the Universe since very early times -- well before photon decoupling. In the CDM limit, the only measurable parameter related to DM is its relic non-relativistic density today, $\rho_{\rm cdm}$, which can be expressed in terms of a dimensionless density parameter $\omega_{\rm cdm} := \Omega_{\rm cdm}h^2$, where 
$\Omega_{\rm cdm}$ is the fractional density of CDM (relative to the critical density) and 
$h := H_0/(100\,\kmsMpc)$ 
is the reduced Hubble parameter. However, in non-minimal scenarios, DM could have several other parameters of possible relevance for fitting cosmological observations, such as: a non-negligible velocity dispersion \citep{Bond:1983aaa,Bode:2000gq}, a lifetime not considerably larger than the age of the Universe \citep[e.g.,][]{Ichiki:2004vi,Audren:2014bca}, and cross-sections describing either its self-interaction \citep{Spergel:1999mh,Feng:2009mn} or its feeble interaction with other species \citep{Boehm:2000gq,Cyr-Racine:2015ihg}.

From a high-energy physics point of view, non-standard DM models are easy to motivate. The logic pursued successfully by high-energy physicists for almost a century consists of postulating additional symmetries (rather than adding individual particles)  in order to explain unaccounted experimental results. The current standard model of particle physics is known to be incomplete \citep{ParticleDataGroup:2022pth} and the assumption of new symmetries usually comes together with a rich dark sector, i.e., several new particles with new interactions, with potentially more than one population surviving until today and contributing to DM or dark radiation. From this point of view, having just one decoupled, stable, and cold relic in our Universe does not sound much more natural than being surrounded by one or more dark species with potentially non-trivial properties. High-energy physicists often suggest that, given the richness of the visible sector, there is no obvious reason for the dark sector to reduce to a single CDM relic particle.

The astrophysics community is sometimes reluctant to investigate the possible consequences of non-minimal particle-physics assumptions in cosmology as long as the minimal $\Lambda$CDM model has not been ruled out. The situation is however evolving given the accumulation of tensions or unresolved questions in cosmological observations (like the small-scale CDM crisis, Hubble tension or $S_8$ tension), see for instance \cite{Verde:2019ivm}, \cite{Abdalla:2022yfr}. In this context, it sounds at least reasonable to investigate the possibility of detecting some effects induced by non-minimal DM models.
Of course, it is still possible that future observations only provide bounds on these models and leave us with plain CDM as a preferred case. Even in this case, it would be extremely interesting for particle physics model-builders to have such bounds, since constraints from accelerators or astroparticle experiments usually probe a different regime or different model assumptions than cosmological data.

Non-minimal DM properties may affect the growth of structures in the Universe in different ways, at different times, and on different scales. Thus, they can leave several types of signatures in the 2-point correlation function of matter fluctuations in Fourier space, called the matter power spectrum. This spectrum can be reconstructed from several types of cosmological observations at different redshifts. The modified growth of structure induced by non-minimal DM models could also affect other statistical probes of structure formation (higher-order correlation functions, halo mass function, peak and void statistics), but in this work we only consider its impact on the matter power spectrum.

\Euclid 
\citep{EuclidSkyOverview}
%\citep{EUCLID:2011zbd, Euclid:2021icp} 
is a medium-class mission of the European Space Agency, which will map the local Universe to improve our understanding of the expansion history and of the growth of structures. The satellite will observe roughly $15\,000\,{\rm deg^2}$ of the sky through two instruments, a visible imager %\citep[VIS,][]{2016SPIE.9904E..0QC} 
\citep[VIS,][]{EuclidSkyVIS}
and Near-Infrared Spectrometer and Photometer
%\citep[NISP,][]{2022SPIE12180E..1KM}
\citep[NISP,][]{EuclidSkyNISP}, delivering the images of more than one billion galaxies and the spectra of tens of millions of galaxies out to redshift of about $2$. The combination of spectroscopy and photometry will allow us to reconstruct the matter power spectrum up to $1\%$ accuracy.

Since the matter power spectrum could be strongly affected by the nature of DM, \Euclid{} is a perfect tool for testing non-minimal DM properties. It may either confirm the standard CDM paradigm or discover some new DM features. The goal of this work is precisely to estimate the sensitivity of \Euclid{} to different DM parameters beyond its mere relic density. Given the wide range of possible alternatives to standard CDM, we cannot explore all possibilities. We will concentrate on four examples of non-minimal scenarios that are still compatible with current data and could be either constrained or detected by \Euclid{}. Our choice of models is dictated by simple considerations. First, we should select some representative cases. Since in non-minimal models, DM particles are expected to either free-stream (with some velocity dispersion) and/or decay (with some rate) and or scatter (with some cross-sections), we pick up examples in each of these three categories. A well-motivated example of DM with a velocity dispersion is warm DM; some simple examples of decaying DM consist of particles with a constant decay rate, decaying either into relativistic or non-relativistic daughter particles; and a representative case of scattering DM  is that of DM interacting with dark radiation. Second, we are interested in models such that galaxy redshift surveys could provide stronger bounds than other observables, and in particular, than cosmic microwave background (CMB) and/or Lyman-$\alpha$ forest data. For reasons detailed in the next sections, this would not be the case for pure warm DM or pure decaying DM. Thus, going to the next level of complexity, we will assume a mixture of either cold and warm dark matter, or of stable and unstable particles. In conclusion, we will focus on four interesting and representative models: a mixture of cold and warm dark matter, a mixture of stable and unstable particles decaying either into relativistic or non-relativistic particles, and dark matter interacting with dark relativistic relics.

\Euclid{} will deliver several types of observations. Among these, the weak lensing (WL) survey and the galaxy clustering (GC) photometric survey will be ideal to constrain DM properties, since they will both provide a measurement of the matter power spectrum down to small scales and up to high redshift. These two surveys will return maps in tomographic bins that can be analysed all together (taking into account cross-correlations between WL and galaxy density maps). In addition to this joint data set, called the photometric probe, \Euclid{} will provide other observations. The \Euclid{}  spectroscopic galaxy redshift survey will play an essential role for constraining several cosmological models and parameters. Cluster number counts will also convey very useful information. However, these surveys will not provide information on such small scales as WL, and their implementation in sensitivity forecasts relies on a different methodology than for the photometric probe. In particular, they require a different approach to model non-linear effects for each non-minimal DM model. Thus, for simplicity, we choose to concentrate only on the \Euclid{} photometric probe in this work. 

In Sect.\,\ref{sec:pdm_models} of this work, we review the four DM models that we will investigate, with a brief discussion of their foundations, their free parameters and their effects on the linear matter power spectrum. In Sect.\,\ref{sec:non-linear}, we explain how to model the effect of these scenarios at the level of the non-linear power spectrum, using emulators trained on dedicated $N$-body simulations. In Sect.\,\ref{sec:method}, we summarise the assumptions and numerical pipelines used in our parameter sensitivity forecasts for the \Euclid{} photometric probe. We present our results for each model in Sect.\,\ref{sec:results} and provide final conclusions in Sect.\,\ref{sec:discussion}.

\section{Non-minimal particle dark matter Models\label{sec:pdm_models}}

Many particle DM properties can be tested with cosmology \citep{Gluscevic:2019yal}. As mentioned in the introduction, we only focus here on four particular models. On the one hand, these models constitute representative samples of the three most plausible properties of non-minimal particle DM: a non-negligible velocity dispersion, some decay rate, or a non-negligible scattering rate. On the other hand, within their respective category, they account for the simplest scenarios that can be constrained better by weak lensing and galaxy surveys than CMB and Lyman-$\alpha$ data.

\subsection{Cold plus warm dark matter\label{sec:theo_CWDM}}

In this model, a fraction $f_{\rm wdm}$ of the total DM fractional density $\Omega_{\rm dm}$ is assumed to be warm, so that $\Omega_{\rm dm}=\Omega_{\rm cdm}+\Omega_{\rm wdm}=(1-f_{\rm wdm}) \, \Omega_{\rm dm}+f_{\rm wdm}\,\Omega_{\rm dm}$. The warm dark matter (WDM) component possesses a thermal (root mean square) velocity $v_{\rm rms}$ that depends on the temperature-to-mass ratio $T_{\rm wdm}/m_{\rm wdm}$. WDM would revert to CDM in the limit $v_{\rm rms}\to0$, or equivalently $m_{\rm wdm}\to\infty$. 

This mixed cold plus warm dark matter (CWDM) model has been studied previously, for instance, in \cite{Boyarsky:2008xj}, \cite{Schneider:2014rda}, \cite{Murgia:2017lwo}, \cite{Murgia:2018now}, \cite{Parimbelli_CWDM}, or \cite{Hooper:2022byl}. It may account either for cosmologies with two distinct DM components, or also, effectively, for cosmologies with a single DM component with a non-thermal distribution, such as resonantly-produced sterile neutrinos \citep{Boyarsky:2008mt}. This model has been often invoked as a possible solution to the small-scale CDM crisis \citep{Anderhalden:2012jc,Maccio:2012rjx}. Current best constraints come from Lyman-$\alpha$ forest surveys \citep{Hooper:2022byl}, Milky Way satellites \citep{Diamanti:2017xfo}, and WL surveys \citep{Peters_2023}, see Sect.\,\ref{sec:res_cwdm}.

The thermal velocity of WDM defines its maximum free-streaming scale, reached at the time of its non-relativistic transition during radiation domination. On larger wavelengths, cosmological fluctuations have the same evolution as in a model in which all the DM would be cold. On smaller scales, the perturbations of the WDM component are negligible and the growth of CDM density fluctuations is suppressed. Thus, at the level of linear perturbations, the overall effect of WDM is to induce a step-like suppression in the matter power spectrum. The amplitude of the step is controlled by $f_{\rm wdm}$.\footnote{The step-like amplitude can be approximated as $(1-f_{\rm wdm})^2 \left[{D(a_0)}/{D(a_{\rm eq})}\right]^{-(3/2)f_{\rm wdm}}$, where $D(a)$ is the scale-independent linear growth factor of CDM density fluctuations in a pure $\Lambda$CDM universe, $a_{\rm eq}$ is the scale factor at radiation-to-matter equality, and $a_0$ is the scale factor today \citep{Boyarsky:2008xj}.}
The shape of the step is universal for all models in which the WDM phase-space distribution has a thermal shape up to a rescaling factor. This covers two well-known limits: on the one hand, thermal WDM, for which the phase-space distribution is thermal (with no rescaling factor) but the WDM temperature $T_{\rm wdm}$ is reduced compared to the active neutrino temperature, due to its earlier decoupling time; and the Dodelson--Widrow (DW) model~\citep{Dodelson:1993je,Colombi:1995ze}, for which the phase-space distribution is identical to that of active neutrinos (with $T_{\rm wdm}=T_\nu$) up to a rescaling factor $\chi \ll 1$ accounting for the efficiency of active-sterile neutrino oscillations in the early Universe with a small mixing angle, $\chi \sim \sin^2 \theta$. For this broad category of models, the location of the step-like suppression is controlled by the thermal velocity, i.e., by the temperature-to-mass ratio $T_{\rm wdm}/m_{\rm wdm}$.

It is convenient to parameterise the location of the step in terms of the rescaled mass
\begin{align}
x := m_{\rm wdm} \frac{T_{\nu}}{T_{{\rm wdm}}}\,,
\end{align}
where $T_{\nu}$ is the current value of the active neutrino temperature computed in the instantaneous decoupling limit, i.e., such that $T_{\nu}/T_{\gamma}=(4/11)^{1/3}$. For the class of models described above, the effect of WDM is entirely described by the two parameters $(f_{\rm wdm}, x)$, independently of the chosen model (thermal WDM or DW). In the DW case, $x$ coincides with $m_{\rm wdm}^{\rm DW}$. In the thermal case, one has 
\begin{align}
m_{\rm wdm}^{\rm thermal} =
\left( 94.1 \,\Omega_{\rm wdm} h^2\right)^{1/4} \left(\frac{x}{1\,{\rm eV}}\right)^{3/4}\,{\rm eV}\,,
\label{eq:m_wdm_thermal}
\end{align}
where we used the fact that for Fermi--Dirac thermal relics $X$ with a temperature $T_X=T_\nu$ one gets 
$m_X = 94.1 \, \Omega_{X} h^2 \, {\rm eV}$.\footnote{Since our definition of the reference temperature $T_\nu$ applies to neutrinos in the instantaneous decoupling limit, in order to be consistent, we need to stick to the same limit when computing the factor $m_X / (\Omega_{X} h^2 {\rm eV})$. Thus, for this factor, we must use the value 94.1 rather than the slightly smaller value 93.1 that accounts for the mass-to-density ratio of active neutrino.}

\begin{figure*}[t]
    \centering
    \includegraphics[width=.99\textwidth]{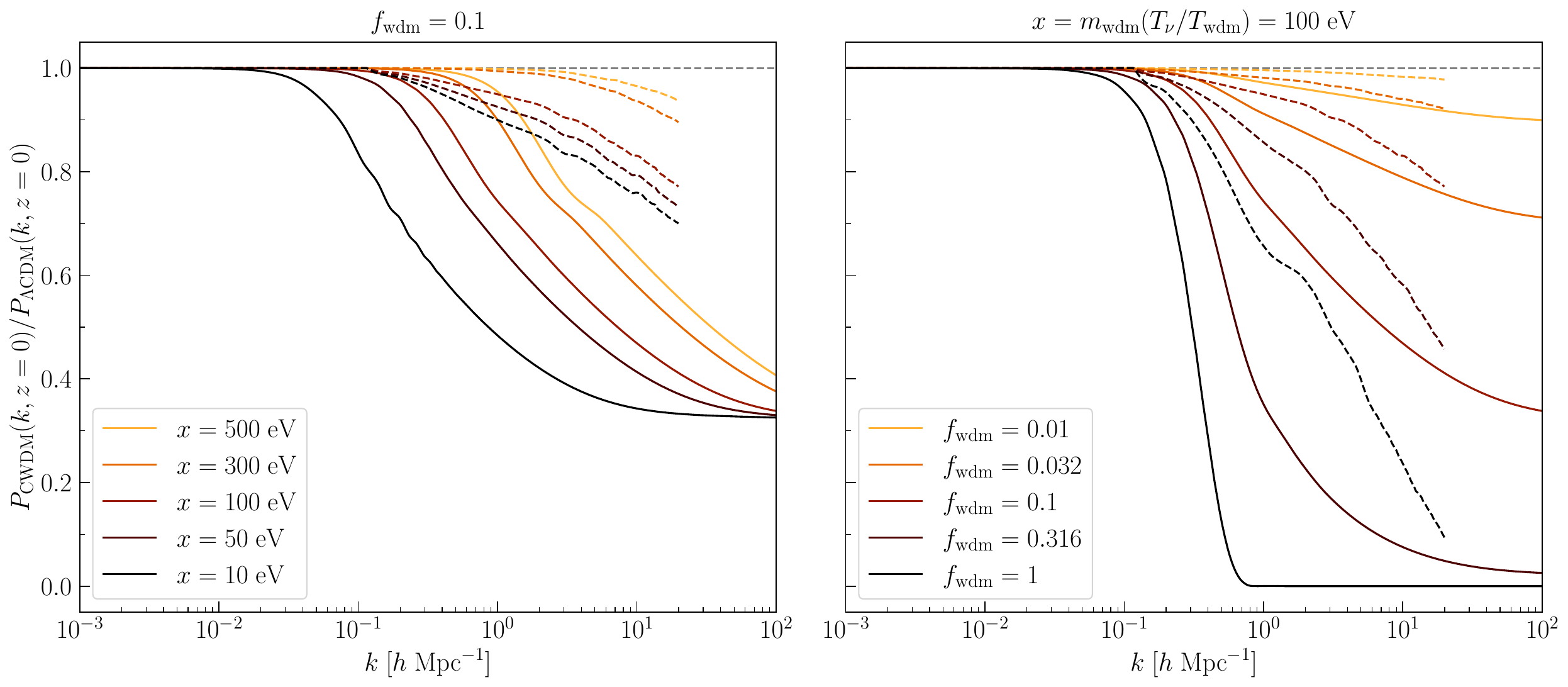}
    \caption{Ratio of the linear (solid lines) and non-linear (dashed lines) power spectra of several CWDM models to that of a pure $\Lambda$CDM model with the same cosmological parameters, parameterised by the fraction $f_{\rm wdm}$ and the rescaled mass $x$. The other parameters ($\Omega_{{\rm dm}}$, $\Omega_{{\rm b}}$, $h$, $A_{\rm s}$, $n_{\rm s}$) are kept fixed. All spectra are computed today ($z=0$). These plots cover all the cases in which WDM has a Fermi--Dirac distribution possibly rescaled by a factor $\chi$, including the limits of the thermal WDM ($\chi=1$) and Dodelson--Widrow ($T_{\rm wdm}=T_\nu$) models. In the latter case $x$ coincides with the physical mass. The non-linear spectra are predicted by the emulator introduced in Sect.\,\ref{sec:nl_cwdm} and plotted up to the maximum wavenumber at which this emulator is trusted.
    \label{fig:lin_cwdm}
    }
\end{figure*}

In this model, the evolution of linear cosmological perturbation can be computed with the public version of \class{}. Then, to account for the thermal warm dark matter case, we pass to the code the parameters 
$\Omega_{\rm wdm} h^2 = f_{\rm wdm}\, \Omega_{\rm dm} h^2$,
$m_{\rm wdm}^{\rm thermal}$, and finally
${T_{\rm wdm}}/{T_\gamma} = (4/11)^{1/3} (m_{\rm wdm}^{\rm thermal}/x)$ with $x$ inferred from Eq.~\eqref{eq:m_wdm_thermal}.\footnote{Here we use \class{} {\tt v3.2.0}. In practise, we fix the number of non-cold dark matter species to one, 
$\mathtt{N\_ncdm}=1$,
and we pass to the code
$\mathtt{omega\_ncdm}= f_{\rm wdm}\, \Omega_{\rm dm} h^2$, 
$\mathtt{m\_ncdm}=m_{\rm wdm}^{\rm thermal}$, and
\begin{equation}
\mathtt{T\_ncdm}= \frac{T_{\rm wdm}}{T_\gamma} = \left(\frac{4}{11}\right)^{1/3} \!\! \left( 94.1 \, \Omega_{\rm wdm} h^2 \right)^{1/3} \left(\frac{m_{\rm wdm}^{\rm thermal}}{1\,{\rm eV}}\right)^{-1/3}\,. \nonumber
\end{equation}
}
In principle one could use a different set of parameters for the equivalent DW model and find the exact same linear power spectra \citep{Lesgourgues:2011re,Blas:2011rf,Lesgourgues:2011rh}. Figure\,\ref{fig:lin_cwdm} shows the power spectrum at redshift zero for several CWDM models rescaled by that of a pure $\Lambda$CDM model, for various values of ($f_{\rm wdm}, x$) but fixed values of the usual $\Lambda$CDM parameters ($\Omega_{{\rm m}}$, $\Omega_{{\rm b}}$, $h$, $A_{\rm s}$, $n_{\rm s}$) accounting respectively for the fractional density of total non-relativistic matter (baryonic plus dark),  the fractional density of baryonic matter, the reduced Hubble parameter, and the amplitude and spectral index of the primordial spectrum of scalar (curvature) perturbations. In the left panel we vary $x$ (or equivalently $m_{\rm wdm}^{\rm DW}$) with fixed $f_{\rm wdm}$ to check that $x$ only controls the location of the step. In the right panel we do the opposite to show that $f_{\rm wdm}$ controls its amplitude.

In Sect.\,\ref{sec:nl_cwdm}, we will show how to compute the impact of CWDM on the non-linear matter spectrum. In Sect.\,\ref{sec:res_cwdm}, we will perform \Euclid{} forecasts on the parameter of the CWDM model. For that purpose, we will use a Bayesian MCMC approach to fit the CWDM model to mock \Euclid{} data, assuming a logarithmic prior on $f_{\rm wdm} \in [2\times 10^{-3},\,1]$ and a flat prior on $m_{\rm wdm}^{\rm thermal} \in [10 \, {\rm eV},\,1\,{\rm keV}]$. 

Such a logarithmic prior on $f_{\rm wdm}$ will allow us to assess precisely the constraining power of \Euclid{} even when $f_{\rm wdm}$ is very small (e.g., in the range from $10^{-3}$ to $10^{-1}$). This limit is the most interesting in the case of the \Euclid{} probes since, in this case, the data may be compatible with a relatively small WDM mass, and thus a small step located on relatively large wavelengths, in the range probed by WL and GC surveys in the linear and mildly non-linear regime. Large values of $f_{\rm wdm}$ (e.g., in the range from 0.1 to 1) imply a strong suppression of the power spectrum that is already excluded by Lyman-$\alpha$ forest data unless the mass is really large -- a limit in which, from the point of view of \Euclid{} data, CWDM would be indistinguishable from pure CDM.  
%Another advantage of using a logarithmic prior on $f_{\rm wdm}$ is to remove the limit $f_{\rm wdm}=0$ from the parameter space. In this limit, any value of $x$ or $m_{\rm wdm}$ should be allowed, which would compromise the convergence of MCMC chains. 

\subsection{Dark matter with one-body decay \label{sec:theo_1bddm}}

If DM particles are unstable, they may decay in different ways into lighter particles. Cosmological observables are not sensitive to all details concerning the nature of the decay products, but they depend on simple considerations like the fact that these decay products could be relativistic or non-relativistic. In the simplest scenario, all decay products are assumed to be ultra-relativistic and can be considered as a single species, dubbed dark radiation (DR). This simple model of decaying dark matter (DDM) is often called one-body decaying DM and abbreviated as 1b-DDM.

In this section, we assume that DM is made up of two cold species: a fraction $1-f_{\rm ddm}$ of stable dark matter (CDM) and a fraction $f_{\rm ddm}$ of 1b-DDM decaying into DR. For simplicity, we assume a constant decay rate $\Gamma_{\rm ddm}=1/\tau_{\rm ddm}$, where $\tau_{\rm ddm}$ is the lifetime of the decaying species. The current value of the fractional dark radiation density, $\Omega_{\rm dr}$, is not an independent parameter of the model: it can be computed consistently for each value of $f_{\rm ddm}$ and $\Gamma_{\rm ddm}$. 

This model has been studied previously, for instance, in \cite{Ichiki:2004vi}, \cite{Audren:2014bca}, \cite{Berezhiani:2015yta}, \cite{Chudaykin:2016yfk}, \cite{Oldengott:2016yjc}, \cite{Poulin:2016nat}, \cite{Chudaykin:2017ptd}, \cite{Pandey:2019plg}, \cite{Xiao:2019ccl}, \cite{Nygaard:2020sow}, \cite{Schoneberg:2021qvd}, \cite{Simon:2022ftd}, \cite{Holm:2022kkd}, or \cite{bucko_2022_1bddm}. It has been often invoked as a possible solution to the Hubble and/or $S_8$ tension. The best constraints at the moment come from CMB plus baryon acoustic oscillation (BAO) data \citep{Nygaard:2020sow}, galaxy surveys \citep{Simon:2022ftd}, and WL surveys \citep{bucko_2022_1bddm}, see Sect.\,\ref{sec:res_1bddm}.

\begin{figure*}[t]
    \centering
    \includegraphics[width=.99\textwidth]{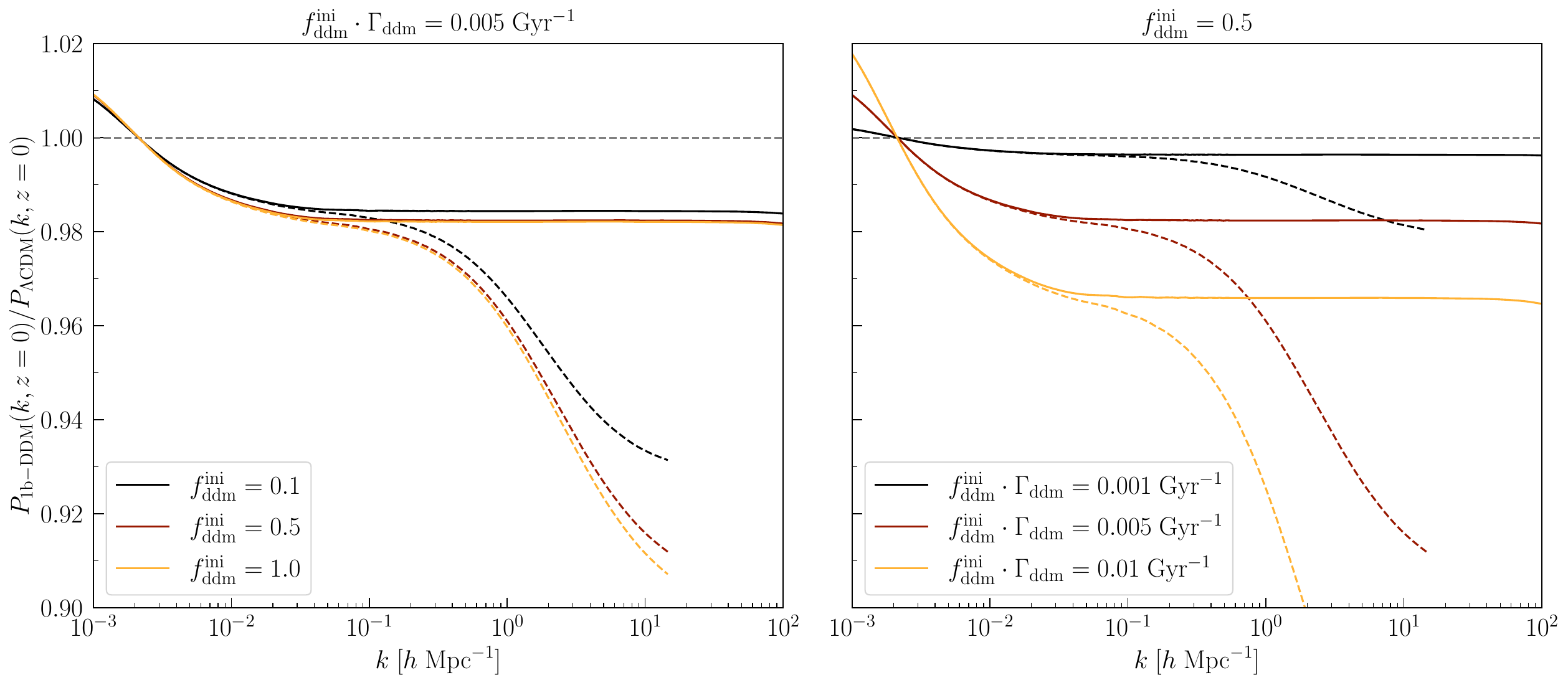}
    \caption{
    Ratio of the linear (solid lines) and non-linear (dashed lines) power spectra of several 1b-DDM models to that of a pure $\Lambda$CDM model with the same cosmological parameters, parameterised by the fraction $f_{\rm ddm}^{\rm ini}$ and the decay rate $\Gamma_{\rm ddm}$. We work in the basis $(f_{\rm ddm}^{\rm ini},\Gamma_{\rm ddm} \, f_{\rm ddm}^{\rm ini})$ to show that only the product of the two DDM parameters affects the linear power spectrum. The other parameters ($\Omega_{\rm dm}^{\rm ini}$, $\Omega_{{\rm b}}$, $h$, $A_{\rm s}$, $n_{\rm s}$) are kept fixed, and the spectra are computed today ($z=0$). The non-linear spectra are predicted by the emulator introduced in Sect.\,\ref{sec:nl_1bddm} and plotted up to the maximum wavenumber at which this emulator is trusted.
    \label{fig:lin_1b}
    }
\end{figure*}

In this model, the evolution of linear cosmological perturbations can be computed with the public version of \class.\footnote{Here we use \class{} {\tt v3.2.0}.} The code accepts two possible definitions of the decaying DM fraction: one can either pass the value of $f_{\rm ddm}$ today, taking the effect of decay into account, or the value $f_{\rm ddm}^{\rm ini}$ evaluated at some initial time $\tau^{\rm ini} \ll \tau_{\rm ddm}$, before any significant decay has occurred,
\begin{equation}
    f_{\rm ddm}^{\rm ini} = \frac{\rho_{\rm ddm}^{\rm ini}}{\rho_{\rm dm}^{\rm ini}}
    = \frac{\rho_{\rm ddm}^{\rm ini}}{\rho_{\rm cdm}^{\rm ini} + \rho_{\rm ddm}^{\rm ini}}\, .
    \label{eq:def_f_ddm_ini}
\end{equation}
Here we choose to report results on $f_{\rm ddm}^{\rm ini}$, for the purpose of easier comparison with previously published bounds. Some related parameters are $\Omega_{\rm ddm}^{\rm ini}$ (respectively $\Omega_{\rm dm}^{\rm ini}$), the fractional density that DDM (respectively total DM) would have today if DDM did not decay. The free parameters of the 1b-DDM model are then ($\Gamma_{\rm ddm}$, $f_{\rm ddm}^{\rm ini}$, $\Omega_{\rm dm}^{\rm ini}$, $\Omega_{b}$, $h$, $A_{\rm s}$, $n_{\rm s}$), while the cosmological constant parameter $\Omega_{\Lambda}$ is adjusted to match the budget equation in a flat universe.\footnote{To be precise, for each 1b-DDM model, we pass to the \class{} code the decay rate expressed in units of ${\rm km}\,{\rm s}^{-1}\,{\rm Mpc}^{-1}$, $\mathtt{Gamma\_dcdm}=977.792 \, (\Gamma_{\rm ddm}/1\,{\rm Gyr}^{-1}) \,{\rm km}\,{\rm s}^{-1}\,{\rm Mpc}^{-1}$, the DDM density parameter  $\mathtt{Omega\_ini\_dcdm}=f_{\rm ddm}^{\rm ini} \, \Omega_{\rm dm}^{\rm ini}$, the CDM density parameter $\mathtt{Omega\_cdm}=(1-f_{\rm ddm}^{\rm ini}) \, \Omega_{\rm dm}^{\rm ini}$, and the remaining four parameters following the usual syntax.}

If one varies the two decaying DM parameters ($\Gamma_{\rm ddm}$, $f_{\rm ddm}^{\rm ini}$) while fixing the other parameters ($\Omega_{\rm dm}^{\rm ini}$, $\Omega_{{\rm b}}$, $h$, $A_{\rm s}$, $n_{\rm s}$), one changes the predicted age of the Universe, which controls the amplitude of the matter power spectrum on all scales, as well as the redshift of radiation-to-matter equality, which determines the scale of the overall peak in the spectrum. These effects cause an enhancement of the matter power spectrum on scales larger than those crossing the Hubble radius around the time of equality, corresponding to comoving wavenumbers $k <  2$--$3 \times 10^{-3}\,h\,{\rm Mpc}^{-1}$, and a suppression on smaller scales. For wavenumbers $k \geq 10^{-1}\,h\,{\rm Mpc}^{-1}$, the power spectrum is suppressed by a constant factor with respect to the $\Lambda$CDM case. A larger fraction $f_{\rm ddm}^{\rm ini}$ or a higher rate $\Gamma_{\rm ddm}$ both imply a smaller amplitude of the power spectrum on these scales. Actually, the suppression factor is found to depend essentially on the product  $\Gamma_{\rm ddm} \, f_{\rm ddm}^{\rm ini}$, as illustrated in Fig.\,\ref{fig:lin_1b}. In the left panel, we vary $f_{\rm ddm}^{\rm ini}$ while keeping the product $\Gamma_{\rm ddm} \, f_{\rm ddm}^{\rm ini}$ fixed to $0.005\,{\rm Gyr}^{-1}$ (a value representative of the constraints found in the result Sect.\,\ref{sec:res_1bddm}). Then, the power spectrum of the 1b-DDM model is found to be independent of $f_{\rm ddm}^{\rm ini}$ up to the order of one per mille. Thus, we anticipate that the parameter $f_{\rm ddm}^{\rm ini}$ alone is difficult to constrain with \Euclid{} data. Instead, in the right panel of Fig.\,\ref{fig:lin_1b}, we vary the product $\Gamma_{\rm ddm} \, f_{\rm ddm}^{\rm ini}$ while keeping $f_{\rm ddm}^{\rm ini}$ fixed. We clearly see a change in the suppression factor for $k \geq 10^{-1}\,h\,{\rm Mpc}^{-1}$ and in the slope of the power spectrum for $k \sim 10^{-2}\,h\,{\rm Mpc}^{-1}$, potentially detectable using \Euclid{} probes.

In Sect.\,\ref{sec:nl_1bddm}, we compute the impact of the 1b-DDM model on the non-linear matter spectrum. In Sect.\,\ref{sec:res_1bddm}, we fit the 1b-DDM model to mock \Euclid{} data. In order to obtain fast-converging MCMC chains, we adopt some flat priors on $f_{\rm ddm}^{\rm ini}$ and $\Gamma_{\rm ddm} \, f_{\rm ddm}^{\rm ini}$, with prior edges detailed in Sect.\,\ref{sec:res_1bddm}, but we expect interesting constraints only on the second parameter.

\subsection{Dark matter with two-body decay\label{sec:theo_2bddm}}

In the next-to-simplest cosmological model of DDM, a cold DDM particle with a large mass $m_{\rm ddm}$ and a constant decay rate $\Gamma_{\rm ddm}$ is assumed to decay into a first massless daughter particle and a second massive daughter particle with mass $m_{\rm daughter}$. This model is dubbed two-body decaying DM (2b-DDM). The parent particle is assumed to account for a fraction $f_{\rm ddm}^{\rm ini}$ of the initial CDM budget, defined in the same way as for one-body decay -- see Eq.\,\eqref{eq:def_f_ddm_ini}, with the remaining fraction $1-f_{\rm ddm}^{\rm ini}$ corresponding to ordinary stable CDM. In each decay, the fraction of energy transferred from the parent particle to the first massless daughter particle, $\varepsilon$, can be related to the mass ratio:
\begin{equation}
    \varepsilon = \frac{1}{2} \left(1-\frac{m^2_{\rm daughter}}{m^2_{\rm ddm}}\right)\, .  
\end{equation}
In the limit $m_{\rm daughter} \rightarrow m_{\rm ddm}$, all the energy goes into the second massive daughter, but since this corresponds to the conversion of one CDM particle into another one, the model is indistinguishable from the standard $\Lambda$CDM model. In the opposite limit $m_{\rm daughter} \rightarrow 0$, the two daughter particles are ultra-relativistic and share the same amount of energy, which corresponds to $\varepsilon=0.5$: this limit is equivalent to the 1-body decay model introduced in the previous section. However, in the more interesting range $0<\varepsilon<0.5$, the second daughter particle can behave as WDM. \cite{Aoyama:2014tga} have shown that for the purpose of computing cosmological observables one only needs to specify the three parameters ($f_{\rm ddm}$, $\Gamma_{\rm ddm}$, $\varepsilon$) on top of the usual $\Lambda$CDM parameters.

This model has been studied previously, for instance, in \cite{Aoyama:2014tga}, \cite{Vattis:2019efj}, \cite{Haridasu:2020xaa}, \cite{FrancoAbellan:2020xnr}, \cite{FrancoAbellan:2021sxk}, \cite{Schoneberg:2021qvd}, \cite{Simon:2022ftd}, or \cite{Bucko:2023twobody}. It has also been invoked as a possible solution to the $H_0$ and/or $S_8$ tension. The best constraints at the moment come from CMB plus BAO data \citep{Schoneberg:2021qvd}, galaxy surveys \citep{Simon:2022ftd}, and WL surveys \citep{Bucko:2023twobody}, see Sect.\,\ref{sec:res_2bddm}.

\begin{figure*}[t]
    \centering
    \includegraphics[width=.99\textwidth]{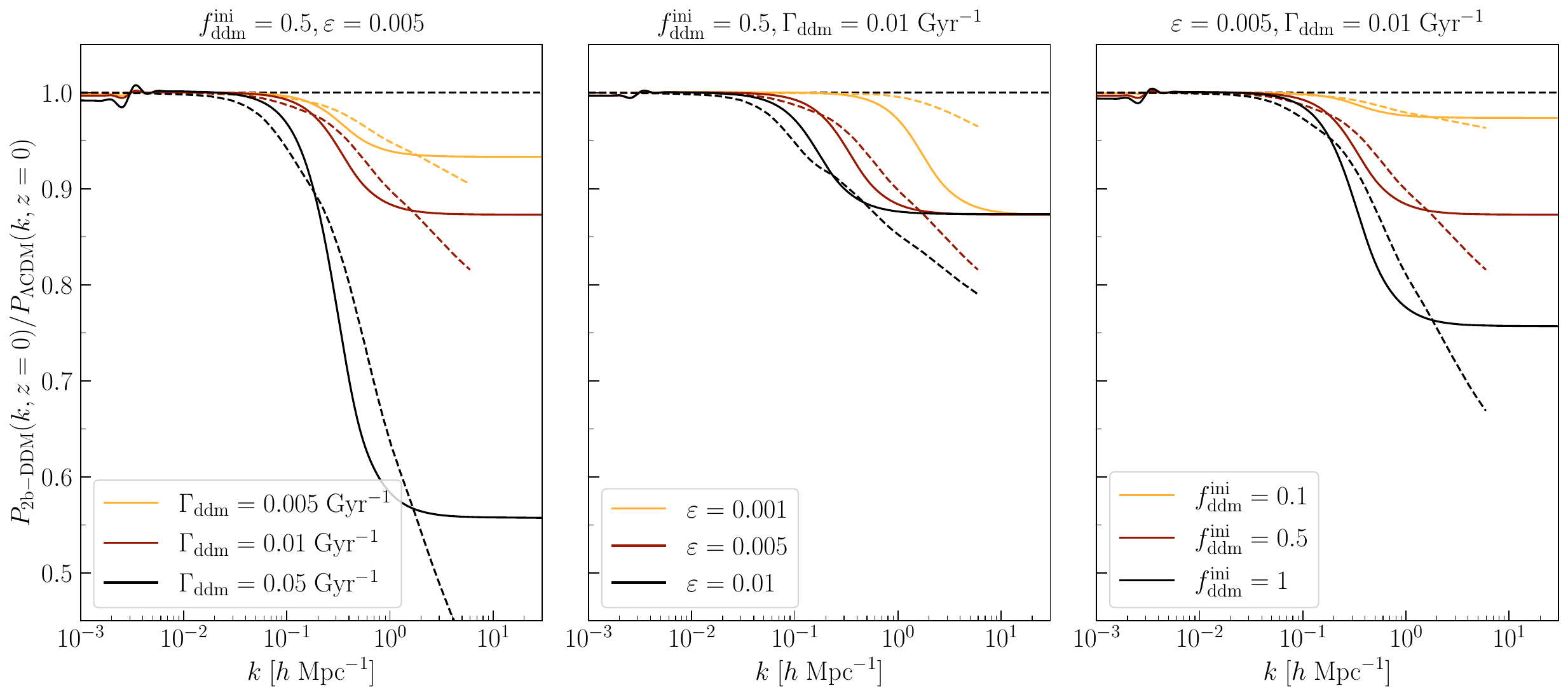}
    \caption{Ratio of the linear (solid lines) and non-linear (dashed lines) power spectra of several two-body DDM models to that of a pure $\Lambda$CDM model with the same cosmological parameters, parameterised by the fraction $f_{\rm ddm}^{\rm ini}$, the decay rate $\Gamma_{\rm ddm}$, and the fraction of energy $\varepsilon$ going into the ultra-relativistic daughter particle at each decay. The parameters ($\Omega_{\rm dm}^{\rm ini}$, $\Omega_{{\rm b}}$, $h$, $A_{\rm s}$, $n_{\rm s}$) are kept fixed, and the spectra are computed today ($z=0$). The non-linear spectra are predicted by the emulator introduced in Sect.\,\ref{sec:nl_2bddm} and plotted up to the maximum wavenumber at which this emulator is trusted.
    \label{fig:lin_2b}
    }
\end{figure*}

At the level of linear perturbation theory, this model is implemented in a branch of \class{} developed and publicly released\footnote{We use the branch called ${\tt merging\_with\_master}$ of the GitHub repository \url{https://github.com/PoulinV/class_decays}. This branch is an extension of \class{} {\tt v2.7.1.}} by the authors of \cite{FrancoAbellan:2021sxk,FrancoAbellan:2020xnr}. Figure~\ref{fig:lin_2b} shows the effect on the linear power spectrum of a variation of the parameters ($\Gamma_{\rm ddm}$, $\varepsilon$, $f_{\rm ddm}^{\rm ini}$) for fixed $\Lambda$CDM parameters. We see that this model leads to a step-like suppression of the matter power spectrum, which is not surprising since, in this case, DM is split between a CDM and a WDM component. The shape of the step is however different from the CWDM case, because the warm component gets produced progressively and affects different scales at different times. Figure~\ref{fig:lin_2b} focuses on cases with $\varepsilon \ll 0.5$ for which, in each decay, most of the energy is transferred from one non-relativistic dark matter species to another one. Thus, while the universe expands, the energy density of total matter evolves almost like in the case of stable DM, $\rho_{\rm m} \propto a^{-3}$, and the age of the universe is not significantly affected by the DDM parameters. This explains why in Fig.~\ref{fig:lin_2b} we do not see any effect of the 2b-DDM parameters on the matter power spectrum on very large scales ($k \ll 10^{-1}\,h\,{\rm Mpc}^{-1}$), as it was the case for 1b-DDM. As a side note, one can observe tiny oscillations in the linear power spectrum ratios of Figure~\ref{fig:lin_2b}. This is most likely a numerical artefact caused by the use of a fluid approximation for the perturbations of the warm species within a fixed range of scales. The same figure shows that these spurious oscillations are smoothed out by the emulator introduced in Sect.\,\ref{sec:nl_2bddm}. Thus, they cannot affect our results.\footnote{Even without such smoothing, these oscillations would be innocuous since they only occur on huge scales (larger than the scale of the broad peak of the matter power spectrum, with $k \ll 10^{-2} \, h \, {\rm Mpc}^{-1}$), which are hardly constrained by \Euclid{}.}

The parameter $\varepsilon$ controls the velocity of the daughter particle just after the decay, which reads $v_{\rm k} = c \, \varepsilon / \sqrt{1-2\varepsilon}$ in the centre of mass frame (the subscript k refers to `kick', since the daughter particles get a velocity kick). Thus, by analogy with WDM, $\varepsilon$ determines the free-streaming scale and the location of the step in the power spectrum. The parameters  ($\Gamma_{\rm ddm}$, $f_{\rm ddm}^{\rm ini}$) both control the abundance of 2b-DDM as a function of time and thus the linear growth rate of the total DM density fluctuation $\delta_{\rm dm}(a)$. Hence these parameters both control the amplitude of the step. The $\Lambda$CDM limit is recovered for $\varepsilon=0$ and/or $f_{\rm ddm}^{\rm ini}=0$ and/or $\Gamma_{\rm ddm}=0$.

In Sect.\,\ref{sec:nl_2bddm}, we show how to compute the impact of the 2b-DDM model on the non-linear matter spectrum. In Sect.\,\ref{sec:res_2bddm}, we fit the 2b-DDM model to mock \Euclid{} data. We perform our sensitivity forecast with flat priors on $\{\log_{10} f_{\rm ddm}^{\rm ini}, \log_{10} (\Gamma_{\rm ddm} / {\rm Gyr}^{-1}), \log_{10} \varepsilon \}$, with prior edges detailed in that section.

\subsection{ETHOS $n=0$
\label{sec:theo_ethos}}

The ETHOS framework \citep{Cyr-Racine:2015ihg} is a general attempt to parameterise physically plausible interactions in a dark sector featuring at least one type of non-relativistic relics (playing the role of cold interacting dark matter, IDM) and one type of ultra-relativistic relics (playing the role of interacting dark radiation, IDR). The theory provides a mapping between phenomenological parameters describing the relevant interaction rates and fundamental parameters appearing in the Lagrangian of the dark sector. In particular, the ETHOS index $n$ describes to the power-law dependence of the IDR-IDM interaction rate $\Gamma_{\rm idr\mhyphen idm}$ on the temperature of the dark sector. 

The case $n=0$ is of particular interest, because it corresponds to an IDM-IDR momentum exchange rate $\Gamma_{\rm idm-idr}$ scaling like the Hubble radius during radiation domination \citep{Buen-Abad:2015ova,Cyr-Racine:2015ihg,Becker:2020hzj}. Thus, in this model, the ratio $\Gamma_{\rm idm-idr}/H$ (where both $\Gamma_{\rm idm-idr}$ and $H$ depend on time) remains constant during radiation domination and decreases slowly during matter domination. This means that IDM and IDR can remain in a regime of feeble but steady interactions until equality. The IDR-IDM interactions then become gradually irrelevant at the beginning of matter domination and negligible during the formation of non-linear structures. 

This model can be motivated with some concrete and plausible particle physics set up, such as non-Abelian DM  \citep{Buen-Abad:2015ova}. It is interesting from the point of view of cosmology phenomenology because it introduces a very smooth suppression in the matter power spectrum \citep{Lesgourgues:2015wza,Buen-Abad:2017gxg} -- instead of oscillatory patterns or an exponential cut-off as would be the case for ETHOS models with $n > 0$. The power spectrum suppression shape is also very different from the one caused by a hot or warm DM component. This model is often invoked as a solution to the $S_8$ tension \citep{Lesgourgues:2015wza,Buen-Abad:2017gxg} -- or even to the Hubble tension, but this is no longer the case with recent data \citep{Schoneberg:2021qvd}. Current constraints on this model are obtained with CMB data combined with Lyman-$\alpha$ data \citep{Archidiacono:2019wdp,Hooper:2022byl} or with the full-shape power spectrum of the BOSS galaxy redshift survey \citep{Rubira_2023}, see Sect.\,\ref{sec:res_ethos}.

This model can be parameterised in terms of the IDR-IDM scattering amplitude $\Gamma_{\rm idr\mhyphen idm}(z_*)$ at some arbitrary reference redshift $z_*$, of the density of DM (through  $\Omega_{\rm idm} h^2$), and of the density of DR (through $\Omega_{\rm idr} h^2$). Following the rest of the literature, we choose a reference redshift $z_*=10^7$ and express the effective comoving rate of IDR scattering off IDM as 
\begin{equation}
    \Gamma_{\rm idr\mhyphen idm}(z_*) = - \Omega_{\rm idm} h^2 \, c \, a_{\rm dark}\, .
\end{equation}
Assuming IDR with a thermal spectrum and two fermionic degrees of freedom, we can parameterise the IDR density in terms of the IDR-to-photon temperature ratio $T_{\rm idr}/T_\gamma = \xi_{\rm idr}\leq1$, such that $\Omega_{\rm idr} = \frac{7}{8} \xi^4_{\rm idr} \Omega_{\gamma}$. The contribution of IDR to the effective number of neutrinos is then given by $\Delta N_{\rm eff} = (T_{\rm idr}/T_\nu)^4$ with $T_\nu$ defined in the instantaneous neutrino decoupling limit, i.e., $\Delta N_{\rm eff} = (11/4)^{4/3} \xi^4_{\rm idr} \simeq 3.85 \,\xi^4_{\rm idr}$. The ratio $\xi_{\rm idr}$ is a dimensionless parameter. $\Gamma_{\rm idr\mhyphen idm}$ is a rate and $a_{\rm dark}$ is an inverse distance that we express in ${\rm Mpc}^{-1}$ (this definition and choice of units has no other purpose than matching the conventions of the \class{} code and of previous work studying this model).\footnote{Starting from $a_{\rm dark}$ in ${\rm Mpc}^{-1}$, one can obtain the rate $(c\, a_{\rm dark})$ in ${\rm Gyr}^{-1}$ by multiplying with $306\,{\rm Mpc}\,{\rm Gyr}^{-1}$.} Finally, in the ETHOS framework, one needs to specify the self-interaction rate between IDR particles. The non-Abelian DM model and the CMB+Lyman-$\alpha$ constraints of \cite{Lesgourgues:2015wza}, \cite{Buen-Abad:2017gxg}, \cite{Archidiacono:2019wdp}, or \cite{Hooper:2022byl} assumed a strongly self-interacting IDR fluid. One may assume instead free-streaming IDR, and \cite{Rubira_2023} consider the two cases. These two different assumptions are expected to have a small impact on CMB constraints (due to the effect of IDR fluctuations dragging the photons fluctuations before decoupling) but a negligible impact on constraints from large-scale structure (because IDR self-interactions are irrelevant for the growth rate of IDM). Here we stick to the assumption of free-streaming IDR.

\begin{figure*}[t]
    \centering
    \includegraphics[width=.99\textwidth]{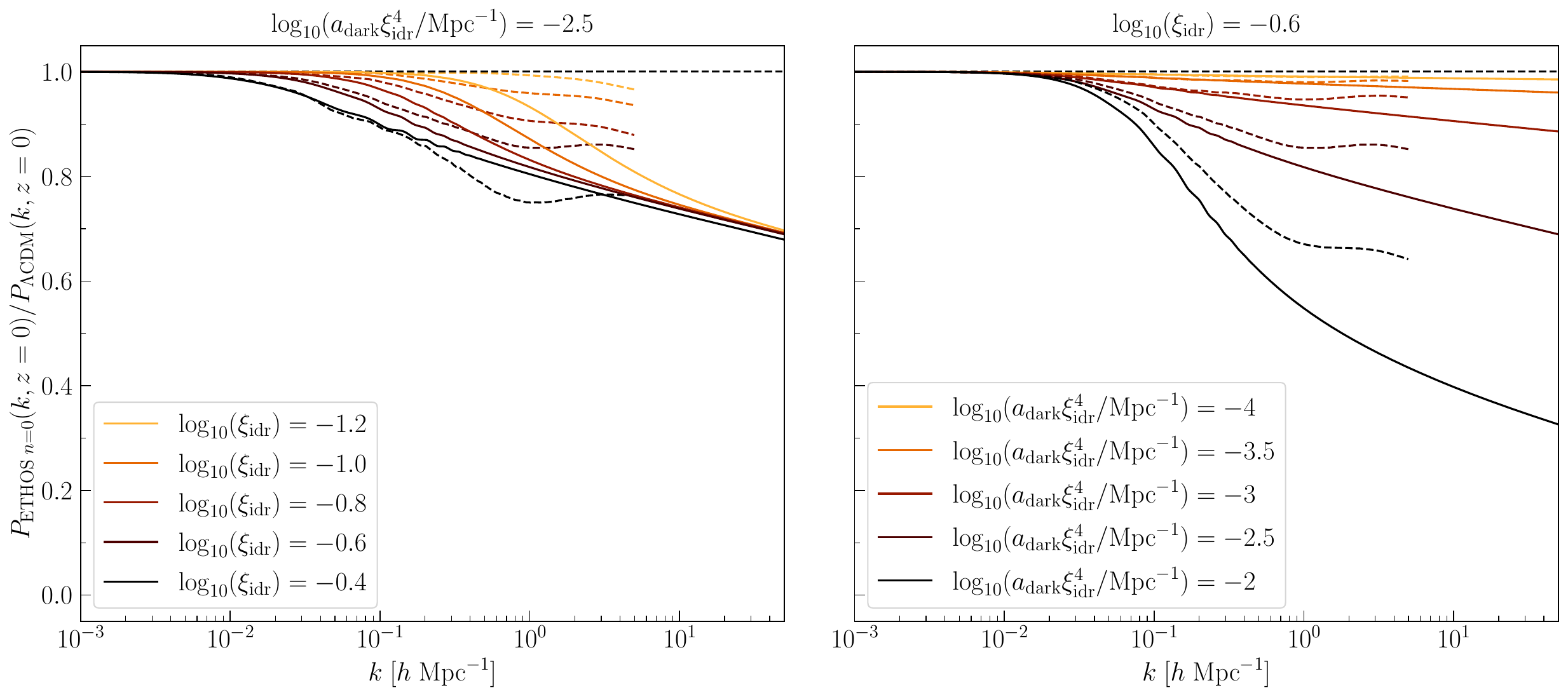}
    \caption{Ratio of the linear (solid lines) and non-linear (dashed lines) power spectra of several free-streaming ETHOS $n=0$ models to that of a pure $\Lambda$CDM model with the same cosmological parameters, parameterised by the dark-radiation-to-photon temperature ratio $\xi_{\rm idr}$ and interaction strength $a_{\rm dark}$. The effects are displayed in the basis ($\xi_{\rm idr}$, $a_{\rm dark} \, \xi_{\rm idr}^4$) to show that the combination $a_{\rm dark} \, \xi_{\rm idr}^4$, which gives the scattering rate of IDR off IDM, controls the amplitude of the small-scale suppression of the linear matter power spectrum. The other parameters ($\Omega_{\rm dm}^{\rm ini}$, $\Omega_{{\rm b}}$, $h$, $A_{\rm s}$, $n_{\rm s}$) are kept fixed, and the spectra are computed today ($z=0$). The non-linear spectra are predicted by the emulator introduced in Sect.\,\ref{sec:nl_ethos} and plotted up to the maximum wavenumber at which this emulator is trusted.
    \label{fig:lin_ethos}
    }
\end{figure*}

The most important physical effect of this model on the matter power spectrum comes from the fact that the IDR-IDM interactions tend to slow down the growth rate of DM fluctuations on sub-Hubble scales during radiation domination, and to suppress the power spectrum on small scales at all subsequent times \citep{Lesgourgues:2015wza,Buen-Abad:2015ova}. Actually, as mentioned in \cite{Archidiacono:2019wdp}, the power spectrum suppression is mainly sensitive to the effective comoving scattering rate of IDR off IDM, which is given by
\begin{equation}
    \Gamma_{\rm idm-idr} = \frac{4 \rho_{\rm idr}}{3 \rho_{\rm idm}} \, \Gamma_{\rm idr\mhyphen idm}\, .
\end{equation}
Since $\rho_{\rm idr}$ is proportional to $\xi_{\rm idr}^4$ while $\rho_{\rm idm}$ is normalised by the measurement of $\Omega_{{\rm idm}}h^2$, this rate is controlled mainly by the parameter combination $a_{\rm dark}\,\xi_{\rm idr}^4$. Therefore, we expect the amplitude of the suppression in the linear matter power spectrum to depend strongly on $a_{\rm dark}\, \xi_{\rm idr}^4$ and weakly on the orthogonal combination, except in the case of sufficiently large $\xi_{\rm idr}^4$, in which the effect of additional radiation with a given $\Delta N_{\rm eff}$ also comes into play. Indeed, an enhancement of $\Delta N_{\rm eff}$ has some well-known effects on the matter and CMB power spectra, explained for instance in \cite{ParticleDataGroup:2022pth_JLLV}, and we expect \Euclid{} to be sensitive to this effect \citep{KP_nu}.

The ETHOS formalism is implemented in the public version of \class{}.\footnote{Here we use \class{} {\tt v3.2.0}, and we set the parameter of the ETHOS sector, described in \cite{Archidiacono:2019wdp}, according to: 
$\mathtt{f\_idm}=1$ to switch on 100\% of IDM;
$\mathtt{nindex\_idm\_dr}=n=0$;
$\mathtt{idr\_nature} = \mathtt{free\_streaming}$;
$\mathtt{a\_idm\_dr}=a_{\rm dark}$ in units of inverse Megaparsecs; and
$\mathtt{xi\_idr}=\xi_{\rm idr}$. Other ETHOS parameters are set to their default value, which means in particular that IDR is assumed to consist of two fermionic degrees of freedom with a statistical factor $\mathtt{stat\_f\_idr} = 0.875$.}
We show in Fig.\,\ref{fig:lin_ethos} the effect of varying the parameters $\xi_{\rm idr}$ or $a_{\rm dark}\,\xi_{\rm idr}^4$ with fixed values of all other cosmological parameters. 

In the left panel, the scattering rate controlled by $a_{\rm dark}\,\xi_{\rm idr}^4$ is fixed, which explains the constant suppression of the linear power spectrum in the large-$k$ limit. When $\log_{10}(\xi_{\rm idr})$ varies from $-1.2$ to $-0.6$, $\Delta N_{\rm eff}$ increases from $6.1\times10^{-6}$ to $0.015$, which is too small to directly affect the matter power spectrum. However, these different values of $\xi_{\rm idr}$ and thus $a_{\rm dark}$ have an impact on intermediate scales: they control the maximum scale at which IDM feels the interaction, and thus the wavenumber at which the matter power spectrum starts to be suppressed. When $\log_{10}(\xi_{\rm idr})$ reaches $-0.4$, the radiation density gets enhanced by a non-negligible amount, $\Delta N_{\rm eff}=0.097$. This results in an additional suppression of the linear power spectrum on small scales.

In the right panel, the amount of IDR is fixed to a small value but the effective scattering rate is increased, leading to more and more suppression. This suppression has a different shape to the case of WDM: it behaves like a transition to a smaller spectral index rather than an exponential cut-off.

In Sect.\,\ref{sec:nl_ethos}, we show how to compute the impact of the ETHOS $n=0$ model on the non-linear matter spectrum. In Sect.\,\ref{sec:res_ethos}, we fit this model to mock \Euclid{} data.
We perform our sensitivity forecast with flat priors on $\{\logaxi$, $\log_{10} \xi_{\rm idr}\}$, with prior edges detailed in that section.

\section{Emulating the non-linear evolution
\label{sec:non-linear}}

To predict observable weak lensing and galaxy correlation spectra, one needs to know the non-linear matter power spectrum for each model. Since $N$-body simulations are computationally too expensive for being run at each point in MCMC chains, it is customary to use a restricted set of $N$-body simulations to build emulators of the non-linear matter power spectrum. These emulators should be accurate compared to the sensitivity of the experiment within the range of model parameters covered by our priors, and fast to evaluate within MCMC runs. In this section, we describe the emulators used in our MCMC forecasts for each of the four non-minimal DM models described in Sect.\,\ref{sec:pdm_models}.

Instead of directly emulating the non-linear power spectrum of the extended cosmological model, $P^{\rm nl}_{\rm model}(k,z)$, it is customary to emulate the ratio 
\begin{equation}
 {\cal S}_{\rm model}(k,z) = \frac{P^{\rm nl}_{\rm model}(k,z)}{P^{\rm nl}_{\Lambda {\rm CDM}}(k,z)}~,  
\end{equation}
and to compute the final observable spectra using
\begin{equation}
    P^{\rm nl}_{\rm model}(k,z) = P^{\rm nl}_{\Lambda {\rm CDM}}(k,z) \,\, {\cal S}_{\rm model}(k,z)~.
    \label{eq:general_S}
\end{equation}
This strategy offers two main advantages. Firstly, it is easier to generate accurate training data for the ratio ${\cal S}_{\rm model}(k,z)$ than for the final spectrum, since several $N$-body simulation artefacts tend to cancel out in the ratio (e.g. resolution effects at small scale, or residual noise from cosmic variance and mesh assignment on large scale). Secondly, the final spectrum depends on all cosmological and dark matter parameters, but the ratio ${\cal S}_{\rm model}(k,z)$ does not in some cases. This ratio depends on course on the DM parameters, but not necessarily on each single parameter of the $\Lambda$CDM model. In each model, one can perform some explicit tests to investigate this dependence and build the emulator on a reduced parameter space.

In this work, we need to decide which tool we should use for  predicting the first factor in Eq.\,(\ref{eq:general_S}), that is, the spectrum $P^{\rm nl}_{\Lambda {\rm CDM}}(k,z)$. In principle, we could use fitting functions like {\tt Halofit} \citep{Smith:2002dz} or \texttt{HMcode 2020} \citep{hmcode2020}, emulators like the {\tt EuclidEmulator2} \citep{Euclid:2020rfv} or {\tt BACCOemulator} \citep{Angulo:2020vky}, etc. In the future, when analysing real data, we will use the best tool available at that time in order to get unbiased results. But for the purpose of the present work, which is to forecast the sensitivity to DM parameters, one could use essentially any of these tools without changing the results on the DM parameter sensitivity, provided that the same tool is used when generating fiducial data and when fitting theoretical predictions. Our choice will be specified in the next sections.

In the context of this work, having accurate predictions for the ratio ${\cal S}_{\rm model}(k,z)$ is more important. With a noisy emulator, one could get slightly wrong predictions for the effect of DM parameters on the final observable spectra, and potentially underestimate degeneracies between these parameters and cosmological or baryonic feedback parameters. In the forecasts presented here, we use emulators designed to achieve per-cent level accuracy up to $k \sim {\cal O}(10) \, h\,{\rm Mpc}^{-1}$ and $z \sim 2.5$ (in the next section we provide further details on the accuracy of each emulator). Given the sensitivity of \Euclid, this is sufficient for robust forecasts. There are some plans to keep training these emulators and improving their accuracy in order to be sure that, when analysing real data, the error coming from the emulator is clearly subdominant in the total systematic error budget.

\subsection{Cold plus warm dark matter
\label{sec:nl_cwdm}}

To predict the non-linear suppression in the matter power spectrum in CWDM scenarios, we use an improved version of the emulator already described in \cite{Parimbelli_CWDM}.
Such an emulator is trained on a large set of $N$-body simulations, covering a large parameter space, for a total of 100 models with different WDM fractions $f_{\rm wdm}$ and WDM masses. The simulations explicitly assume thermal WDM, but this assumption is not relevant in the final analysis: as long as one performs the mass conversion described in Sect.\,\ref{sec:theo_CWDM} before calling the emulator, the latter still applies to all models in which WDM has a Fermi--Dirac distribution possibly rescaled by a factor $\chi$. The simulations cover masses down to $m_{\rm wdm}^{\rm thermal} = 0.03\,{\rm keV}$, but we have checked that the emulator provides a consistent extrapolation down to $m_{\rm wdm}^{\rm thermal} = 0.01\,{\rm keV}$ for small fractions $f_{\rm ddm}^{\rm ini}$ \citep[see][]{Peters_2023}. 
For each model, four realisations are run with fixed amplitudes: two with different random phases and two with the opposite phases.
The box size is set to $120 \, h^{-1}\,{\rm Mpc}$ in order to reconnect with the linear regime at large scales for all redshifts and without any significant discontinuity and to obtain percent-level convergence up to $k\approx 10 \,h\,{\rm Mpc}^{-1}$.
The (fixed) cosmological parameters are $\Omega_{\mathrm {m}} =0.315$, $\Omega_{{\rm b}} =0.049$, $h =0.674$, $n_{\rm s}=0.965$, and a value of $A_{\rm s}$ that would give $\sigma_8=0.811$ in the pure $\Lambda$CDM limit (where $\sigma_8$ is the square root of the variance of matter fluctuations in spheres of radius $8\,h^{-1}\,{\rm Mpc}$).

Initial conditions are set at $z=99$ with a modified version of the \texttt{N-GenIC} code \citep{Ngenic}, using a linear power spectrum obtained from \class{} \citep{Blas:2011rf}.
The simulations are run with the tree-particle mesh (TreePM) code \texttt{GADGET-III} \citep{Ngenic} and follow the gravitational evolution of 512$^3$ particles.
Snapshots are taken starting from $z=3.5$ down to $z=0$, linearly spaced with $\Delta z=0.5$.
Once the power spectra from these snapshots are measured, we take their ratio with respect to the corresponding $\Lambda$CDM spectrum and build the emulator following the exact same procedure as in \cite{Parimbelli_CWDM}. This new tool emulates the first 20 principal components of the power spectrum suppression using Gaussian processes. It is trained on the redshift range $z\in[0-3.5]$ and in the range of scales $k\in [0.07-25] \ h$ Mpc$^{-1}$. The performances are found to be comparable to the ones stated in \cite{Parimbelli_CWDM}, that is, the difference between the emulated and the simulated suppressions never exceeds $\sim 1.5\%$.
All in all, the non-linear matter power spectrum in the presence of CWDM is given by
\begin{equation}
    P^{\rm nl}_{\rm \Lambda CWDM}(k,z) = P^{\rm nl}_{\rm \Lambda CDM}(k,z) \, \mathcal{S}_{\rm CWDM}(k,z) \, ,
\end{equation}
where the last term is precisely what the emulator predicts and $P^{\rm nl}_{\rm \Lambda CDM}(k,z)$ is computed with the version of {\tt Halofit} revisited by \cite{Takahashi:2012em} and \cite{Bird:2011rb}. 

We plot a few examples of predictions for the non-linear spectrum at $z=0$ (compared to the linear predictions of \class{}) in Fig.\,\ref{fig:lin_cwdm}. We can clearly see that the suppression of power induced by the WDM component on small scales is much smaller in the non-linear (rather than linear) power spectrum. This is a well-known effect of mode-mode coupling when perturbations become non-linear. The smaller is the redshift, the less pronounced is the power spectrum suppression on scales smaller than the maximum free-streaming scale.

\begin{figure*}[t]
    \centering
    \includegraphics[width=.52\textwidth]{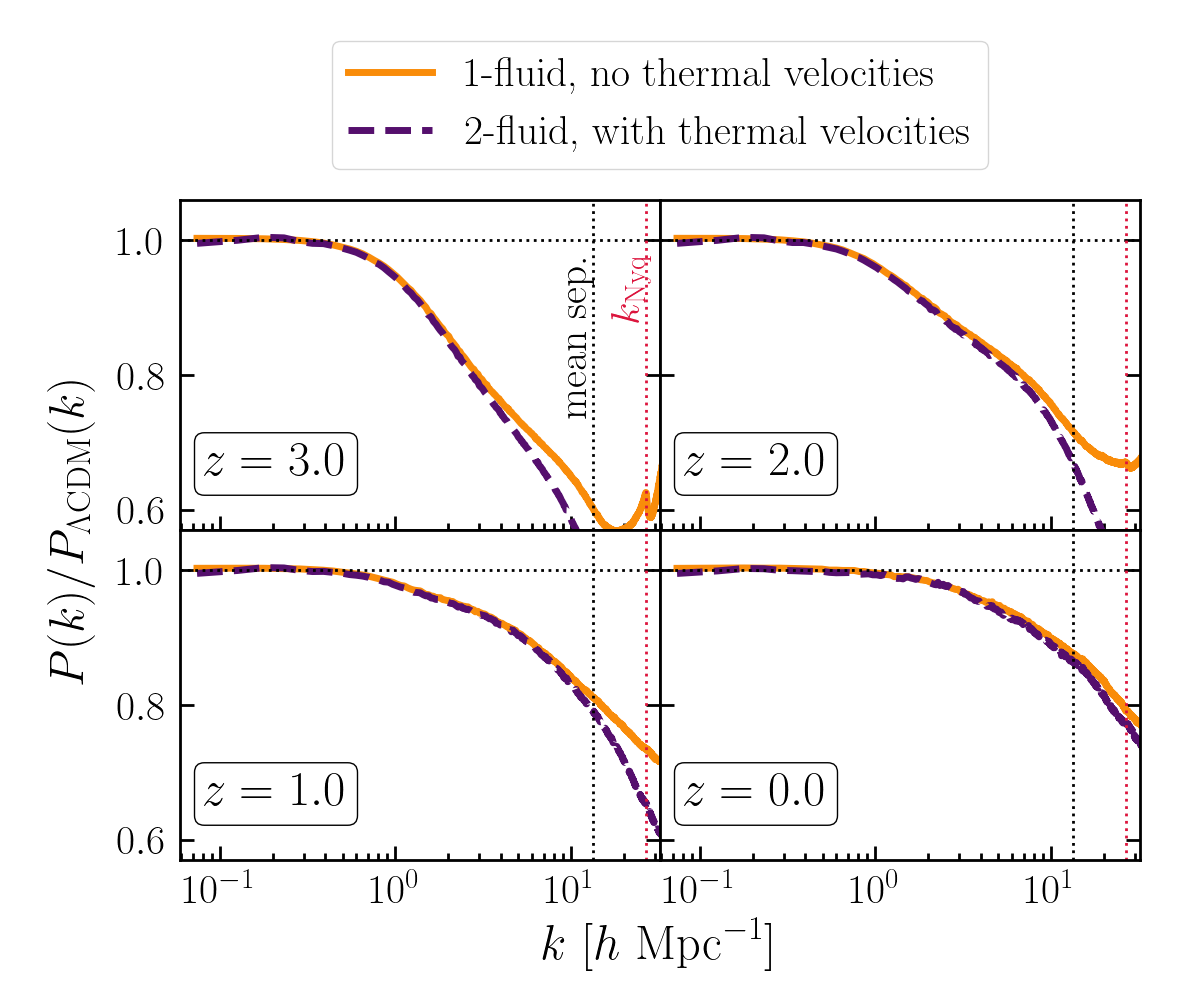}
    \includegraphics[width=.47\textwidth]{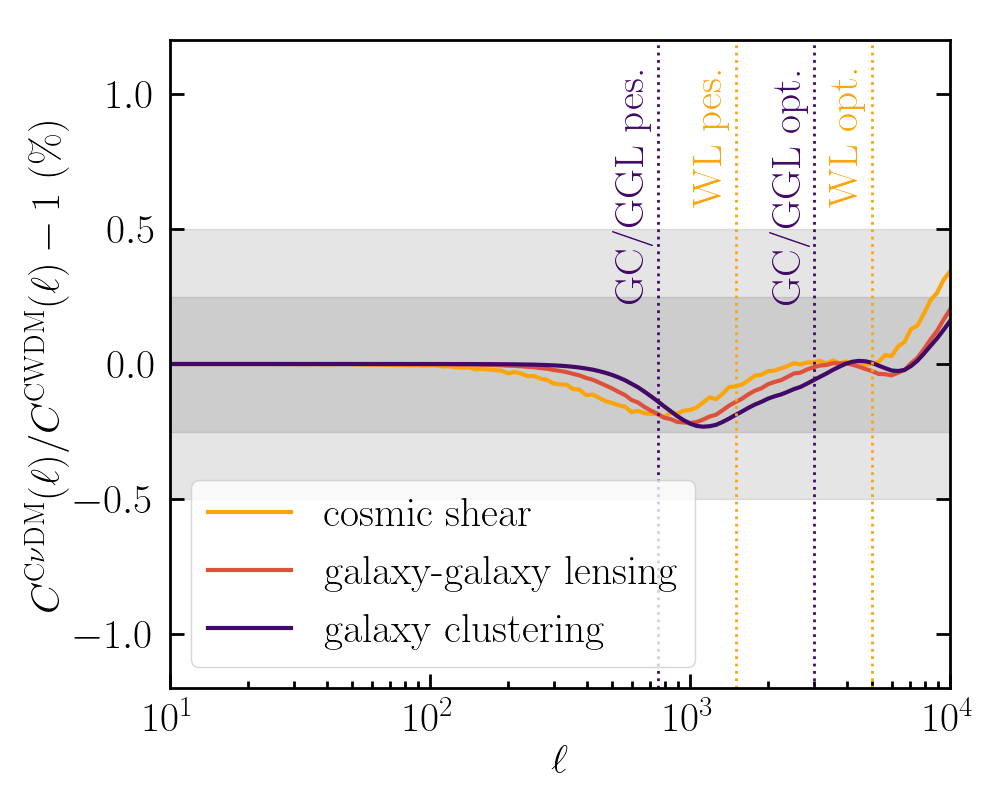}
    \caption{\textit{Left:} effect of neglecting the WDM thermal velocities in CWDM simulations with $f_{\rm wdm}=0.2$ and $m_{\rm wdm}^{\rm thermal}=0.13\,{\rm keV}$. In each panel, solid orange lines represent the suppression in the non-linear matter power spectrum when neglecting WDM thermal velocities; dashed violet lines do the same when implementing WDM as a second fluid in the simulation, with its own thermal velocity field. We plot as vertical lines the mean interparticle separation in blue and the Nyquist frequency in red. \textit{Right:} ratio of angular power spectra $C(\ell)$ for cosmic shear (orange), the cross-correlation of galaxy clustering and galaxy lensing (red), and galaxy clustering (purple), defined in Eq.\,\eqref{eq:cl_integral}, and computed using either the power spectra that neglect or consider thermal velocities. These $C(\ell)$ are computed for simplicity in a single redshift bin ranging from $z=0$ to 3.5 with the galaxy distribution of Eq.\,\eqref{eq:nz}.} We show the 0.25\% and 0.5\% regions as dark and light shaded areas. The maximum $\ell$ value corresponding to the optimistic and pessimistic settings for \Euclid{} are drawn as vertical lines for each probe (cosmic shear or equivalently WL in dotted yellow, GC and cross-correlation in dotted violet).
    \label{fig:thermal_velocities_effect_CWDM}
\end{figure*}

A few considerations about the simulations must be made here. For the sake of computational efficiency, all the particles in all the realisations are initialised as cold particles, even in the runs containing WDM.
This assumption has a twofold implication.
First, we assume that the differences between a CWDM model and $\Lambda$CDM reside in the initial conditions and in their linear power spectra; second, we are neglecting WDM thermal velocities.
We tested the impact of these two assumptions by running a further realisation, with $f_{\rm wdm}=0.2$ and $m_{\rm wdm}^{\rm thermal}=0.13\,{\rm keV}$, in which we initialise 512$^3$ CDM particles as well as $512^3$ more particles as \texttt{Type2}, with the correct thermal velocities.\footnote{Notice that, in \texttt{N-GenIC}, \texttt{Type2} particles are assigned thermal velocities as if they were standard neutrinos with three species degenerate in mass. Therefore, in order to correctly account for thermal velocities of WDM, one needs to rescale the mass to assign to the \texttt{NU\_PartMass\_in\_ev} key in the parameter file. 
The renormalised mass $m_{\rm wdm}^{\rm ren}$ can be computed through \citep{Bode:2000gq,Lesgourgues_Pastor_2006}
\begin{equation}
\frac{m_{\rm wdm}^{\rm ren}}{3} = \frac{150 \, {\rm km/s}}{120 \, {\rm km/s}} \, \left(\frac{\Omega_{\rm wdm}}{0.3}\right)^{-1/3} \! \left(\frac{h}{0.65}\right)^{-2/3} \! \left(\frac{m_{\rm wdm}}{1\,{\rm eV}}\right)^{4/3} \rm{eV} \, . \nonumber
\end{equation}
In our specific case the renormalised mass value is $4290.7\,{\rm eV}$.
}
This value of $f_{\rm wdm}$ has been chosen because, below this fraction, current data are compatible with any value for $m_{\rm wdm}$; the value of the mass has been chosen in order to have a $\sim 50\%$ suppression in the linear power spectrum at $k\sim 5 \,h\,{\rm Mpc}^{-1}$.
We show the results of this test in Fig.\,\ref{fig:thermal_velocities_effect_CWDM}.
In the left plot, we compare the matter power spectrum suppression at various redshifts when neglecting thermal velocities (solid orange lines) and when fully considering them (dashed violet lines).
As can be noted, differences between the two treatments are only relevant at $z\gtrsim 2$ and for $k\gtrsim 5 \,h\,{\rm Mpc}^{-1}$.
The right plot shows instead the ratios between the angular power spectra of cosmic shear (or equivalently WL, orange), the cross-correlation between galaxy clustering and galaxy lensing (red), and GC (purple), computed according to the prescriptions described in Sect.\,\ref{sec:method}, using each of the two sets of power spectra in the left plot.
We use a single bin here for simplicity, ranging from $z=0$ to $z=3.5$, and neglect intrinsic alignment.
Differences are well below percent level; for comparison, at $\ell=10^4$, the \Euclid{} sample variance is expected to be $\sim 1.6\%$.
We can conclude that our assumptions do not introduce any systematic effects in the analysis.

\subsection{Dark matter with one-body decay\label{sec:nl_1bddm}}

We employ the fitting functions found by \cite{Hubert_2021} to model the non-linear matter power spectrum in the presence of one-body decay. These fits are inspired by fitting functions published in \cite{Enqvist:20151bddm} and built upon $N$-body simulations implementing DDM into the \texttt{PKDGRAV3} code \citep{Potter:2017pkdgrav3}. 

We have seen in Sect.\,\ref{sec:theo_1bddm} that  1b-DDM induces a suppression in the linear matter power spectrum that is asymptotically constant on intermediate and small scales, with a suppression factor proportional to $\Gamma_{\rm ddm} \, f_{\rm ddm}^{\rm ini}$, or to $f_{\rm ddm}^{\rm ini} / \tau_{\rm ddm}$. The amplitude and redshift dependence of this suppression factor is given by 
\begin{equation}
\label{eq:eps_lin}
\varepsilon_{{\rm lin}}(z) = \alpha \,\, f_{\rm ddm}^{\rm ini} \,\,  \left(\frac{\rm Gyr}{\tau_{\rm ddm}}\right)^{\beta}\left(\frac{1}{0.105 \, z+1}\right)^{\gamma} \, ,
\end{equation}
where $\alpha$, $\beta$, $\gamma$ are functions of $\omega_{\rm b}:= \Omega_{\rm b} h^2$, $h$, and $\omega_{\rm m} := \Omega_{\rm b} h^2 + \Omega_{\rm  dm} h^2$. We refer to \cite{Hubert_2021} and \cite{bucko_2022_1bddm} for their detailed form. Note that the suppression functions $\varepsilon_{\rm lin}(z)$ and $\varepsilon_{\rm nonlin}(k,z)$ introduced respectively in Eqs.\,(\ref{eq:eps_lin}, \ref{fit}) should not be confused with the parameter $\varepsilon$ of the 2b-DDM model. The non-linear evolution imprints an additional suppression 
that can be inferred from $N$-body simulations. \cite{Enqvist:20151bddm} provided a fit to the non-linear suppression function $\varepsilon_{\rm nonlin}(k,z)$ in the case $f_{\rm ddm}^{\rm ini}=1$ that \cite{Hubert_2021} generalised to arbitrary values of the DDM fraction. The suppression function is estimated from $N$-body simulations for a fixed cosmology. Since only late-time DM decays are of interest, the initial conditions of such $N$-body simulations are identical to those in a $\Lambda$CDM scenario. However, to account for the 1b-DDM, the particle masses are being gradually decreased in the simulation as a function of the rate $\Gamma_{\rm ddm}$, the fraction $f_{\rm ddm}^{\rm ini}$ and the simulation time, mimicking the decay process \citep[for more details, see][]{Hubert_2021}. The suite of $N$-body simulations used to construct the fitting functions was run with a box size of $500 \,h^{-1}\,{\rm Mpc}$ evolving $1024^3$ particles. The cosmological parameters were fixed to fiducial values $\Omega_{{\rm m}} = 0.307$, $\Omega_{{\rm b}} = 0.048$, $10^9A_{\rm s} = 2.43$, $h = 0.678$, and $n_{\rm s} = 0.96$. The convergence of the 1b-DDM $N$-body simulations was studied in \cite{Hubert_2021} with the conclusion that the implementation of the model is trustworthy at least up to $k\simeq 6.4\,h\,{\rm Mpc}^{-1}$. Finally, \cite{Hubert_2021} argue that $\varepsilon_{\rm nonlin}(k,z)$ is nearly cosmology-independent and can be extrapolated to cosmologies well beyond those probed in our work. 

\begin{figure*}[t]
    \centering
    \includegraphics[width = 0.99\textwidth]{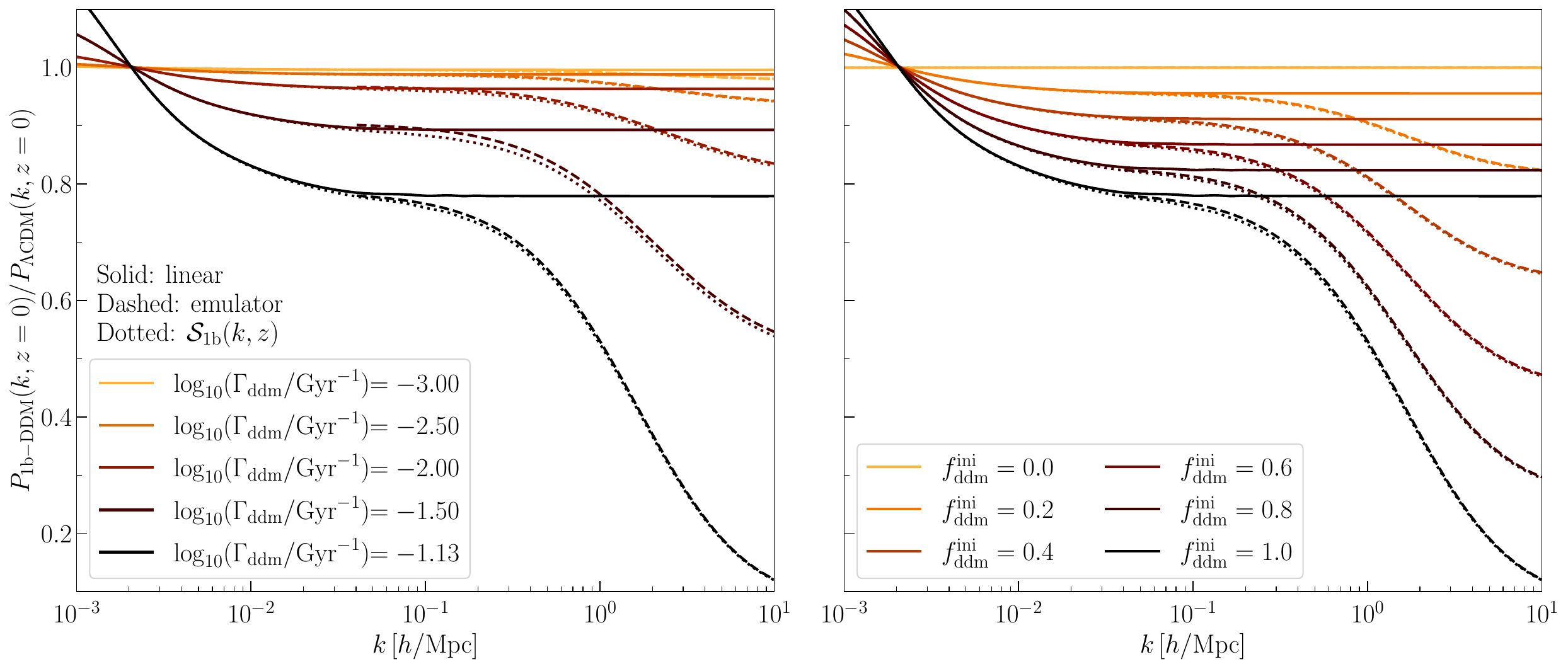}
    \caption{Effect of 1b-DDM parameters on the linear (solid) and non-linear (dashed) matter power spectrum. \emph{Left}: effect of varying the decay rate $\Gamma_{\rm ddm}$ with a fixed fraction $f_{\rm ddm}^{\rm ini}=1$. \emph{Right}: effect of varying the fraction $f_{\rm ddm}^{\rm ini}$ with a fixed decay rate $\Gamma_{\rm ddm} = (1/13.5)\,{\rm Gyr}^{-1}$. The other parameters ($\Omega_{\rm dm}^{\rm ini}$, $\Omega_{{\rm b}}$, $h$, $A_{\rm s}$, $n_{\rm s}$) are kept fixed, and the spectra are computed today ($z=0$). Dashed lines show the predictions of the emulator of \cite{Hubert_2021}. Our pipeline relies on the prescription of Eq.\,\eqref{eq:boost_1bddm}, shown as a dotted line, which smoothly interpolates from the linear to non-linear behaviour. 
    }
    \label{fig:emulator_1bddm}
\end{figure*}

The fitting function provides the suppression of the matter power spectrum with respect to the fiducial $\Lambda$CDM cosmology, $P_{\rm 1b\mhyphen DDM}(k,z)/P_{\rm \Lambda CDM}(k,z) = 1 - \varepsilon_{\rm nonlin}(k,z)$, with
\begin{equation}
\label{fit}
\varepsilon_{\rm nonlin}(k,z) = \frac{1+a_1\left(k/{\rm Mpc}^{-1}\right)^{p_1}}{1+a_2\left(k/{\rm Mpc}^{-1}\right)^{p_2}}\,\, \varepsilon_{{\rm lin}}(z) \, .
\end{equation}
The suppression function interpolates from the linear behaviour on intermediate scales, $\varepsilon_{\rm nonlin}(k,z)\longrightarrow \varepsilon_{{\rm lin}}(z)$, to a power-law suppression on small scales with $\varepsilon_{\rm nonlin}(k,z) \propto k^{p_1-p_2}$. The factors $a_1$, $a_2$, $p_1$, and $p_2$ are given for each lifetime $\tau_{\rm ddm}$ and redshift $z$ by
\begin{align}
a_1 &= 0.7208 + 2.027\left(\frac{{\rm Gyr}}{\tau_{\rm ddm}}\right) + \frac{3.031}{1+1.1\,z} - 0.18 \, ,\nonumber\\
a_2 &= 0.0120 + 2.786\left(\frac{{\rm Gyr}}{\tau_{\rm ddm}}\right) + \frac{0.6699}{1+1.1\,z} - 0.09 \, ,\nonumber\\
p_1 &= 1.045 + 1.225\left(\frac{{\rm Gyr}}{\tau_{\rm ddm}}\right) + \frac{0.2207}{1+1.1\,z} - 0.099 \, ,\nonumber\\
p_2 &= 0.992 + 1.735\left(\frac{{\rm Gyr}}{\tau_{\rm ddm}}\right) + \frac{0.2154}{1+1.1\,z} - 0.056 \,.\label{eq:1bddm_emu_fits}
\end{align}
These fitting functions are publicly available as a part of the {\tt DMemu} package,\footnote{\url{https://github.com/jbucko/DMemu}}
and designed to reproduce the results of $N$-body simulations with a precision better than 1\%\ up to $k=13\,h\,{\rm Mpc}^{-1}$. Note that, at a given redshift, the fitting functions of Eq.\,\eqref{eq:1bddm_emu_fits} depend only on $\tau_{\rm ddm}$ (or $\Gamma_{\rm ddm}$), while  $\varepsilon_{{\rm lin}}$ depends only on $\Gamma_{\rm ddm} \, f_{\rm ddm}^{\rm ini}$. Thus, the non-linear evolution lifts the degeneracy between $\Gamma_{\rm ddm}$ and $f_{\rm ddm}^{\rm ini}$ observed at the level of the linear power spectrum.

In order to match the linear predictions of \class{} on large and intermediate scales with those of the  fitting functions on intermediate and small scales without introducing any discontinuity, we use the following ansatz to calculate the non-linear matter power spectrum of the 1b-DDM model:
\begin{align}
P_{\rm 1b\mhyphen DDM}(k,z) &= 
P_{\rm 1b\mhyphen DDM, lin}(k,z) 
\,  
\frac
{P_{\Lambda{\rm CDM}}(k,z)}
{P_{\Lambda{\rm CDM,lin}}(k,z)}
\nonumber \\ 
& \times 
\frac{1-\varepsilon_{\rm nonlin}(k,z)}{1-\varepsilon_{\rm lin}(z)} \, ,
\label{eq:1bddm_supp}
\end{align}
with the non-linear $\Lambda$CDM spectrum evaluated with the version of {\tt Halofit} revisited by \cite{Takahashi:2019hth} and \cite{Bird:2011rb}. 
Then, firstly, on intermediate (linear) scales, the second and third factor in the right-hand side of Eq.\,\eqref{eq:1bddm_supp} go to one, and one recovers $P_{\rm 1b\mhyphen DDM}(k,z) \longrightarrow P_{\rm 1b\mhyphen DDM, lin}(k,z)$. Secondly, on smaller (non-linear) scales, after noticing that we can rewrite Eq.~\eqref{eq:1bddm_supp} as
\begin{align}
P_{\rm 1b\mhyphen DDM}(k,z) 
&=
\frac{P_{\rm 1b\mhyphen DDM, lin}(k,z)}{1-\varepsilon_{\rm lin}(z)} 
\,  
\frac
{P_{\Lambda{\rm CDM}}(k,z)}
{P_{\Lambda{\rm CDM,lin}}(k,z)}
\nonumber \\ 
& \times
[1-\varepsilon_{\rm nonlin}(k,z)] \, ,
\end{align}
and that the first fraction tends towards
$P_{\rm \Lambda\rm CDM, lin}(k,z)$, we get
\begin{equation}
P_{\rm 1b\mhyphen DDM}(k,z) 
\longrightarrow
P_{\rm \Lambda\rm CDM}(k,z) \, [1-\varepsilon_{\rm nonlin}(k,z)] \, ,
\end{equation}
that is, the approximation to the non-linear 1b-DDM power spectrum provided by the emulator. Equation\,\eqref{eq:1bddm_supp} is designed to provide a smooth transition between these two limits. Note that, according to this ansatz,
the ratio $P_{\rm 1b\mhyphen DDM}(k,z)/P_{\rm \Lambda CDM}(k,z)$ is given by the boost factor
\begin{equation}
\label{eq:boost_1bddm}
    \mathcal{S}_{\rm 1b}(k,z) = \frac{P_{\rm 1b\mhyphen DDM,lin}(k,z)}{P_{\Lambda\rm CDM,lin}(k,z)}\,  \frac{ 1 - \varepsilon_{\rm nonlin}(k,z)}{ 1 - \varepsilon_{\rm lin}(z)} \, .
\end{equation}
%To evaluate the total matter power spectrum in the presence of decays, we apply the above boost to the DM-only $\Lambda$CDM power spectrum using 
%\begin{equation}
%\label{eq:pktot_1bddm}
%    P_{\rm tot}(k,z) = \mathcal{S}_{\rm 1b}(k,z) \, P_{\rm DM}(k,z).
%\end{equation}

We already saw in Fig.\,\ref{fig:lin_1b} the ratio of 1b-DDM-to-$\Lambda$CDM linear power spectra, as well as the ratio of non-linear spectra given by Eq.\,\eqref{eq:boost_1bddm}. Figure\,\ref{fig:emulator_1bddm} is similar to Fig.\,\ref{fig:lin_1b} but shows additionally the raw result of the emulator, i.e., the ratio $P_{\rm 1b\mhyphen DDM}(k,z)/P_{\rm \Lambda CDM}(k,z) \simeq 1 - \varepsilon_{\rm nonlin}(k,z)$ above $k\gtrsim 0.05\,h\,{\rm Mpc}^{-1}$ (dashed lines).
%In Fig.\,\ref{fig:emulator_1bddm} we plot the effects of 1b-DDM on the matter power spectrum at redshift $z=0$, varying the decay rate $\Gamma_{\rm ddm}$ (left panel) and the fraction $f_{\rm ddm}^{\rm ini}$ for fixed
%($\Omega_{\rm dm}^{\rm ini}$, $\Omega_{{\rm b}}$, $h$, $A_{\rm s}$, $n_{\rm s}$). 
%Solid lines show the prediction from \class{} for the ratio $P_{\rm 1b\mhyphen DDM,lin}(k,z=0)/P_{\rm \Lambda CDM,lin}(k,z=0)$ already discussed in Sect.\,\ref{sec:theo_1bddm}. Dashed lines (defined only above $k\gtrsim 0.05$~h/Mpc) correspond to the emulator fit to the ratio of non-linear spectra $P_{\rm 1b\mhyphen DDM}(k,z=0)/P_{\rm \Lambda CDM}(k,z=0)$ \citep{Hubert_2021}. 
The linear prediction (solid lines) and the raw emulator (dashed lines) match each other quite well around $k = 0.05\,h\,{\rm Mpc}^{-1}$, but switching abruptly from one to the other at a given wavenumber would introduce a small discontinuity in the spectrum. Dotted lines show the boost factor defined in Eq.\,\eqref{eq:boost_1bddm} and used in our pipeline. This factor provides a very smooth interpolation from the prediction of \class{} to that of the emulator.

\subsection{Dark matter with two-body decay\label{sec:nl_2bddm}}
\label{sec:2bDDM_emulator}

To model the two-body decays up to non-linear scales, we use the emulator published in \cite{Bucko:2023twobody}, which can provide the 2b-DDM-to-$\Lambda$CDM non-linear power spectrum ratio
\begin{equation}
\mathcal{S}_{\rm 2b}(k,z) = \frac{P_{\rm 2b\mhyphen DDM}(k,z)}{P_{\Lambda{\rm CDM}}(k,z)}
\end{equation}
up to $z\simeq2.3$ and $k\simeq 6\,h\,{\rm Mpc}^{-1}$. The emulator was trained on approximately 100 \texttt{PKDGRAV3} $N$-body simulations directly implementing the late-time DM decays, while starting from $\Lambda$CDM-like initial conditions at $z_{\rm ini} = 49$. \cite{Bucko:2023twobody} set $L_{\rm box} = 125,250,512\,h^{-1}\,{\rm Mpc}$ and $N = 256^3,512^3,1024^3$ depending of each specific DDM configuration, in such way to achieve converged simulations up to $k_{\rm max} = 6\,h\,{\rm Mpc}^{-1}$. \cite{Bucko:2023twobody} argue that the suppression $\mathcal{S}_{\rm 2b}(k,z)$ is approximately independent of cosmology and fix the standard cosmological parameters to $\Omega_{{\rm m}} = 0.307$, $\Omega_{{\rm b}} = 0.048$,
$10^9A_{\rm s} = 2.43$, $h = 0.678$, and $n_{\rm s} = 0.96$ in the simulations. At each simulation time step, a number of DM particles is randomly selected for decay. The decay into a lighter daughter particle is accounted for through a velocity kick with amplitude $v_{\rm k}$ and random direction. In the limit $\varepsilon \ll 0.5$ considered here, $v_{\rm k}$ is approximately given by $c\,\varepsilon$. These kicks lead to suppression in the matter power spectrum below the free-streaming length of the massive daughter particles controlled by $v_{\rm k} \sim c \, \varepsilon$. 

The emulator predicts $\mathcal{S}_{\rm 2b}(k,z)$ using a combination of a `Principal Component Analysis' (PCA) with feed-forward `sinusoidal representation networks' (SIRENs), see \cite{sitzmann:2019siren}. Within the emulation process, the PCA is used to compress the power spectrum ratios $\mathcal{S}_{\rm 2b}(k,z)$, taking into account 5 principal components. Then, the SIREN architecture is trained in a supervised fashion to predict these principal components given the input parameters of 2b-DDM model and the redshift of interest. The loss function of the network is the square distance of the input and output 2b-DDM-to-$\Lambda$CDM ratio, reconstructed from the PCA components predicted by the network. The emulator covers the case of an arbitrary fraction $f_{\rm ddm}^{\rm ini} \in [0,\,1]$ of long-lived DDM particles with $\tau_{\rm ddm} := \Gamma^{-1}_{\rm ddm} \geq 13.5\,{\rm Gyr}$ decaying into non-relativistic daughters with 
$v_{\rm k} \lesssim 5000\,{\rm km}\,{\rm s}^{-1}$, corresponding to $\varepsilon < 0.017$. The emulator can predict ratios of 2b-DDM and $\Lambda$CDM nonlinear matter power spectra up to $z=2.3$ and $k\simeq 6\,h\,{\rm Mpc}^{-1}$, with a precision better than 1\% at the 68\% CL. It is implemented inside the publicly available {\tt DMemu} package introduced after Eq.\,\eqref{eq:1bddm_emu_fits}. We already compared the emulator result to the linear 2b-DDM-to-$\Lambda$CDM linear power spectrum ratio  in Fig.\,\ref{fig:lin_2b}.

Like in other cases, the final non-linear power spectrum of the 2b-DDM model is obtained by mutiplying the non-linear power spectrum of the $\Lambda$CDM model (computed using {\tt Halofit}) with the emulated ratio $\mathcal{S}_{\rm 2b}(k,z)$.

\subsection{ETHOS $n=0$\label{sec:nl_ethos}}

The non-linear matter power spectrum of the ETHOS $n=0$ model is predicted by a dedicated emulator that will be presented in Bucko et al. (2024, in preparation). Like for the 1b-DDM and 2b-DDM cases, this emulator will be released within the {\tt DMemu} package. It assumes the particular case in which IDR consists of two free-streaming fermionic degrees of freedom. It predicts the ETHOS-to-$\Lambda$CDM non-linear power spectrum ratio
\begin{equation}
\mathcal{S}_{{\rm ETHOS}\,n=0}(k,z) = \frac{P_{\rm {\rm ETHOS}\,n=0}(k,z)}{P_{\Lambda{\rm CDM}}(k,z)}
\end{equation}
up to $z=3$ and $k\simeq5\,h\,{\rm Mpc}^{-1}$. The architecture used to train the ETHOS emulator is similar to the one used in the 2b-DDM scenario, described in Sect. \ref{sec:2bDDM_emulator}, with slight modifications. First of all, only 4 PCA components are used to compress the input ETHOS-to-$\Lambda$CDM matter power spectra, while the SIREN architecture involves two dense hidden layers with 256 neurons each. The emulator provides below 1\% errors at the aforementioned scales and redshifts, within the range of ETHOS models defined by the $N$-body simulations discussed in the next paragraphs.

The emulator is built upon a suite of $N$-body simulations which have been run using \texttt{PKDGRAV3} with $L_{\rm box} = 325\,h^{-1}\,{\rm Mpc}$ and $N = 512^3$, assuming a fiducial cosmology with $\omega_{\rm idm} := \Omega_{\rm idm} h^2 = 0.1202$, $\omega_{\rm b} = 0.02236$, $h = 0.6727$, $n_{\rm s} = 0.9649$, and $10^9A_{\rm s} = 2.101$. One massive neutrinos species with $m_\nu = 0.06\,{\rm eV}$ was included. Instead of using a typical back-scaling approach to generate the initial conditions, Bucko et al. (2024, in preparation) follow an alternative method described in \cite{Tram_2019}. The ``true'' initial conditions are generated using the \texttt{C0NCEPT} code \citep{Dakin_2022}. In combination with \class, \texttt{C0NCEPT} also computes the linear evolution of all species (photons, metric, neutrinos, IDR, IDM). This information is used to calculate the gravitational potential at each time step in the \texttt{PKDGRAV3} simulation. In this way, the DM particles of the simulation feel their own gravity, taken into account at the non-linear level, plus the gravity from other species, modelled at the linear level.

The simulations used to train the emulator implement the ETHOS $n=0$ only through modified initial conditions at $z_{\rm ini}=49$. The effect of IDM-IDR scattering at $z<z_{\rm ini}$ is neglected. This assumption is valid only for model parameters such that the scattering rate $\Gamma_{\rm idm-idr}$ is negligible compared to the Hubble rate at $z=z_{\rm ini}$. Since the rate $\Gamma_{\rm idm-idr}$ is computed with respect to conformal time, it should be compared to the conformal Hubble rate ${\cal H}=aH$. \cite{Rubira_2023} provide an analytical approximation for the (redshift-dependent) interaction rate to Hubble rate ratio,
\begin{align}
    \label{eq:ethos_gamma_h}\frac{\Gamma_{\rm idm-idr}}{\mathcal{H}} & \simeq 0.0152\left( \frac{a_{\rm dark}}{1000\, {\rm Mpc^{-1}}} \right)\left(\frac{\xi_{\rm idr}}{0.1} \right)^4 
    \nonumber \\
    &\times
    \frac{(1+z)^2}{\left[\Omega_{{\rm m}}(1+z)^3 + \Omega_{{\gamma}}(1+z)^4(1+\xi^4_{\rm idr}) + \Omega_\Lambda  \right]^{1/2}}\, .
\end{align}
Assuming $\Omega_{{\rm m}}=0.27$ and $\xi^4_{\rm idr}\ll 1$, this gives approximately ${\Gamma_{\rm idm-idr}}/{\mathcal{H}}\simeq a_{\rm dark} \xi_{\rm idr}^4 / (0.48\,{\rm Mpc}^{-1})$ at $z=49$.
Bucko et al. (2024, in preparation) suggest to trust the simulations and the emulator as long as this ratio is smaller than 0.1. In first approximation, this is the case for $a_{\rm dark} \xi_{\rm idr}^4<0.05\,{\rm Mpc}^{-1}$. We will see in Sect.\,\ref{sec:res_ethos} that this region is appropriate to study the sensitivity of \Euclid{} to ETHOS parameters, at least when the fiducial model is assumed to be $\Lambda$CDM (or close to it). 

\begin{figure*}[t]
    \centering
    \includegraphics[width = \textwidth]{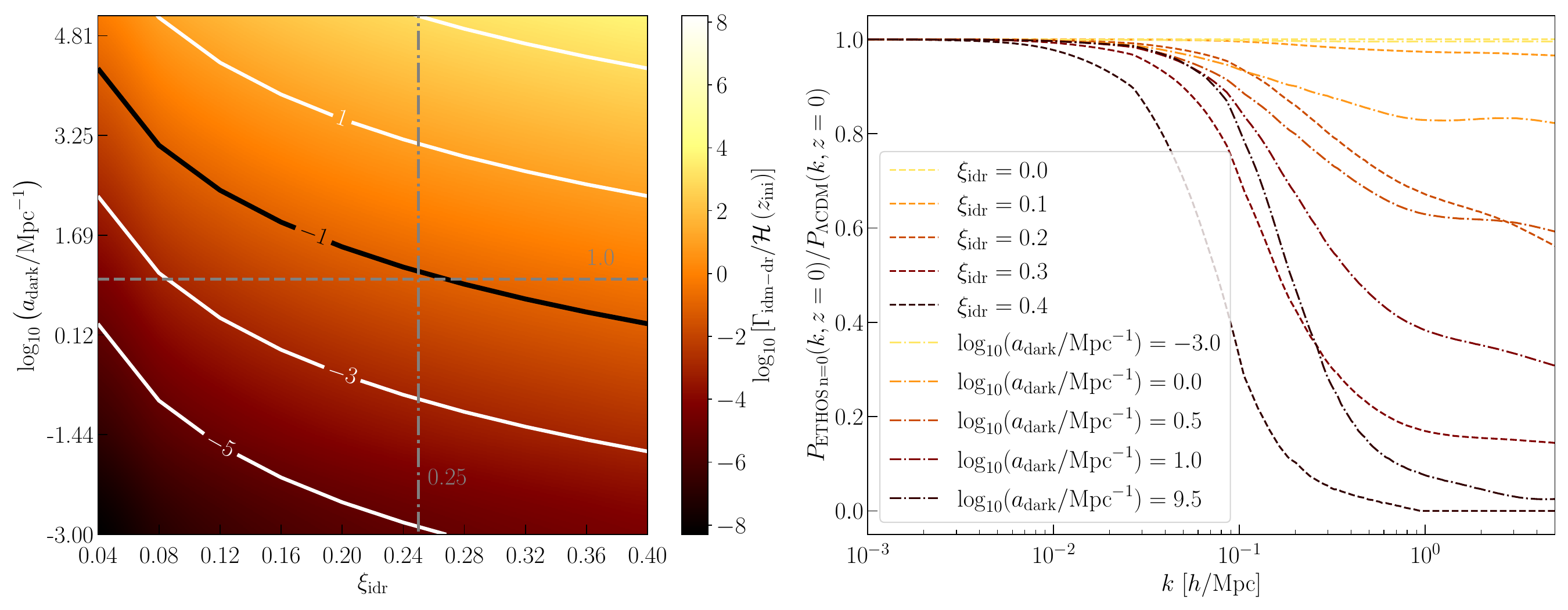}
    \caption{\emph{Left}: ratio of the interaction rate between IDM and IDR ($\Gamma_{\rm idm-dr}$) and comoving Hubble rate ($\mathcal{H}$) as a function of the dark-radiation-to-photon temperature ratio $\xi_{\rm idr}$ and interaction strength $a_{\rm dark}$, computed at the redshift $z_{\rm ini}=49$ at which the $N$-body simulations used to construct the ETHOS emulator are initialised. We display the contours of equal ratio as solid white lines and highlight the threshold value of 0.1 in black. We further depict the region with $a_{\rm dark} = 10\,{\rm Mpc}^{-1}$ (dashed grey) and $\xi_{\rm idr} = 0.25$ (dash-dotted grey). \emph{Right}: power spectrum suppression $\mathcal{S}_{\rm ETHOS}(k,z)$ predicted by the emulator for parameters chosen along each of the two grey lines of the left panel.
    \label{fig:emulator_ethos}
    }
\end{figure*}

We plot the magnitude of the ratio given by Eq.\,\eqref{eq:ethos_gamma_h} at $z=49$ as a function of $(\xi_{\rm idr}, a_{\rm dark})$ in the left panel of Fig.\,\ref{fig:emulator_ethos}. The region where the emulator is to be trusted lays below the solid black line. Dashed lines in the left panel of Fig.\,\ref{fig:emulator_ethos} correspond to models with either $\loga = 1.0$ (dashed) or $\xi_{\rm idr} = 0.25$ (dash-dotted), for which we show the emulator predictions in the right panel of the same figure. Namely, the dashed curves show the power spectrum suppression as a function of $\xi_{\rm idr}$, ranging from the $\Lambda$CDM limit for $\xi_{\rm idr} = 0.0$ up to $\xi_{\rm idr} = 0.4$, while fixing $\loga$ to 1.0. Similarly, the dash-dotted lines show the emulator output for $\xi_{\rm idr} = 0.25$ varying the coupling strength $a_{\rm dark}$ from $10^{-3}\,{\rm Mpc}^{-1}$ to $10^{9.5}\,{\rm Mpc}^{-1}$. Note that some of the cases shown in the right panel stand outside of the region ${\Gamma_{\rm idm-idr}}/{\mathcal{H}}\leq 0.1$ in which the emulator is to be fully trusted. 

The effect of the ETHOS parameters on the non-linear power spectrum is also shown in Fig.\,\ref{fig:lin_ethos} in the ($\xi_{\rm idr}$, $a_{\rm dark}\xi_{\rm idr}^4$) basis. While at the linear level the suppression of the matter power spectrum is mainly controlled by the combination $a_{\rm dark}\xi_{\rm idr}^4$, we see that at the non-linear level  $\xi_{\rm idr}$ also plays a significant role for fixed $a_{\rm dark}\xi_{\rm idr}^4$.

\subsection{Baryonic feedback\label{sec:baryon}}

Baryonic feedback processes can alter the gas distribution around DM halos, causing a deviation between the total matter distribution and the distribution of DM \citep[e.g.,][]{chisari_2018, vanDaalen:2019pst}. These processes induce a suppression of the total matter power spectrum $P_{\rm m}$ on small scales that may be somewhat similar to the effect of non-minimal DM and induce degeneracies between baryonic and DM parameters \citep[see e.g.,][]{Hubert_2021}. Hence, for our purposes, it is crucial to incorporate these processes into our modelling framework. In this study, we use the \bcemu{} framework\footnote{The code is available at \url{https://github.com/sambit-giri/BCemu}.} \citep{Giri2023BCemu} to address this concern. The \bcemu{} framework serves as an emulator for the suppression ${\cal S}_{\rm bf} (k)$ caused by baryonic feedback. Consequently, the total non-linear matter power spectrum in a given cosmological model, $P_{\rm m}$, can be expressed as
\begin{eqnarray}
P_{\rm m} (k,z) = {\cal S}_{\rm bf} (k,z) \, P_{\rm m, no\,bf} (k,z) \, ,
\end{eqnarray}
where $P_{\rm m, no\, bf}(k,z)$ is the total matter power spectrum neglecting baryonic feedback effects at redshift $z$. 

\bcemu{} has been used in several recent  WL studies \cite[e.g.,][]{schneider_constraining_2022,Grandis:2023qwx}. It is based on the baryonic correction modelling framework of \cite{schneider_new_2015}, \cite{Schneider:2018pfw}, and \cite{giri_emulation_2021}. This framework parameterises the stellar and gas profiles at a given redshift with seven baryonic parameters. \citet{giri_emulation_2021} analysed these baryonic parameters and found that three parameters are enough to model the suppression seen in hydrodynamical simulations at scales $k\lesssim 10\,h\,{\rm Mpc^{-1}}$ and at a given redshift. We will use this 3-parameter model in this work. 
%Two of them are gas parameters ${\rm log_{10}}M_c$ and $\theta_{\rm ej}$, which impact the strength and largest scale of suppression respectively. 

Two of these parameters describe the gas profile in halos of given virial radius $r_\mathrm{vir}$ and virial mass $M_\mathrm{vir}$, modelled as
\begin{equation}
\label{rhogas}
\rho_{\rm gas}(r)\propto\frac{\Omega_{\rm b}/\Omega_{\rm m} -f_{\rm star}(M_{\rm vir})}{\left[1+10\frac{r}{r_{\rm vir}}\right]^{\beta(M_{\rm vir})}\left[1+\frac{r}{\theta_{\rm ej}r_{\rm vir}}\right]^{\frac{2}{5}[7-\beta(M_{\rm vir})]}} \, ,
\end{equation}
with a total stellar fraction $f_{\rm star}(M_{\rm vir})$
and a mass-dependent index
\begin{equation}
\beta(M_{\rm vir}) = \frac{3\, M_{\rm vir}/M'_{\rm c}}{1+M_{\rm vir}/M'_{\rm c}} \, .
\end{equation}
The former function is assumed to be known,
\begin{equation}\label{fstar}
f_{\rm star}(M_{\rm vir}) = 0.055\left(\frac{10^{11.3} \,h^{-1}\, M_\odot}{M_{\rm vir}}\right)^{0.2} \ .
\end{equation}
Thus, in this model, the gas profile only depends on two free parameters: a critical mass $M'_{\rm c}$ such that small halos with $M_{\rm vir} \ll M_{\rm c}$ have a gas profile shallower than the Navarro--Frenk--White profile, and an ejection factor $\theta_{\rm ej}$ giving the ratio of the gas ejection radius to the virial radius. The \bcemu{} model also involves assumptions concerning the stellar profile of the central galaxy. The fraction of stars in the central galaxy, $f_{\rm cga}(M_{\rm vir})$, is given by a relation similar to  $f_{\rm star}(M_{\rm vir})$, but with a different exponent,
\begin{equation}
    f_{\rm cga}(M_{\rm vir}) = 0.055\left(\frac{10^{11.3} \,h^{-1}\, M_\odot}{M_{\rm vir}}\right)^{0.2+\eta_\delta} \, ,
\end{equation}
where the index $\eta_{\delta}$ is an additional free parameter.

In summary, the minimal \bcemu{} model relies on three free parameters ($M'_{\rm c}$, $\theta_{\rm ej}$, $\eta_\delta$) impacting, respectively, the overall suppression induced by baryonic feedback, the maximum scales affected by the suppression, and the upturn of ${\cal S}_{\rm bf} (k,z)$ at large $k$. In order to deal only with dimensionless parameters, \cite{schneider_constraining_2022} redefine the first one as $M_c := {M'_c} / \left(1\,h^{-1}\,M_\odot\right)$. 
Figure\,2 in \citet[][]{Schneider:2018pfw} shows the impact of these parameters on the matter power spectrum. In the \bcemu{} model, the only cosmology dependence of ${\cal S}_{\rm bf} (k,z)$ comes through the baryon fraction $\Omega_{\rm b}/ \Omega_{\rm m}$. 

In particular, we assume no explicit dependence of the baryonic feedback suppression function ${\cal S}_{\rm bf}(k,z)$ on the parameters describing non-standard DM models. This assumption was shown to be valid at least for $k<5 \, h\,{\rm Mpc}^{-1}$ in the CWDM scenario, see section 3.4 in \cite{Parimbelli_CWDM}. This conclusion is expected to apply also to the other DM scenarios studied here in which, like in the CWDM case, DM particles are decoupled at low redshift and behave either as CDM or WDM. In our analysis, smaller scales with $k>5 \, h\,{\rm Mpc}^{-1}$ only have a small contribution to the spectra $C_{ij}^{\rm XY}(\ell)$ involving the first two weak lensing redshift bins. Thus, the findings of \cite{Parimbelli_CWDM} suggest that we can safely neglect the impact of non-standard DM on baryonic feedback. More generally, we can think of the effect of non-standard DM and of BF on the non-linear matter power spectrum as two leading-order effects, of a few percents each within the range of scales relevant in our analysis, and that of non-standard DM on BF as a next-to-leading order effect of a few percents squared, that is, a few per mille. It is thus reasonable to neglect this correction in a first analysis.

We refer interested readers to \citet{giri_emulation_2021} for a more detailed description of the \bcemu{} parameters.

The redshift evolution of ${\cal S}_{\rm bf} (k,z)$ is modelled by making each of the three baryonic parameters $b$ redshift dependent as
\begin{equation}
    b(z) = b(0) (1+z)^{-\nu_b} \, , \qquad 
    b \in \{ {\rm log_{10}}M_c, \theta_{\rm ej}, \eta_{\delta}\} \, ,
\end{equation}
where $\nu_b$ is a free parameter. This leads to a total of six parameters to model the baryonic feedback. In our choice of fiducial values and priors, we restrict the values of $\nu_b$ such that the baryonic parameters $\{ {\rm log_{10}}M_c, \theta_{\rm ej}, \eta_{\delta}\}$ remain within the range of the parameter space where \bcemu{} is trained.

While it is known that the WL signal is modified at small scales by baryonic feedback effects, the situation is much less clear regarding the GC signal. Since galaxies act as tracers of the underlying DM distribution, they are not directly affected by the ejection of gas via feedback processes. We rather expect an indirect effect caused by the relaxation of the DM potential reacting to the ejection of gas. Since we do not know the true amplitude of this indirect effect, we consider two extreme cases where the GC is either unchanged by baryonic feedback or it is affected in the same way as the weak-lensing. We expect the truth to lie somewhere  between these two cases.

\section{Forecast methodology \label{sec:method}}

\subsection{Likelihood}

We use a standard formalism to describe the \Euclid{} photometric likelihood already presented, for instance, in \cite{Audren:2012vy}, \cite{Blanchard:2019oqi}, \cite{Casas23}, and \cite{KP_nu}. The galaxy images of the WL survey and the galaxy positions of the GC photometric survey are binned into $N$ redshift bins. In each bin, the raw data can be processed into two-dimensional spherical maps of either the lensing potential field density in the WL case or the galaxy density field in the GCph case.
The maps are decomposed into spherical harmonics with coefficients $a_i(\ell,m)$ for each redshift bin $i$. Each $a_i$ is assumed to obey a Gaussian distribution with covariance matrix $C_{ij}(\ell)=\frac{1}{2\ell+1}\sum_ma_i(\ell,m)[a_i(\ell,m)]^*$.
The $C_{ij}(\ell)$ are the observed power spectra of WL or GCph in harmonic space and can be compared to theoretical predictions.

In our forecasts, we assume that the power spectra observed by \Euclid{} coincides with the theoretical predictions of a given \textit{fiducial} cosmology with spectra $C_{ij}^{\rm fid}(\ell)$ arising from multipoles $a_i^{\rm fid}$ such that $C_{ij}^{\rm fid}(\ell)=\frac{1}{2\ell+1}\sum_m a_i^{\rm fid}(\ell,m)[a_i^{\rm fid}(\ell,m)]^*$. The likelihood $\mathcal{L}$ of the observed data given a theoretical model with spectrum $C_{ij}^{\rm th}(\ell)$ is then given by
\begin{align}
    \mathcal{L}
    &=
    \mathcal{N} \prod_{\ell,m} \left[ \det \tens{C}^{\rm th}(\ell)\right]^{-1/2} 
    \nonumber \\
    & \times \exp\left\{-f_{\rm sky}\frac{1}{2}\sum_{ij}a_i^{\rm fid}(\ell,m)
    \,\,
    (C_{ij}^{\rm th})^{-1}(\ell)
    \,\,
    [a_j^{\rm fid}(\ell,m)]^*\right\}\, ,
\end{align}
where ${\cal N}$ is a normalisation factor, and partial sky coverage is approximately accounted for through multiplication with the sky fraction $f_{\rm sky}$. This can be rewritten as \citep{Audren:2012vy}
\begin{align}
    \chi^2 
    &:= 
    -2\ln\frac{\mathcal{L}}{\mathcal{L}_{\rm max}}
    \nonumber \\
    & = f_{\rm sky}\sum_\ell(2\ell+1)\left\{
    {\rm Tr}[(\tens{C}^{\rm th})^{-1}(\ell)\,\,\tens{C}^{\rm fid}(\ell)]
    \right.
    \nonumber \\
    &~~~~~~~~~~~~~~~~~~~~~~~~~~~+
    \left.
    \ln\frac{\det \tens{C}^{\rm th}(\ell)}{\det \tens{C}^{\rm fid}(\ell)}
    -N\right\} 
    \nonumber \\
    & = f_{\rm sky}\sum_\ell(2\ell+1)\left\{\frac{d_\ell^{\rm mix}}{d_\ell^{\rm th}}+\ln\frac{d_\ell^{\rm th}}{d_\ell^{\rm fid}}-N\right\} \, ,
\end{align}
where $N$ is the size of the matrices $\tens{C}^{\rm th}(\ell)$ and $\tens{C}^{\rm fid}(\ell)$, while
\begin{equation}
    \tens{C}(\ell) := \begin{bmatrix}
            C_{ij}^\mathrm{LL}(\ell) & C_{ij}^\mathrm{GL}(\ell) \\
            C_{ij}^\mathrm{LG}(\ell) & C_{ij}^\mathrm{GG}(\ell)
        \end{bmatrix} \, ,
        \quad \quad
        d_\ell := \det \tens{C}(\ell) 
\end{equation}
for each of the theoretical and fiducial spectra. Finally, the mixed determinant is defined as
\begin{equation}
    d_\ell^{\rm mix} := \sum_{k=1}^N \det \left[ \begin{cases}
        C_{ij}^{\rm th}(\ell) & {\rm for}\;j\neq k \\
        C_{ij}^{\rm fid}(\ell) & {\rm for}\;j=k 
    \end{cases}
    \right]\, ,
\end{equation}
such that in each term of the sum, the determinant is evaluated over a matrix in which the $k$-th column of the theory matrix $\tens{C}^\mathrm{th}$ has been substituted by the $k$-th column of the fiducial matrix $\tens{C}^\mathrm{fid}$.

We then perform MCMC forecasts \citep{Audren:2012vy,Casas23} using this likelihood. The likelihood is incorporated into the \montepython{} package\footnote{\url{https://github.com/brinckmann/montepython_public}} \citep{Audren:2012wb,Brinckmann:2018cvx} for Bayesian parameter inference. The role of \montepython{} is to fit the fiducial spectra under the assumption of a given theoretical model with a set of free parameters. A few independent Monte Carlo Markov Chains sample the likelihood by exploring the parameter space according to the Metropolis-Hastings algorithm, until some convergence criterium is reached. The best-fit model coincides by construction with the fiducial model, while the marginalised credible interval of each parameter provide an estimate of the sensitivity of \Euclid{} to this parameter. 

\subsection{Observable power spectra}

The model for the spectra $C_{ij}^{XY}$ used in the likelihood, where $X={\rm L}$ (respectively $X={\rm G}$) refers to the WL (respectively GC) probe and $i=1, ..., N_i$ to the bin number, is detailed in \cite{Blanchard:2019oqi}, \cite{Casas23}, and \cite{KP_nu}. The final expression is given by
\begin{align}
    C_{ij}^{XY}(\ell)&=\int_{z_{\rm min}}^{z_{\rm max}}{\rm d}z \, \frac{W_i^X(k(\ell,z),z)\,\,W_j^Y(k(\ell,z),z)}{c^{-1}\,\,H(z)\,\,r^2(z)}
    \nonumber \\
    &  ~~~~~~~~~~~~~~~~~\times P_{\rm m}(k(\ell,z),z) 
    \nonumber \\
    & + N_{ij}^{XY}(\ell) \, ,
    \label{eq:cl_integral}
\end{align}
where $W_i^X(k,z)$ are the window functions of the $X$ probe, $H(z)$ is the Hubble rate at redshift $z$, $r(z)$ the comoving distance to an object at redshift $z$, $P_{\rm m}(k,z)$ the matter power spectrum evaluated at wavenumber $k$, and $N_{ij}^{XY}(\ell)$ the noise spectrum. The boundaries $z_{\rm min}$ and $z_{\rm max}$, defined in Table~\ref{tab:photo_parameters}, specify the redshift range covered by the survey. The relation $k(\ell,z)$ is inferred from the Limber approximation \citep{Kaiser:1991qi,Kilbinger:2017lvu},
\begin{eqnarray}
    k(\ell,z)=\frac{\ell+1/2}{r(z)}~,
    \label{eq:k_l}
\end{eqnarray}
which is sufficiently accurate for $\ell>\ell_{\rm min}$, where $\ell_{\rm min}$ is given in Table \ref{tab:photo_parameters}, see however \cite{Tanidis:2019teo}. Assuming a Poissonian distribution of galaxies, the noise spectra read
\begin{equation}
N_{ij}^\mathrm{LL} = \frac{\sigma_\epsilon^2}{\bar{n}_i} \delta_{ij}, \qquad
N_{ij}^\mathrm{GG} = \frac{1}{\bar{n}_i}
\delta_{ij}, \qquad
N_{ij}^\mathrm{LG} = N_{ij}^\mathrm{GL} = 0\, ,
\end{equation}
where $\bar{n}_i$ is the expected average number of galaxies per steradian in the $i$-th bin, and $\sigma_\epsilon^2$ is the variance of the observed ellipticities, also given in  Table \ref{tab:photo_parameters}.
The galaxy field that GCph measures is assumed to be a linear tracer of the underlying matter field, such that the galaxy power spectrum is given by $P_{\rm g}(k,z)=b^2(z)P_{\rm m}(k,z)$ with some bias function $b(z)$. Here, for simplicity, we neglect additional effects on the photometric galaxy power spectrum such as lensing magnification or redshift-space distortions \citep{Yoo:2009au,Bonvin:2011bg,Challinor:2011bk,Yoo:2014sfa}, although these effects are expected to play a non-negligible role in the analysis of real \Euclid data, see \cite{Euclid:2021rez} and \cite{Euclid:2023pyq}.
Sticking to linear bias is conservative as long as we rely on pessimistic assumptions concerning the minimum angular scale or maximal multipole $l_{\rm max}^{\rm GCph}$ described in Table\,\ref{tab:photo_parameters}. In the optimistic case, we should be aware that non-linear biasing may come into play on the smallest scales used in the analysis, and introduce a possible degeneracy with DM parameters that is neglected here.

Then the GC window functions reads
\begin{eqnarray}
    W_i^{\rm G}(z)=\frac{n_i(z)\,\,H(z)\,\,b(z)}{c} \, ,
\end{eqnarray}
where $n_i(z)$ is the observed galaxy density distribution normalised to unit area in redshift bin $i$. Since there is no reliable model for $b(z)$, it is modelled as a step-like function given by $b(z)=b_i$ in the redshift range $z_i^-<z<z_i^+$ of redshift bin $i$. Each $b_i$ is treated as a free nuisance parameter and marginalised over in the forecast. Taking photometric redshift errors into account, the observed distribution of galaxies $n_i(z)$ in bin $i$ is given by the true galaxy distribution,
\begin{eqnarray}
    n(z)= n_0 \left(\frac{z}{z_0}\right)^2\exp\left[-\left(\frac{z}{z_0}\right)^{1.5}\right] \, ,
    \label{eq:nz}
\end{eqnarray}
with $z_0=z_{\rm mean}/\sqrt{2}$, and by the redshift error probability distribution,
\begin{align}
    p_{\rm ph}(z_{\rm p}|z) &=\frac{1-f_{\rm out}}{\sqrt{2\pi}\sigma_{\rm b}(1+z)}\exp\Biggl\{-\frac{1}{2}\left[\frac{z-c_{\rm b}z_{\rm p}-z_{\rm b}}{\sigma_{\rm b}(1+z)}\right]^2\Biggr\}
    \nonumber \\
    &+ \frac{f_{\rm out}}{\sqrt{2\pi}\sigma_0(1+z)}\exp\Biggl\{-\frac{1}{2}\left[\frac{z-c_0z_{\rm p}-z_{\rm b}}{\sigma_0(1+z)}\right]^2\Biggr\} \, .
\end{align}
The normalised distribution $n_i(z)$ then reads \citep{Ma:2005rc,Joachimi:2009fr,Joachimi:2009ez}
\begin{eqnarray}
    n_i(z)=\frac{n(z) \int_{z_i^-}^{z_i^+}{\rm d}z_{\rm p} \, p_{\rm ph}(z_{\rm p}|z)}{\int_{z_{\rm min}}^{z_{\rm max}}{\rm d}\tilde{z} \, n(\tilde{z}) \, \int_{z_i^-}^{z_i^+}{\rm d}z_{\rm p} \, p_{\rm ph}(z_{\rm p}|\tilde{z})} \, .
\end{eqnarray}
The parameters entering this model are listed in Table \ref{tab:photo_parameters}. The WL window functions are given by
\begin{eqnarray}
    W_i^{\rm L}(k,z) = W_i^\gamma(z)-\mathcal{A}_{\rm IA}\mathcal{C}_{\rm IA}\Omega_{{\rm m}}\frac{\mathcal{F}_{\rm IA}(z)}{D(k,z)}W_i^{\rm IA}(z) \, ,
\end{eqnarray}
where the latter term corrects for intrinsic alignment (IA) effects, $W_i^{\rm IA}(z)=c^{-1}n_i(z)H(z)$, and $W_i^\gamma(z)$ is the shear-only window function
\begin{align}
    W_i^\gamma(z) &= \frac{3}{2}c^{-2}H_0^2\Omega_{{\rm m}}(1+z)r(z)
    \nonumber \\
    & \times \int_z^{z_{\rm max}}{\rm d}z'n_i(z')\left[1-\frac{r(z)}{r(z')}\right] \, .
    \label{eq:W_gamma}
\end{align}
$D(k,z)$ is the linear growth factor, defined as $D(k,z)\coloneqq [P_{\rm m,lin}(k,z)/P_{\rm m,lin}(k,z=0)]^{1/2}$. In linear $\Lambda$CDM cosmology, perturbations grow independently of scale and the $k$ dependence exactly cancels out. Instead, the particle DM models considered in this work lead to some scale-dependent linear growth, such that the function $D$ is a function of $(k,z)$. The factor $\mathcal{F}_{\rm IA}$ is modelled as
\begin{eqnarray}
    \mathcal{F}_{\rm IA}(z)=(1+z)^{\eta_{\rm IA}}[\langle L\rangle(z)/L_*(z)]^{\beta_{\rm IA}}
    \label{eq:ia}
\end{eqnarray}
and depends on the mean galaxy luminosity divided by a characteristic luminosity $\langle L\rangle(z)/L_*(z)$, which is read from the file \texttt{scaledmeanlum$\_$E2SA.dat} provided by the authors of \cite{Blanchard:2019oqi}. In practice, we vary $\eta_{\rm IA}$ and ${\cal A}_{\rm IA}$ as nuisance parameters but fix $\beta_{\rm IA}$ and $\mathcal{C}_{\rm IA}$ due to the strong degeneracies between the former and the latter.

\begin{table}[ht]
    \centering
    \caption{Specifications used in our mock \Euclid{} photometric likelihood in the pessimistic (pess.) and optimistic (opt.) cases.\label{tab:photo_parameters}}
    \begin{tabular}{l c c}
        \hline
         Type & Name & Value (pess./opt.) \\
         \hline
         \noalign{\vskip 2pt}
         Redshift bins & $N_{\rm bin}$ & $10$ \\
         Redshift bins & $z_{\rm min}=z_0^-$ & $0.001$ \\
         Redshift bins & $z_1^+, z_2^+, z_3^+$ & $0.418$, $0.560$, $0.768,$ \\
         Redshift bins & $z_4^+, z_5^+, z_6^+$ & $0.789$, $0.900$, $1.019$ \\
         Redshift bins & $z_7^+, z_8^+, z_9^+$ & $1.155$, $1.324, 1.576$ \\
         Redshift bins & $z_{\rm max}=z_{10}^+$ & $2.5$ \\
         Redshift bins & $z_{\rm mean}$ & $0.9$ \\
         %\hline
         Photometric error & $c_0$ & $1.0$ \\
         Photometric error & $c_{\rm b}$ & $1.0$ \\
         Photometric error & $z_0$ & $0.1$ \\
         Photometric error & $z_{\rm b}$ & $0.0$ \\
         Photometric error & $\sigma_0$ & $0.05$ \\
         Photometric error & $\sigma_{\rm b}$ & $0.05$ \\
         Photometric error & $f_{\rm out}$ & $0.1$ \\
         %\hline
         %Intrinsic alignment & $\mathcal{A}_{\rm IA}$ & $1.72$ \\
         Intrinsic alignment & $\mathcal{C}_{\rm IA}$ & $0.0134$ \\
         Intrinsic alignment & $\beta_{\rm IA}$ & $2.17$ \\
         %Intrinsic alignment & $\eta_{\rm IA}$ & $-0.41$ \\
         %\hline
         Noise & $\sigma_\epsilon$ & $0.3$ \\
         Noise & $n_{\rm gal}$ & $30$ arcmin$^{-2}$ \\
         %\hline
         Multipoles & $\ell_{\rm min}$ & $10$ \\
         Multipoles & $\ell^{\rm WL}_{\rm max}$ & $1500 / 5000$ \\
         Multipoles & $\ell^{\rm GCph}_{\rm max}$ & $750 / 1500$ \\
         %\hline
         Sky coverage & $f_{\rm sky}$ & $0.3636$ \\
         \noalign{\vskip 2pt}
         \hline
    \end{tabular}
\end{table}

Note that Eq.~\eqref{eq:W_gamma} is derived under the assumption that the non-relativistic matter density scales like $a^{-3}$: the factor $(1+z)$ in front of the integral actually comes from the product $\rho_{\rm m}(z) \,a^{2}(z)$. In the $\Lambda$CDM, CWDM, and ETHOS models, the assumption $\rho_{\rm m} \propto a^{-3}$ is excellent (as long as massive neutrino effects are neglected). In the 2b-DDM case, it is still excellent since we are only interested in the limit $\varepsilon \ll 1$ in which the decays convert a negligible fraction of the non-relativistic energy density $\rho_{\rm m}$ into relativistic energy density $\rho_{\rm r}$. However, in the 1b-DDM case, the product $\rho_{\rm m} \,a^{3}$ decreases slightly between the highest and lowest redshift probed by the survey, which spans an interval of proper time $\Delta t$. The relative variation of $\rho_{\rm m} \,a^{3}$ over this interval is given by $f_{\rm ddm}^{\rm ini} \, \Gamma_{\rm ddm} \, \Delta t$, and remains below a few percents for the 1b-DDM models studied in the next sections. Thus we neglect this sub-dominant effect and stick to Eq.~\eqref{eq:W_gamma}.\footnote{If this effect was not negligible and was implemented in Eq.~\eqref{eq:W_gamma}, it would lead to a redshift-dependent rescaling of $W_i^\gamma(z)$, which could only increase the sensitivity of observations to 1b-DDM parameters. Thus our approximation stays on the conservative side.}

In our forecast we rely either on a pessimistic or optimistic assumption concerning the range of scales at which our model is trusted and data is included. In the pessimistic case, we truncate the data at $\ell_{\rm max}^{\rm WL}=1500$ for WL and $\ell_{\rm max}^{\rm GC}=750$ for GCps.
In the optimistic case, we use
$\ell_{\rm max}^{\rm WL}=5000$ for WL and $\ell_{\rm max}^{\rm GC}=1500$ for GCps.

The matter power spectrum $P_{\rm m}(k,z)$ that appears in Eq.\,\eqref{eq:cl_integral} is usually computed in four steps. First, we call a Boltzmann code to compute the linear matter power spectrum of a $\Lambda$CDM model with the same cosmological parameters as the non-standard DM model we are interested in. Second, we ask the same Boltzmann code to use a standard algorithm to infer the non-linear power spectrum for this $\Lambda$CDM model. In the forecasts of this work, for simplicity, we use the version of {\tt Halofit} revisited by \cite{Takahashi:2012em} and \cite{Bird:2011rb} as a baseline, or \texttt{HMcode 2020} \citep{hmcode2020} in the case where neutrinos are assumed to have a mass of $0.06\,{\rm eV}$. Third, we use one of the emulators described in Sects.\,\ref{sec:nl_cwdm} to \ref{sec:nl_ethos} to transform this into a non-linear power spectrum for the non-standard DM model of interest. Fourth, when baryonic feedback corrections need to be taken into account, we call the emulator described in Sect.\,\ref{sec:baryon} to add baryonic corrections. For the WL auto-correlation spectra, $C^{\rm LL}_{ij}$, the matter power spectrum of Eq.\,\eqref{eq:cl_integral} always includes baryonic feedback. For the GCph auto-correlation spectra, $C^{\rm GG}_{ij}$, we will consider the two cases in which the power spectrum incorporates such corrections or not. Note that for the cross-correlation spectra, when baryonic feedback is included in WL but not GCph, we use for
$C_{ij}^{\rm LG}(\ell)=C_{ji}^{\rm GL}$,
\begin{align}
    C_{ij}^{\rm LG}(\ell)&=\int_{z_{\rm min}}^{z_{\rm max}}{\rm d}z\frac{W_i^{\rm L}(k(\ell,z),z)W_j^{\rm G}(k(\ell,z),z)}{c^{-1}H(z)r^2(z)}
    \nonumber \\
    & ~~~~~~~~~~~~~~~~\times
    \sqrt{P_{\rm m}^{\rm BF}(k(\ell,z),z)P_{\rm m}^{\rm no\,BF}(k(\ell,z),z)}
    \nonumber \\
    &
    + N_{ij}^\mathrm{LG}(\ell)\, .
\end{align}

\subsection{Boltzmann code \label{sec:boltzmann}}

We need to call a Boltzmann code for two purposes: first, to compute the comoving distance-redshift relation $r(z)$ and the (scale-dependent) growth factor $D(k,z)$; and second, to compute the non-linear matter power spectrum $P_{\rm m}(k,z)$. However, the strategy of the emulators described in Sects.\,\ref{sec:nl_cwdm} to \ref{sec:nl_ethos} implies a calculation of the non-linear matter power spectrum for the equivalent $\Lambda$CDM sharing the same value of the standard cosmological parameters $\{\omega_{\rm b}, \omega_{\rm cdm}$, $h$, $A_{\rm s}$, $n_{\rm s}\}$ as the non-standard DM model of interest. Instead, the distance-redshift relation and the scale-dependent growth factor should be computed according to the background and linear theory equations describing the true non-standard DM model. We solve this issue by calling the Boltzmann code twice at each point in parameter space: first for the non-standard DM model with linear output, to infer $r(z)$ and $D(k,z)$; and second for the  equivalent $\Lambda$CDM with {\tt Halofit} or \texttt{HMcode 2020} corrections switched on, to get the desired non-linear matter power spectrum.

We choose to use \class{} {\tt v 3.2} \citep{Lesgourgues:2011re,Blas:2011rf} as our Boltzmann code since the CWDM, 1b-DDM and ETHOS models are implemented in the main public branch of the code,\footnote{\url{https://github.com/lesgourg/class}} while 2b-DDM is implemented in a public but separate branch {\tt class\_decays}.\footnote{\url{https://github.com/PoulinV/class_decays}}

%---------------
\begin{figure*}[t]
    \centering
    \includegraphics[width=0.99\textwidth]{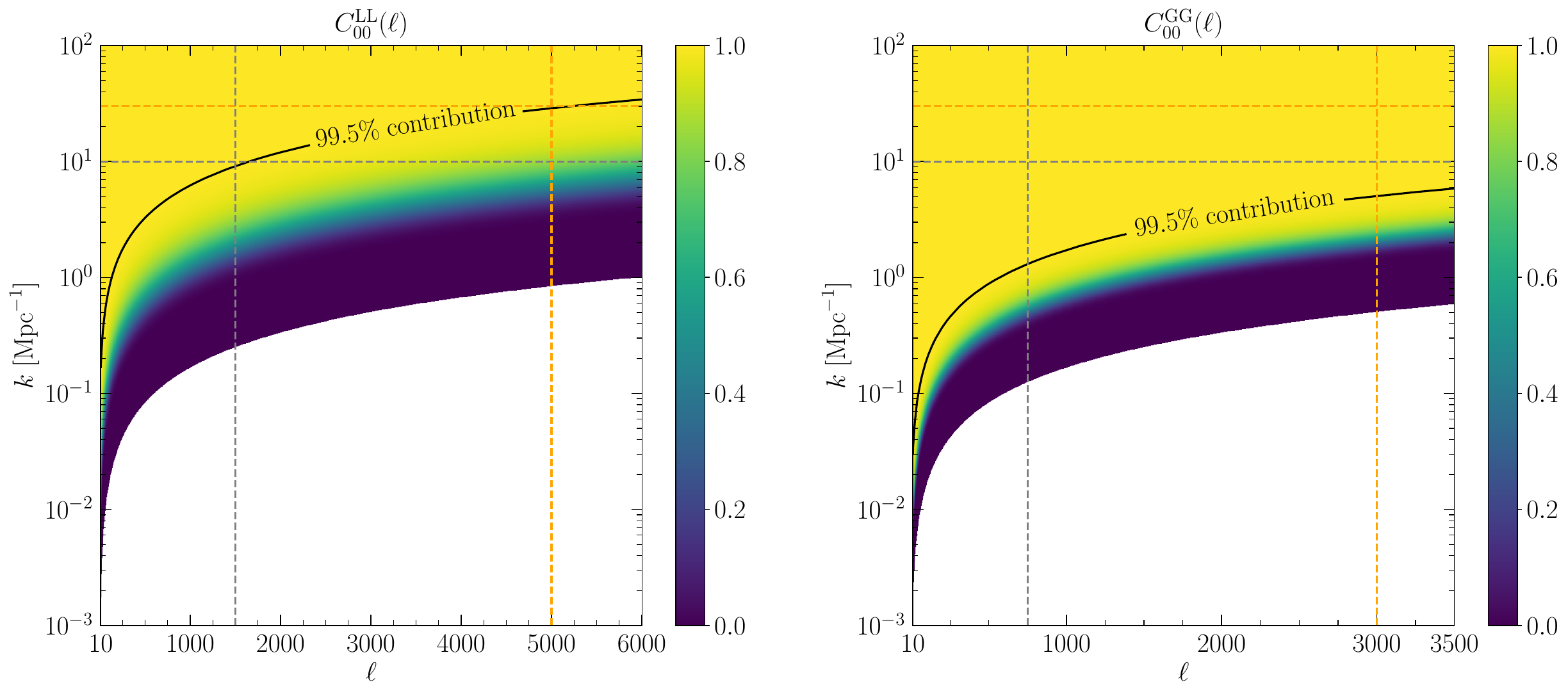}
    \caption{Cumulative contribution of different $k$ values to $C^{XY}_{ij}(\ell)$ for a given $\ell$ in the nearest redshift bin ($ij=00$). \emph{Left}: case of weak lensing, $XY={\rm LL}$.
    \emph{Right}: case of galaxy clustering, $XY={\rm GG}$. For each $\ell$, 99.5\% of the contribution stands below the black isocontour. In the pessimistic case, we include all values of $(\ell,k)$ on the left of and below the dashed grey lines in the calculation of our observables; in the optimistic case, on the left and below the dashed orange line. Thus, we always include at least 99.5\% of the contribution to each $C^{XY}_{ij}(\ell)$.
    \label{fig:k_l_contribution}
    }
\end{figure*}
%--------------- 

When calling \class{}, we should specify a maximum wavenumber $k_{\rm max}$. A given Fourier mode $k$ of a field observed at redshift $z$ projects under a given angle $\theta$ contributing mainly to a mutipole $\ell$, with the relation between $k$, $\ell$, and $z$ given by Eq.\,\eqref{eq:k_l}. Thus the choice of $k_{\rm max}$ should reflect the maximum multipole $\ell_{\rm max}$ and minimum redshift $z_{\rm min}$ contributing to the power spectra in a given analysis. Using the relation $k(\ell,z)$ at fixed $\ell$, one can express each $C_{ij}^{XY}(\ell)$ as an integral over $k$ rather than $z$, and plot the cumulative contribution of different $k$ values to $C_{ij}^{XY}(\ell)$. To make a robust choice for $k_{\rm max}$, we show in Fig.\,\ref{fig:k_l_contribution} the contribution of different $k$ values to the power spectra of the first redshift bins, $C^\mathrm{LL}_{00}(\ell)$ and $C^\mathrm{GG}_{00}(\ell)$. The figure shows that in the pessimistic case ($\ell_{\rm max}^{\rm WL}=1500$), 
choosing $k_{\rm max}=10\,{\rm Mpc}^{-1}$ is sufficient to include $99.5\%$ of the contribution the $C^\mathrm{LL}_{00}(\ell)$, and a fortiori to all other spectra. For the optimistic case ($\ell_{\rm max}^{\rm WL}=5000$), we find that $k_{\rm max}=30\,{\rm Mpc}^{-1}$ is sufficient.

The $C^{XY}_{ij}(\ell)$ are computed with a trapezoidal integral over $200$ values of $z$ on a linear grid between $z_{\rm min}$ and $z_{\rm max}$. The integral is performed for discrete values of $\ell$, with a logarithmically spaced grid of $100$ values of $\ell$ between $\ell_{\rm min}$ and $\ell_{\rm max}$. Finally, a second-order spline interpolation is used to get the spectra for every integer $\ell$.

\subsection{Parameters and priors}

\begin{table}[ht]
\centering
\caption{List of free parameter, fiducial values, and top-hat prior ranges used in all our runs (not including the DM parameters which are different in each model and specified in Sect.\,\ref{sec:pdm_models}). This list includes five $\Lambda$CDM cosmological parameters, six baryonic feedback parameters, ten bias parameters for GC, and two intrinsic alignment parameters for WL. The acronym n.i. means `non-informative prior' (see the text for details). \label{tab:priors}}
\begin{tabular}{ccc}
\hline
Parameter & Fiducial value & Range \\
\hline
$\Omega_{\rm b}$ & 0.049199 & n.i. \\
$h$ & 0.67370 & n.i. \\
$n_{\rm s}$ & 0.96605 & n.i. \\
$\ln(10^{10} A_{\rm s})$ & 3.0447 & n.i. \\
$\Omega_{\rm m}$ & 0.31457 & n.i. \\
%$m_{\rm wdm}^{\rm thermal}\,\,(eV)$ & $\infty$ & $[10 , 1000]$ \\
%$\log_{10} f_{\rm wdm}$ & $-\infty$ & $[-3, 0]$ \\
$\log_{10} M_c$ & 13.25 & $[11,\,15]$ \\
$\theta_{\rm ej}$ & 4.711 & $[2,\,8]$ \\
$\eta_\delta$ & 0.097 & $[0.05,\,0.4]$ \\
$\nu_{\log_{10} M_c}$ & 0.038 & n.i. \\
$\nu_{\theta_{\rm ej}}$ & 0 & n.i. \\
$\nu_{\eta_\delta}$ & 0.06 & n.i. \\
$b_1$ & 1.0998 & n.i. \\
$b_2$ & 1.2202 & n.i. \\
$b_3$ & 1.2724 & n.i. \\
$b_4$ & 1.3166 & n.i. \\
$b_5$ & 1.3581 & n.i. \\
$b_6$ & 1.3998 & n.i. \\
$b_7$ & 1.4446 & n.i. \\
$b_8$ & 1.4965 & n.i. \\
$b_9$ & 1.565 & n.i. \\
$b_{10}$ & 1.7430 & n.i. \\
$A_{\rm IA}$ & 1.72 & $[0,12.1]$ \\
$\eta_{\rm IA}$ & $-0.41$ & $[-7, \,6.17]$\\
\hline
\end{tabular}
\end{table}

We list in Table\,\ref{tab:priors} the free parameters used in our forecasts (not including the DM parameters specific to each model, which will be specified in each Sect.\,\ref{sec:results}). The table also provides the assumed fiducial values and priors. We remind that forecast errors are anyway nearly independent of the chosen fiducial values, especially for the $\Lambda$CDM parameters, which have nearly Gaussian posteriors. The first five parameters are the cosmological parameter of the standard $\Lambda$CDM model. The following six parameters describe the baryonic feedback model as implemented in \bcemu. The last 12 nuisance parameters account for linear bias in each bin and intrinsic alignment parameters.

We use a flat prior on each of these parameters. For the first three \bcemu{} parameters and the two intrinsic alignment parameter, we pass explicit prior edges to ensure that these parameters remain a range making sense physically. For the other parameters, as long as we only perform parameter inference with the Metropolis-Hastings algorithm, it is strictly equivalent to pass to \montepython{} some very remote prior edges -- such that the chains never reach the prior boundaries -- or to require the code to use non-informative priors (abbreviated as n.i. in Table\,\ref{tab:priors}), in which case the code achieves the same feature automatically. Note that our chains never reach the prior edges passed for the intrinsic alignement parameters, so we are effectively using  non-informative priors also for ${\cal A}_{\rm IA}$, $\eta_{\rm IA}$.

Finally, theoretical predictions depend on a few additional parameters that are usually kept fixed because the set of free nuisance parameters from Table\,\ref{tab:priors} are sufficient to account for the uncertainty on the model. For intrinsic alignment, we fix the parameter $\beta_{\rm IA}$ defined in Eq.\,\eqref{eq:ia} to $\beta_{\rm IA}=2.17$; for baryonic feedback, we fix the \bcemu{} parameters $\mu=1$, $\gamma=2.5$,
$\delta=7$, $\eta=0.2$.

\section{Results and discussion\label{sec:results}}

\subsection{Cold plus warm dark matter
\label{sec:res_cwdm}}

{\it Main results.} For the CWDM model, we perform forecasts using the free parameters $m_{\rm wdm}^{\rm thermal}$ and $f_{\rm wdm}$ introduced in Sect.\,\ref{sec:theo_CWDM}. Our fiducial values and priors are summarised in Table\,\ref{tab:priors_cwdm} for these parameters and Table\,\ref{tab:priors} for all other free parameters. The fiducial model is chosen to be a pure $\Lambda$CDM model. As already stated in Sect.\,\ref{sec:theo_CWDM}, we use a linear prior on the mass and a logarithmic prior on the WDM fraction (that is, a flat priors on its logarithm). The latter choice allows us to explore the constraining power of \Euclid{} for very small WDM fractions \citep[see also][]{Schneider:2019xpf}. This limit is particularly interesting to study with \Euclid{} since, in this case, \Euclid{} bounds can be competitive with respect to Lyman-$\alpha$ bounds \citep{Hooper:2022byl}. 

\begin{table}[ht]
\centering
\caption{List of free parameters names, fiducial values, and top-hat prior ranges (in addition to those listed in Table\,\ref{tab:priors}) for the CWDM model. The fiducial values correspond to the pure $\Lambda$CDM limit.\label{tab:priors_cwdm}}
\begin{tabular}{ccc}
\hline
Parameter & Fiducial value & Range\\
\hline
\noalign{\vskip 2pt}
$m_{\rm wdm}^{\rm thermal}\,\,[{\rm eV}]$ & $\infty$ & $[10 , \, 1000]$ \\
$\log_{10} f_{\rm wdm}$ & $-\infty$ & $[-3,\,  0]$\\
\noalign{\vskip 2pt}
\hline
\end{tabular}
\end{table}

Indeed, Lyman-$\alpha$ data probe smaller scales than WL and GC surveys, and can in principle better constrain models with a large mass. However, when the WDM fraction is small, the effect of WDM on the power spectrum is also small. Then, it could be unconstrained by Lyman-$\alpha$ data, and if WDM is light enough its effects can manifest themselves on relatively large scales. In this case, the precise measurement of the power spectrum at larger scales with \Euclid{} remains decisive.

\begin{figure*}[t]
    \centering
    \includegraphics[width=0.9\textwidth]{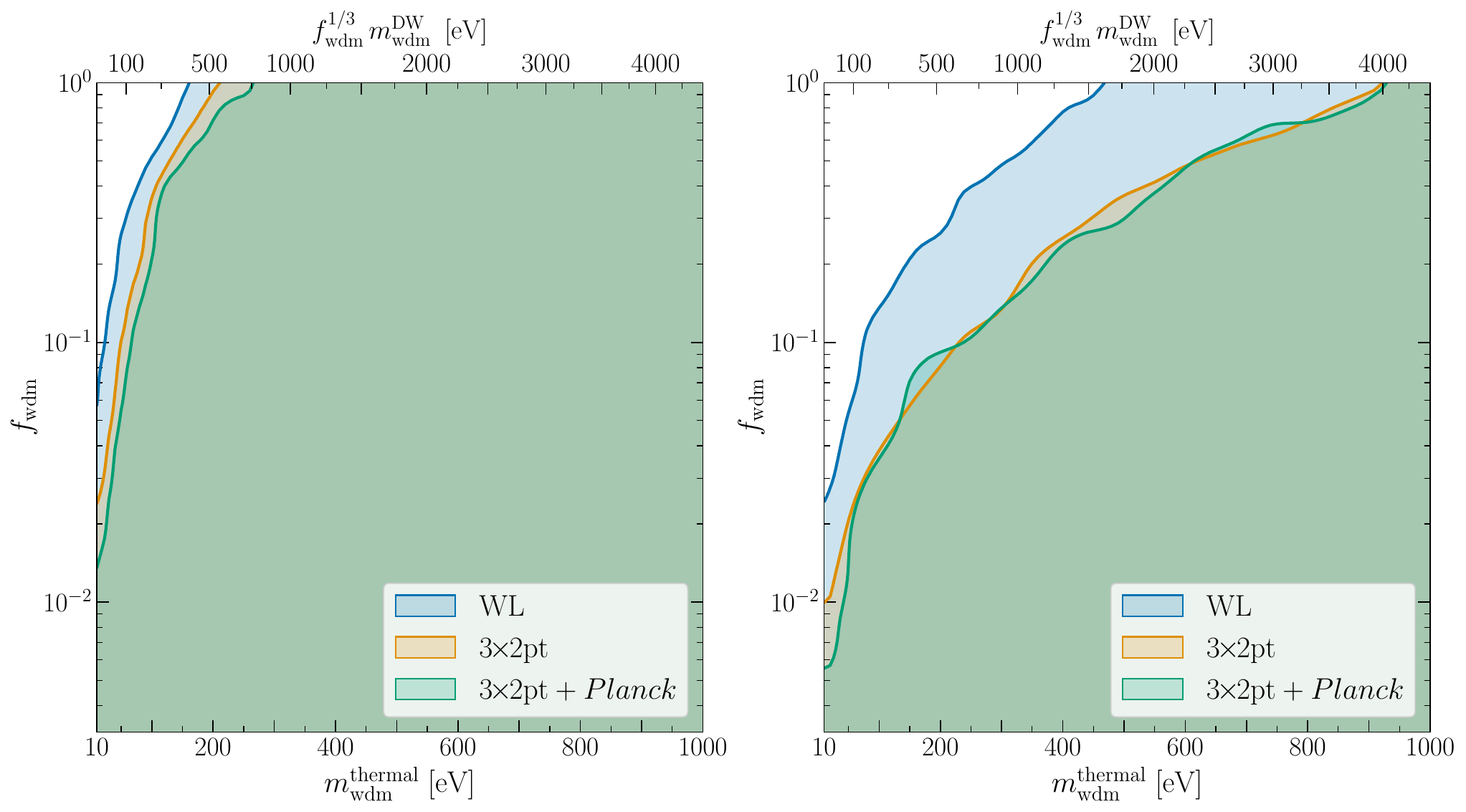}
    \caption{\emph{Left}: edges of the 95\% credible interval on the WDM mass $m_{\rm wdm}$ and fraction $f_{\rm wdm}$ for the CWDM model, with pessimistic assumptions and three data combinations: weak lensing (WL) alone, weak lensing plus galaxy clustering from the photometric survey (3\texttimes2pt), and 3\texttimes2pt combined with \Planck{} CMB data. For the 3\texttimes2pt and 3\texttimes2pt\,+\,\Planck{} data sets, baryonic feedback has been assumed to affect the WL power spectrum but not the GC power spectrum. The posterior is marginalised over other cosmological parameters, baryonic feedback parameters, and nuisance parameters (accounting for bias uncertainty and intrinsic alignment). The model is equivalent to pure $\Lambda$CDM towards the lower horizontal axis (small $f_{\rm wdm}$) and right vertical axis (large $m_{\rm wdm}$). The forecast assumes a flat prior on the mass of thermal WDM (lower axis) and a logarithmic prior on the WDM fraction (left axis), but we show the relation to Dodelson--Widrow masses in the upper axis (see Sect.\,\ref{sec:theo_CWDM} for definitions).
    \emph{Right}: same with optimistic assumptions.
    \label{fig:CWDM_pess_opt}
    }
\end{figure*}

Our main results are summarised in Fig.\,\ref{fig:CWDM_pess_opt}, where we show the marginalised 95\% credible intervals for $f_{\rm wdm}$ and $m_{\rm wdm}$. The horizontal axis can be interpreted as the thermal WDM mass (bottom axis) or as the Dodelson--Widrow mass (top axis, see Sect.\,\ref{sec:theo_CWDM} for details). The different colours of the contours mark the different probes that have been used to obtain the constraints. These contours are already marginalised over all other cosmological and nuisance parameters (including baryonic feedback parameters). For each of the six cases shown in Fig.\,\ref{fig:CWDM_pess_opt}, we ran 36 chains summing up to $\sim1.4$ millions of steps (MS) in each optimistic case or $\sim2.6$ MS in each pessimistic case. The Gelman-Rubin convergence criterium \citep{GelmanRubin} reached about $|R-1| \sim 0.002$ for most parameters, with a worse value of $\sim 0.008$ for a few parameters.

The constraints from WL are considerably looser (by about a factor five) in the pessimistic rather than optimistic case. Indeed, in the pessimistic case, the WL data is only fitted up to $\ell_{\rm max}=1500$, while models with a thermal mass of a few hundreds of keV only affect larger multipoles. As long as one sticks to pessimistic assumptions, adding information from GC (in the photometric survey) and on clustering-lensing correlations (3\texttimes2pt) makes a small difference, because in this case the clustering information is taken into account only up to $\ell_{\rm max}=750$. The addition of CMB data from \Planck{} also has a very small impact, given that CMB is sensitive to the clustering properties of pressureless  DM (including WDM) only at second order in perturbations, through CMB lensing effects -- as explained in \cite{Voruz:2013vqa}. Figure\,\ref{fig:CWDM_pess_opt} shows that in the pessimistic case, \Euclid{} 3\texttimes2pt\,+\,\Planck{} data have a potential to rule out masses $m_{\rm wdm}^{\rm thermal}\lesssim 280\,{\rm eV}$ (95\%CL) in the extreme case where these particles make up the totality of DM, or $m_{\rm wdm}^{\rm thermal}\lesssim 75\,{\rm eV}$ for $f_{\rm wdm}=0.1$ (95\%CL). Note that even with pessimistic assumptions, the WDM mass can be constrained even for WDM fractions slightly below 0.1, while current bounds from high-resolution Lyman-$\alpha$ data cannot distinguish models with $f_{\rm wdm}=0.1$ from the pure $\Lambda$CDM limit~\citep{Hooper:2022byl}.

The picture drastically improves when one assumes  optimistic settings with $\ell_{\rm max}=5000$ for WL and $\ell_{\rm max}=1500$ for GC. The data are then able to probe the presence of WDM with a much smaller value of the  maximum free-streaming scale, i.e., a larger mass. For the same WDM fraction, using WL data alone, the mass bounds become approximately five times tighter in the optimistic case. Despite of its limitation to $\ell_{\rm max}\leq1500$, GC data turns out to be very sensitive to the suppression induced by WDM even with a large mass, such that the 3\texttimes2pt probe is about twice more sensitive than the WL probe alone. However, the combination with  \Planck{} data makes no difference also in that case -- at least when the mock data is assumed to account for a pure CDM model. The reason is that, for the large WDM masses that remain compatible with the data, the maximum free-streaming scale of WDM is very low, such that even CMB lensing is unaffected by the suppression induced by WDM. In the optimistic case, the \Euclid{} 3\texttimes2pt probe has a potential to rule out all WDM masses  with $m_{\rm wdm}^{\rm thermal}\lesssim 930\,{\rm eV}$ for $f_{\rm wdm}=1$ and $m_{\rm wdm}^{\rm thermal}\lesssim 230\,{\rm eV}$ for $f_{\rm wdm}=0.1$. It can constrain the mass even when $f_{\rm wdm}$ is as low as a few times $10^{-2}$, i.e., when only a few percents of the total DM is warm. This region of parameter space is far from current Lyman-$\alpha$ bounds, and even future Lyman-$\alpha$ surveys are unlikely to probe such small WDM fractions.

It is still unclear whether the final \Euclid{} sensitivity will be closer to that of our pessimistic or optimistic forecast. At least, we expect that these two forecasts are bracketing the true constraining power of the future data. We will see that, compared to other non-minimal DM models discussed in the next sections, CWDM is particularly sensitive to the choice of a cut-off multipole $\ell_{\rm max}$. This is due to the step-like nature of the effect of WDM on the matter power spectrum: up to a given wavenumber, the $\Lambda$CDM and CWDM models are strictly equivalent, and then the power drops. This means that the constraining power of a data set on the CWDM parameters depends more on the minimum scale (and thus maximum multipole and redshift) included in the analysis than on the actual error bars on the power spectrum. As discussed above, this is particularly true for large values of $f_{\rm wdm}$; for tiny WDM fractions, the precision with which the power spectrum is constrained remains crucial.

\begin{figure*}[t]
    \centering
    \includegraphics[width=0.9\textwidth]{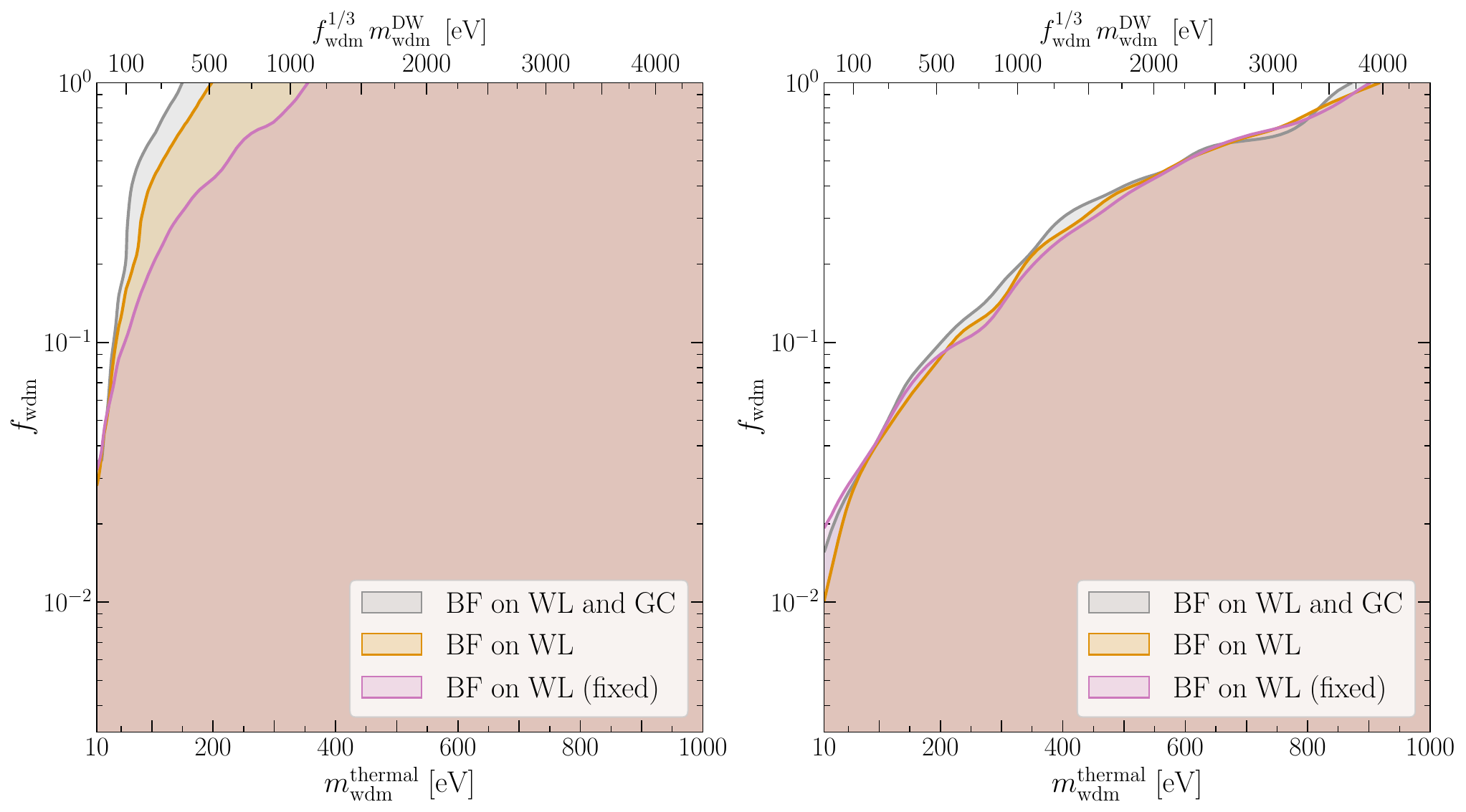}
    \caption{\emph{Left}: same as Fig.\,\ref{fig:CWDM_pess_opt} but only for the 3\texttimes2pt dataset and with different assumptions on baryonic feedback (BF): fixed BF (magenta), BF affecting only the weak lensing (WL) power spectrum (orange), or BF affecting both the WL and galaxy clustering  (GC) power spectra (grey). The ``truth'' is expected to lay between the latter two cases (orange and grey), which give anyway very similar results.
    \emph{Right}: same with optimistic assumptions.
    \label{fig:CWDM_BF_comparison}
    }
\end{figure*}

{\it Importance of baryonic feedback.} In Fig.\,\ref{fig:CWDM_BF_comparison}, we evaluate the impact of different assumptions concerning baryonic feedback effects. We compare the bounds derived from the 3\texttimes2pt probe only under three assumptions. The baseline case (orange contours) is the same as in our previous discussion and in Fig.\,\ref{fig:CWDM_pess_opt}: the six nuisance parameters describing baryonic feedback are marginalised over, and baryonic feedback is assumed to affect only the WL probe, i.e., the total matter power spectrum. The grey contours are derived assuming instead that baryonic feedback affects the two probes (WL and GC) in the same way: in other words, the same \bcemu{} corrections are applied to the total matter and galaxy power spectra. Finally, the magenta contours were obtained with fixed rather than marginalised baryonic feedback parameters: they account for the unrealistic situation in which baryonic feedback effects would be perfectly known, given some independent measurements.

In principle, introducing more freedom in baryonic physics may result in looser constraints on WDM parameters due to parameter degeneracies. On the other hand, it is not obvious that such degeneracies are present due to the different shape of the effects imprinted by either WDM or baryons, not only as a function of scale but also as a function of redshift. In particular, the redshift dependence of the DM-induced suppression is reversed compared to the one from baryonic effects; its amplitude gets smaller at smaller redshift, due to non-linear clustering and mode-mode coupling, while overall the opposite is true for baryons, at least in most of the redshift range probed by \Euclid{}.

We first compare the orange and magenta contours. In the pessimistic case (left panel), we find that introducing more freedom in the baryonic model and marginalising over baryonic feedback parameters does degrade a bit the bounds on WDM parameters. However, this is no longer true in the optimistic case, which proves that the information contained in the data at large mutipoles is sufficient to disentangle between the physical effects of WDM and baryonic feedback.
This underlines the wealth of cosmological information encoded in the deep non-linear regime of the power spectrum.

We now switch to the comparison between the orange and grey contours. The impact of baryonic feedback on the galaxy power spectrum is not understood and modelled as well as its impact on the total matter powers spectrum. However, as discussed in Sect.\,\ref{sec:baryon}, we expect baryonic effects to be smaller in the galaxy powers spectrum -- or at least not bigger. Thus, the true sensitivity of \Euclid{} should lay somewhere between the forecasts corresponding to the grey and orange contours. However, these contours are very close to each other in both pessimistic and optimistic cases. This is likely due to the reverse redshift dependence of WDM and baryonic effects, which allows GC data to discriminate between them. Such a test validates the bounds on CWDM parameters obtained in the previous paragraphs with baryonic feedback included only in the WL probe.

\begin{figure}[ht]
    \centering
    \includegraphics[width=0.45\textwidth]{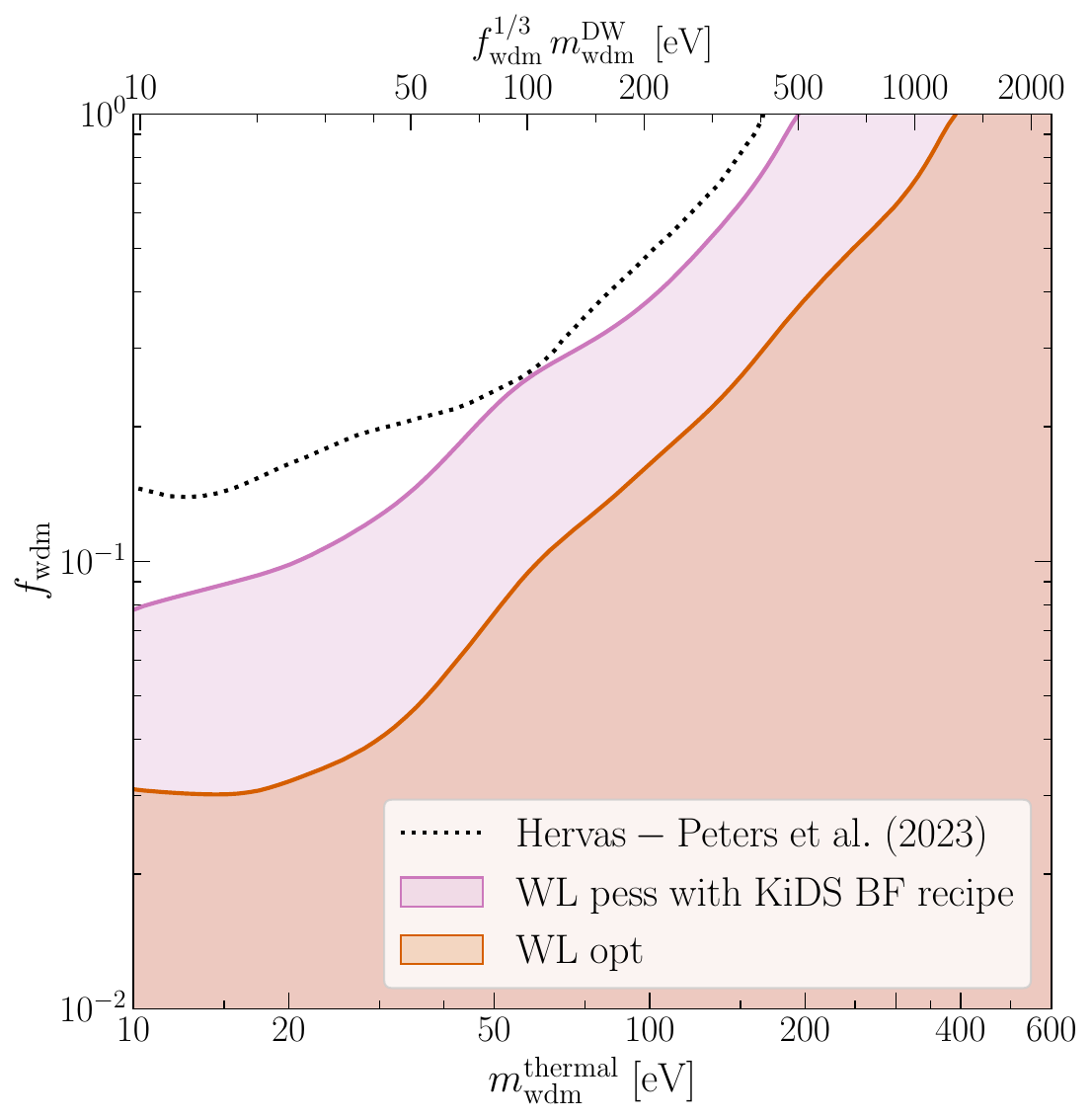}
    \label{fig:wdm_kids_comaprison}
    \caption{Comparison of our \Euclid{} forecasts for the WL-only probe with current bounds from KiDS. In this particular case, we switch to the same top-hat prior on $\log_{10} f_{\rm wdm}^{\rm ini}$ and $\log_{10} (m_{\rm wdm}^{\rm thermal}/{\rm eV})$ as in the KiDS analysis of \cite{Peters_2023}. In the pessimistic case, we also adopt the same baryonic feedback recipe as \cite{Peters_2023} with a marginalisation over three baryonic feedback parameters (instead of six in our baseline treatment). 
    \label{fig:cwdm_kids}
    }
\end{figure}

{\it Comparison with current bounds.} It is interesting to compare the expected sensitivity of \Euclid{} to current bounds on CWDM parameters obtained by \cite{Peters_2023} using an existing WL survey, the Kilo Degree Survey (KiDS) -- see \cite{KiDS:2020suj}. We start by comparing the sensitivity of the \Euclid{} WL probe alone with that of KiDS. A fair comparison requires similar priors. However, the KiDS analysis assumes logarithmic priors on both the WDM mass and fraction, with slightly different prior edges than in our previous analysis. Given that Bayesian credible intervals do depend on priors, especially when constraining some parameters describing a model extension  that is not required by the data, we repeat some of our forecasts with different top-hat priors matching exactly the ones of KiDS: $\log_{10} f_{\rm wdm} \in [\log_{10}(0.005), \, \log_{10}(1)]$ and $\log_{10} (m_{\rm wdm}^{\rm thermal}/{\rm keV}) \in [\log_{10}(10), \, \log_{10}(1.5)]$. The results are shown in Fig.\,\ref{fig:cwdm_kids}.

In the pessimistic case, we find that the \Euclid{} sensitivity is not so different from that of KiDS. This result may sound surprising, given the much larger number of galaxy images expected from \Euclid{}. There are two reasons for this. 

The first reason is that, as explained before, in the case of CWDM, the bounds depend a lot on the minimum scale (or maximum wavenumber $k_{\rm max}$) included in the analysis, more than on the error bar on the measured power spectrum. It also depends on the maximum redshift of the data, since the signature of WDM is more clear at high redshift and partially washed out  at small redshift. Note also that a given multipole $\ell$ probes smaller wavenumbers at high redshift, as shown in Eq.\,\eqref{eq:k_l}. In the KiDS analysis and in our \Euclid{} forecast with pessimistic assumptions, the data is conservatively cut at $\ell_{\rm max}=1500$. Thus, within the redshift range covered by both experiments, the maximum wavenumber at each redshift $k_{\rm max}(\ell_{\rm max}, z)$ is the same, and the sensitivity to WDM is roughly similar despite of the smaller \Euclid{} error bars. For instance,  the highest redshift bin of KiDS peaks around $z\sim 1.1$, probing up to $k \sim 1\,h\,{\rm Mpc}^{-1}$. \Euclid{} adds information at higher redshift, with the highest redshift bin peaking around $z\sim 1.7$, but at such a high redshift the multipole $\ell_{\rm max}=1500$ projects only to $k \sim 0.5\,h\,{\rm Mpc}^{-1}$, such that the information gain on the CWDM model is marginal. Note that the arguments presented in this paragraph are only valid in the case of the \Euclid{} pessimistic case and for the CWDM model, which has the same power spectrum as $\Lambda$CDM up to some large wavenumber $k$ (given by the free-streaming scale). For instance, we shall see that for the 1b-DDM model \Euclid{} is much more constraining than KiDS even with pessimistic assumption because, in that case, the power spectrum contains information on the DM parameters on larger scales / smaller wavenumbers.

A second reason is that, in our analysis, we marginalise over six baryonic feedback parameters, including the three $\nu_{\cal B}$ parameters accounting for a drift of the main feedback parameters ${\cal B}$ with redshift. In the KiDS analysis, the three ${\cal B}$ parameters are instead assumed to be redshift-independent. However, the effect of WDM on the matter power spectrum is redshift-dependent (it decreases when the redshift decreases due to mode-mode coupling). Thus, in our analysis, it is easier to cancel the effect of WDM with a shift in the baryonic feedback parameter, and there is more degeneracy between the WDM and baryonic parameters. This tends to lower our forecasted sensitivity and to compensate the fact that \Euclid{} measurements of the lensing power spectrum are expected to be much more accurate. In Fig.\,\ref{fig:cwdm_kids}, we choose to present the results of the \Euclid{} pessimistic forecast with the same treatment as in KiDS analysis of \cite{Peters_2023}, i.e., with only three free baryonic feedback parameters, while in the \Euclid{} optimistic forecast we stick to our more conservative baseline treatment with a marginalisation over six parameters ${\cal B}$ and $\nu_{\cal B}$.

In the optimistic case, we still find that the sensitivity of the \Euclid{} WL probe is approximately three times bigger than that of KiDS, despite of our more conservative modelling of baryonic feedback. In addition, we have already seen that with the inclusion of all the information from the 3\texttimes2pt probe, we can gain a factor three in sensitivity. All in all, under optimistic assumptions, \Euclid{} could be about ten times more constraining, while providing at the same time some bounds that will be more robust against baryonic feedback effects. Furthermore, note that the results of our optimistic forecast confirm previous findings from \citet{Schneider:2019xpf} using more realistic assumptions for the survey characteristics, especially regarding the tomographic galaxy binning.

\subsection{Dark matter with one-body decay\label{sec:res_1bddm}}

{\it Main results.} For the 1b-DDM model, we perform forecasts using the parameter combinations $f_{\rm ddm}^{\rm ini}$ and $\Gamma_{\rm ddm} \, f_{\rm ddm}^{\rm ini}$ defined in Sect.\,\ref{sec:theo_1bddm}. The fiducial model is chosen to be a pure $\Lambda$CDM model. We showed in Sect.\,\ref{sec:nl_1bddm} that the small-scale suppression induced by 1b-DDM on the linear power spectrum depends only on the product $\Gamma_{\rm ddm} \, f_{\rm ddm}^{\rm ini}$, while the non-linear evolution adds a bit of sensitivity to $\Gamma_{\rm ddm}$ alone. Thus, the data are expected to provide bounds mainly on the product of the two DDM parameters. Our fiducial values and priors are summarised in Table\,\ref{tab:priors_1bddm} for these parameters and Table\,\ref{tab:priors} for all other free parameters. The fiducial model is chosen to be a pure $\Lambda$CDM model. As already stated in Sect.\,\ref{sec:theo_1bddm}, we use linear priors on $f_{\rm ddm}^{\rm ini}$ and $\Gamma_{\rm ddm} \, f_{\rm ddm}^{\rm ini}$. Note that the $N$-body simulations used by \cite{Hubert_2021} to build the emulator were limited to models with a DM lifetime much larger than the age of the Universe, namely $\tau_{\rm ddm}\geq 31.6\,{\rm Gyr}$. We thus conservatively include in our run a prior $\Gamma_{\rm ddm} \leq 0.0316\,{\rm Gyr}^{-1}$. This additional prior excludes a small triangle in the parameter space defined by the priors of Table\,\ref{tab:priors_1bddm}.

\begin{table}[ht]
\centering
\caption{List of free parameters names, fiducial values, and top-hat prior ranges (in addition to those listed in Table\,\ref{tab:priors}) for the 1b-DDM model. The fiducial values correspond to the pure $\Lambda$CDM limit. In addition to the top-hat priors on $f_{\rm ddm}^{\rm ini}$ an $\Gamma_{\rm ddm} \, f_{\rm ddm}^{\rm ini}$ reported in the table, we use an extra prior $\Gamma_{\rm ddm} < 0.0316\,{\rm Gyr}^{-1}$ to remain in the region were the emulator was trained on $N$-body simulations.
\label{tab:priors_1bddm}}
\begin{tabular}{ccc}
\hline
Parameter & Fiducial value & Range\\
\hline
\noalign{\vskip 2pt}
$f_{\rm ddm}^{\rm ini}$ & 0 & $[0 ,\, 1]$ \\
$\Gamma_{\rm ddm} \, f_{\rm ddm}^{\rm ini} \,\,\, [{\rm Gyr}^{-1}]$ & 0 & $[0,\, 10^{-2}]$ \\
\noalign{\vskip 2pt}
\hline
\end{tabular}
\end{table}

\begin{figure*}[t]
    \centering
    \includegraphics[width=0.85\textwidth]{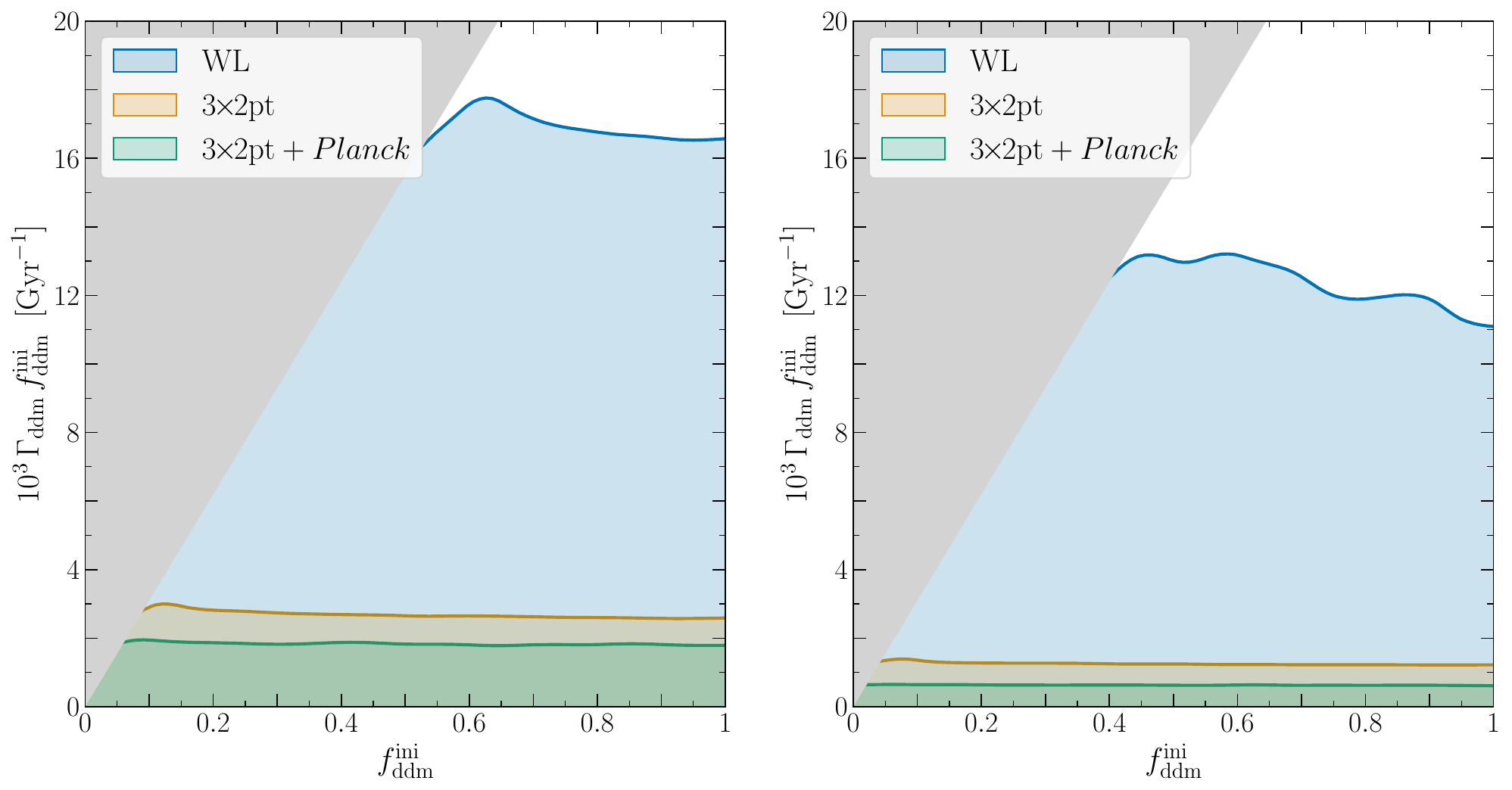}
    \caption{Same as Fig.\,\ref{fig:CWDM_pess_opt} for the 1b-DDM model, parameterised by the DDM fraction $f_{\rm ddm}^{\rm ini}$ and decay rate $\Gamma_{\rm ddm}$. The forecast assumes flat priors on ($f_{\rm ddm}^{\rm ini}$, $\Gamma_{\rm ddm} \, f_{\rm ddm}^{\rm ini}$) because the effect of 1b-DDM on the linear matter power spectrum scales with the product $\Gamma_{\rm ddm} \, f_{\rm ddm}^{\rm ini}$ (see Sect.~\ref{sec:theo_1bddm}). The model is equivalent to pure $\Lambda$CDM in the small $\Gamma_{\rm ddm} \, f_{\rm ddm}^{\rm ini}$ limit. The shaded grey area restricts the parameter space to the region where $\tau_{\rm ddm}= 1 / \Gamma_{\rm ddm} \geq 31.6\,{\rm Gyr}$ in which the emulator was trained.
    \label{fig:1b_pess_opt}
    }
\end{figure*}

Figure\,\ref{fig:1b_pess_opt} presents the 95\% confidence level (CL) isocontours of the marginalised posterior for the 1b-DDM parameters inferred from our forecasts for the \Euclid{} WL probe alone (WL), the full \Euclid{} photometric probe (3\texttimes2pt), and \Euclid{} 3\texttimes2pt combined with \Planck, considering both pessimistic (left panel) or optimistic (right panel) assumptions. The fiducial model, $\Lambda$CDM, spans the lower horizontal axis ($\Gamma_{\rm ddm} \, f_{\rm ddm}^{\rm ini}=0$). 
The shaded grey area restricts the parameter space to the region in which the emulator was trained. The fact that the contour edges remain nearly horizontal is consistent with the fact that 1b-DDM effects depend mainly on the product $\Gamma_{\rm ddm} \, f_{\rm ddm}^{\rm ini}$. The small tilting of the contours comes from the fact that non-linear corrections to the 1b-DDM effects do depend on $\Gamma_{\rm ddm}$ alone. For each of the six cases shown in Fig.\,\ref{fig:1b_pess_opt}, we ran 96 chains summing up to $\sim1.6$ MS in each optimistic case or $\sim3.8$ MS in each pessimistic case. The Gelman-Rubin convergence criterium reached about $|R-1| \sim 0.01$ for most parameters, with a worse value of 0.05 for a few parameters.

In the pessimistic case, the WL-only analysis provides a 95\%CL bound close to $\Gamma_{\rm ddm} \, f_{\rm ddm}^{\rm ini}<8 \times 10^{-3}\,{\rm Gyr}^{-1}$. Incorporating the 3\texttimes2pt data set leads to a substantially stronger bounds,  by approximately a factor of 2, such that $\Gamma_{\rm ddm} \, f_{\rm ddm}^{\rm ini}<4 \times 10^{-3}$. There is an additional factor of 2 improvement when \Planck{} data are integrated into the analysis, requiring $\Gamma_{\rm ddm} \, f_{\rm ddm}^{\rm ini}<1.75 \times 10^{-3}{\rm Gyr}^{-1}$. Switching to optimistic assumptions makes a substantial difference for the WL only bound, which shrinks to $\Gamma_{\rm ddm} \, f_{\rm ddm}^{\rm ini}<6 \times 10^{-3}\,{\rm Gyr}^{-1}$, and an even stronger difference for the 3\texttimes2pt bound, which reaches $\Gamma_{\rm ddm} \, f_{\rm ddm}^{\rm ini}<0.75 \times 10^{-3}\,{\rm Gyr}^{-1}$. \Planck{} further improves this bound down to $\Gamma_{\rm ddm} \, f_{\rm ddm}^{\rm ini}<0.5 \times 10^{-3}\,{\rm Gyr}^{-1}$.

\begin{figure*}[t]
    \centering
    \includegraphics[width=0.99\textwidth]{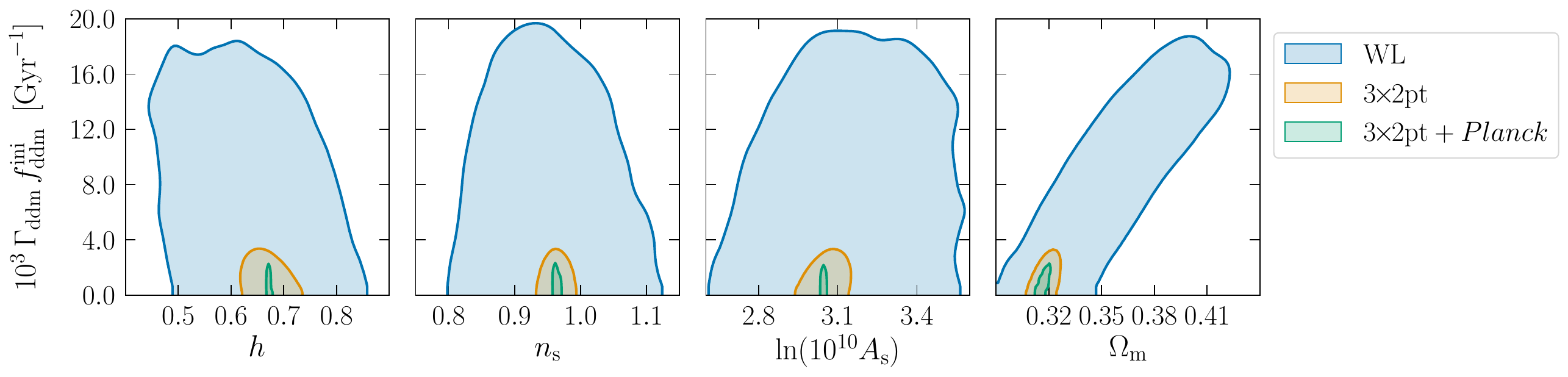}
    \includegraphics[width=0.99\textwidth]{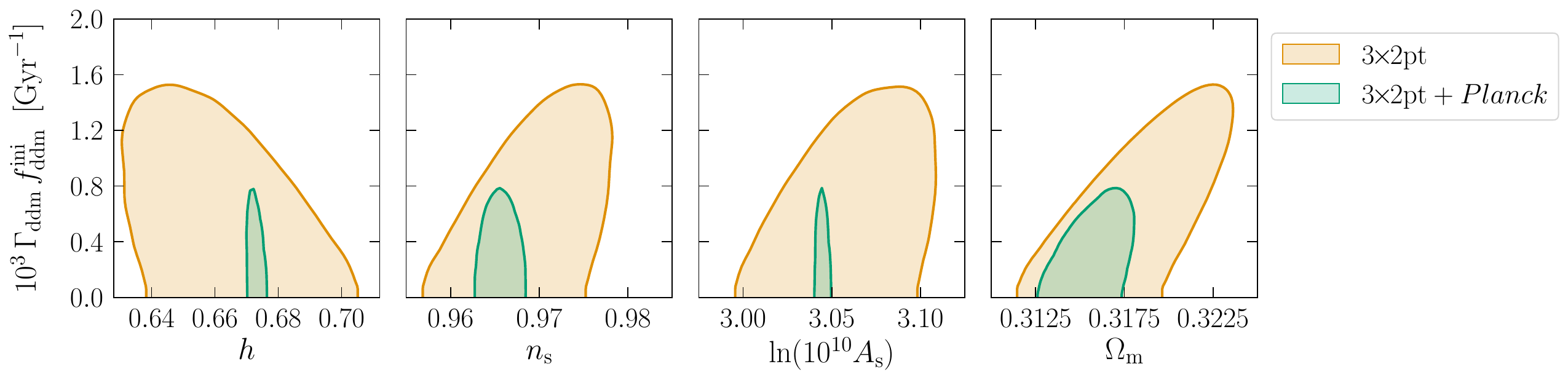}
    \caption{Degeneracies between the 1b-DDM parameter $\Gamma_{\rm ddm} \, f_{\rm ddm}^{\rm ini}$ and four other cosmological parameters for different data sets. \emph{Top}: optimistic case. \emph{Bottom}: pessimistic  case. The addition of 3\texttimes2pt to WL and of \Planck{} to 3\texttimes2pt leads to a better determination of all cosmological parameters and lifts degeneracies.
    \label{fig:1bddm_deg_planck}
    }
\end{figure*}

The first conclusion emerging from these results is that the photometric GC data has a large constraining power compared to the WL data for this particular model. This is illustrated by the factor 8 improvement when switching from WL to 3\texttimes2pt data in the optimistic case. The 2-dimensional likelihood contours shown in the upper panel of Fig.\,\ref{fig:1bddm_deg_planck} show that, with WL data alone, the parameter $\Gamma_{\rm ddm} \, f_{\rm ddm}^{\rm ini}$ is degenerate with cosmological parameters like, for instance, $n_{\rm s}$ or $\Omega_{{\rm m}}$. The addition of GC data is beneficial for two reasons: on the one hand, it adds sensitivity to these parameters and helps removing such degeneracies; on the other hand, it directly probes the 1b-DDM effects on the matter power spectrum, up to smaller wavenumbers $k$ than WL data but with better sensitivity.

Another interesting conclusion is that there is a good synergy between the \Planck{} and \Euclid{} probes for this model.
Note that, using \Planck{} 2018 data alone, \cite{Simon:2022ftd} and \cite{bucko_2022_1bddm} found $\Gamma_{\rm ddm} \, f_{\rm ddm}^{\rm ini}<4 \times 10^{-3}\,{\rm Gyr}^{-1}$. Thus, the \Euclid{} 3\texttimes2pt probe alone is already more constraining than \Planck. In addition, the combination of the two data sets is significantly  more constraining than each data set taken individually. This is usually the consequence of parameter degeneracies being removed by the combination. We get a confirmation of this by looking at the upper and lower panels of Fig.\,\ref{fig:1bddm_deg_planck}. The contours illustrate the existence of correlations between $\Gamma_{\rm ddm} \, f_{\rm ddm}^{\rm ini}$ and other cosmological parameters. The addition of \Planck{} data resolves these degeneracies and pushes the bounds beyond those from \Euclid{} alone -- even if \Planck{} alone is not directly sensitive to such small $\Gamma_{\rm ddm} \, f_{\rm ddm}^{\rm ini}$ values.
 
\begin{figure*}[t]
    \centering
    \includegraphics[width=0.85\textwidth]{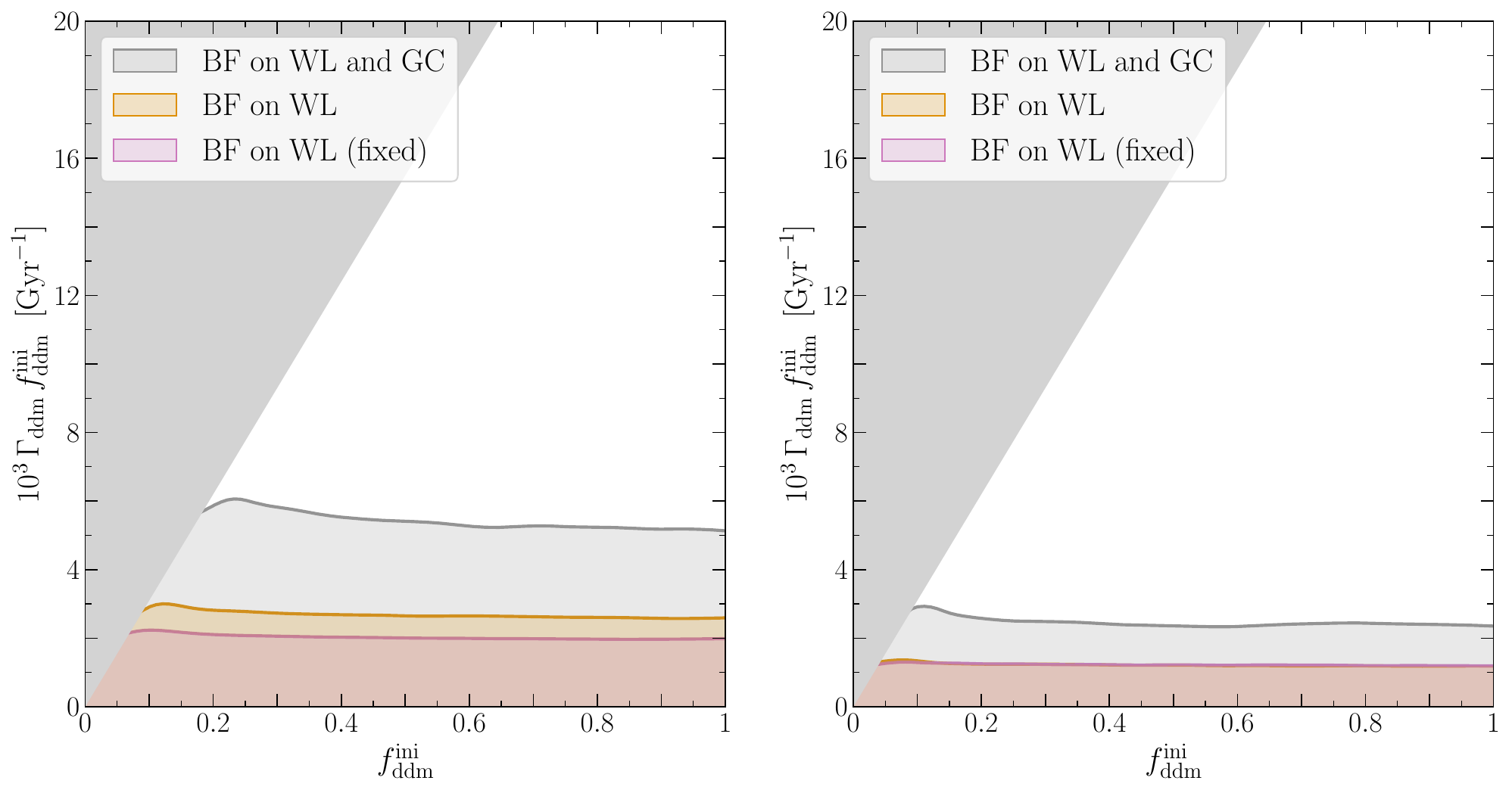}
    \caption{Same as Fig.\,\ref{fig:CWDM_BF_comparison} for the parameters of the 1b-DDM model. Unlike in the case of CWDM, we find that the various assumptions on baryonic feedback have a big impact on the upper bound on $\Gamma_{\rm ddm} \, f_{\rm ddm}^{\rm ini}$.
    \label{fig:1b_BF_comparison}
    }
\end{figure*}

{\it Importance of baryonic feedback.} In Fig.\,\ref{fig:1b_BF_comparison}, we show the impact of marginalisation over baryonic feedback parameters, in the same way as we did in Fig.\,\ref{fig:CWDM_BF_comparison} for CWDM. We compare the bounds derived from the 3\texttimes2pt probe with either marginalised baryonic feedback effects only for the WL probe (orange) or for the full 3\texttimes2pt probe (grey), or with fixed baryonic feedback effects (magenta).

There is a qualitative difference between this model and the CWDM case. In the former case, the suppression of the small-scale matter power spectrum is caused by WDM free-streaming during early cosmological times. However, this suppression tends to be washed out at small redshift by non-linear clustering and mode-mode coupling. As redshift decreases, the CWDM matter power spectrum gets gradually closer to that of $\Lambda$CDM. This is not the case in the 1b-DDM model, since the DM decay occurs mainly at very late times. Then, the modifications to the non-linear matter power spectrum get more and more pronounced as time passes by -- which also tends to be the case for baryonic effects, at least in most of the redshift range probed by \Euclid{}. In principle, this enhances the possibility that 1b-DDM and baryonic effects can compensate each other and that degeneracies are present in parameter space.

As a matter of fact, in the pessimistic case, the bounds are different under the three assumptions. This confirms that baryonic feedback and 1b-DDM effects are partially degenerate, and that the marginalisation over baryonic feedback parameters weakens the bounds. Assuming baryonic feedback effects also on the galaxy clustering spectrum weakens the bound on $\Gamma_{\rm ddm} \, f_{\rm ddm}^{\rm ini}$ by approximately 25\%, such that a conservative estimate in this case is $\Gamma_{\rm ddm} \, f_{\rm ddm}^{\rm ini}<5\times 10^{-3}\,{\rm Gyr}^{-1}$ (95\%CL). 

In the optimistic case, we see that the 3\texttimes2pt data contains enough information to remove the degeneracy between 1b-DDM and baryonic feedback parameters when baryonic feedback is applied to the WL probe only, but not when it is applied also to the GC probe. In this case, the true bound is expected to stand between the one discussed in the previous paragraph, $\Gamma_{\rm ddm} \, f_{\rm ddm}^{\rm ini}<0.75\times 10^{-3}\,{\rm Gyr}^{-1}$, and the more conservative one found here, $\Gamma_{\rm ddm} \, f_{\rm ddm}^{\rm ini}<2\times 10^{-3}\,{\rm Gyr}^{-1}$ (95\%CL). 

\begin{figure}[ht]
    \centering
    \includegraphics[width=0.43\textwidth]{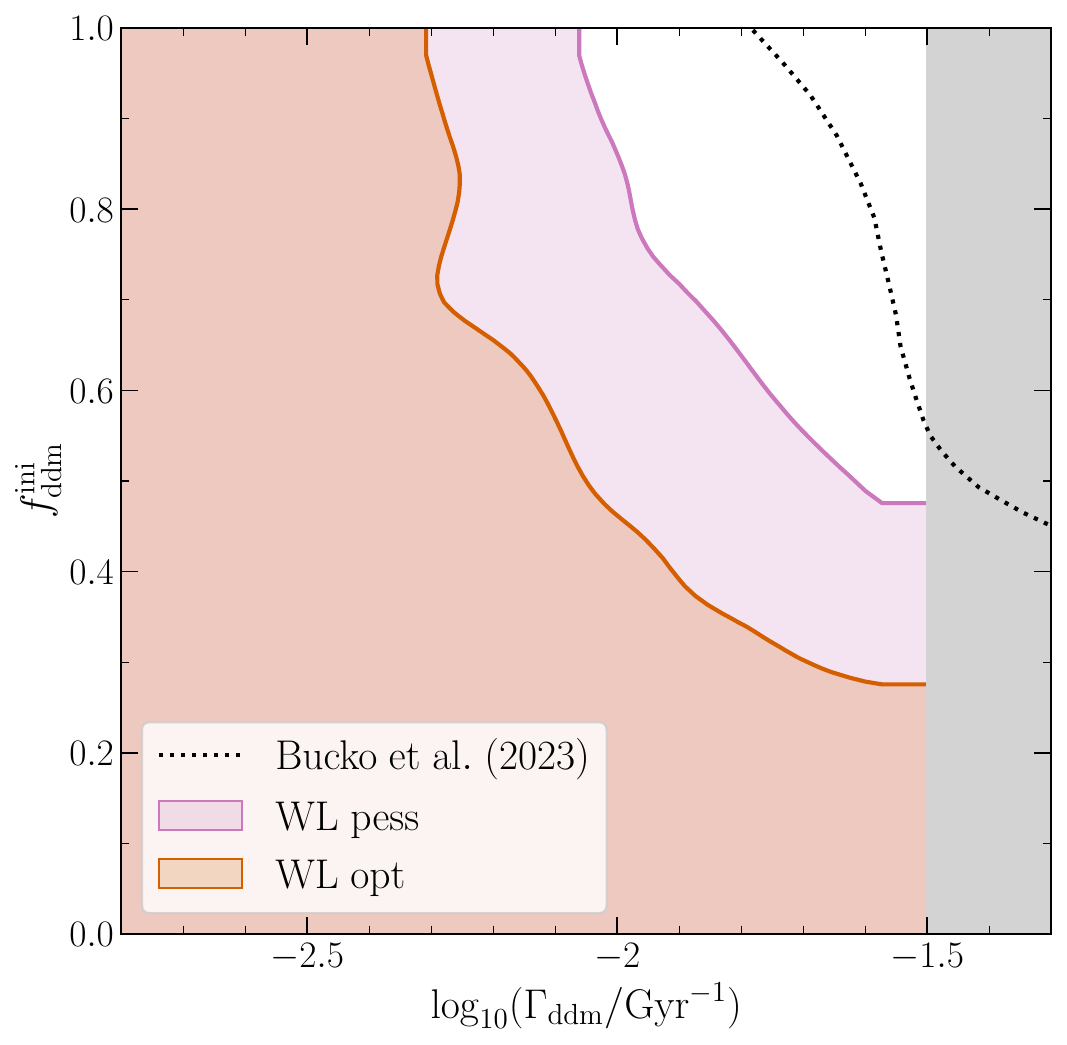}
    \caption{Comparison of bounds from \Euclid{} WL-only (optimistic or pessimistic) and KiDS WL-only \citep{bucko_2022_1bddm} on the 1b-DDM parameters, using the same priors for all three cases (logarithmic on the DDM decay rate, linear on the DDM fraction). The shaded grey area restricts the parameter space to the region where $\tau_{\rm ddm}= 1 / \Gamma_{\rm ddm} \geq 31.6\,{\rm Gyr}$ in which the emulator was trained.
    \label{fig:1bDDM_kids}
    }
\end{figure}

{\it Comparison with current bounds.} For the 1b-DDM model, \cite{Simon:2022ftd} found $\Gamma_{\rm ddm} \, f_{\rm ddm}^{\rm ini}<4\times 10^{-3}\,{\rm Gyr}^{-1}$ when using \Planck{} alone, and no significant improvement when adding information from Type Ia supernovae, baryon acoustic oscillations from a variety of surveys, redshift space distortions from the extended Baryon Oscillation Spectroscopic Survey (eBOSS), and even the full shape of the power spectrum from the Baryon Oscillation Spectroscopic Survey (BOSS). \cite{bucko_2022_1bddm} find no improvement either when adding KiDS data. Additionally, \cite{bucko_2022_1bddm} find that KiDS alone only provides a bound of the order of $\Gamma_{\rm ddm} \, f_{\rm ddm}^{\rm ini}<3\times 10^{-2}\,{\rm Gyr}^{-1}$, almost one order of magnitude weaker than \Planck. This shows that current large-scale structure observations have much less constraining power than current CMB data for this particular model -- a situation very different from that of CWDM. 

In this context, the sensitivity that will be reached by \Euclid{} according to our forecast is remarkable. \Euclid{} WL-only will improve KiDS WL-only bounds by a factor 4 (pessimistic) or 5 (optimistic).\footnote{We cross-checked this statement by running our \Euclid{} WL forecasts with exactly the same top-hat priors on $\log_{10}(\Gamma_{\rm dcdm}/{\rm Gyr}^{-1})$ and $f_{\rm ddm}^{\rm ini}$ as \cite{bucko_2022_1bddm}, see Fig.\,\ref{fig:1bDDM_kids}.} The full \Euclid{} 3\texttimes2pt probe will have the same sensitivity as current CMB data (pessimistic) or improve the bound by a factor 2 to 3 (optimistic). The combined \Euclid{} 3\texttimes2pt\,+\,\Planck{} data will improve over current bounds by a factor 4 (pessimistic) to 8 (optimistic).

We see that \Euclid{} has an even greater potential to improve over current bounds for 1b-DDM than for CWDM. This is related to the shape of the 1b-DDM effects on the matter power spectrum, already displayed in Fig.\,\ref{fig:lin_1b}. The survey probes 1b-DDM effects through the entire shape of the power spectrum over the full range of measured linear and non-linear scales. Unlike the CWDM spectrum, the 1b-DDM spectrum is not identical to the $\Lambda$CDM one up to a given free-streaming scale. Thus, the constraints on 1b-DDM benefit from the unprecedented accuracy expected from \Euclid{} data over the entire range of scales probed by the survey. 

With this model, the sensitivity of \Euclid{} offers an opportunity not only to reconstruct the matter power spectrum accurately over a broad range of scales, but also to disentangle between 1b-DDM and baryonic effects. The strong sensitivity improvement of \Euclid versus KiDS is partly due to the fact that there is a significant degeneracy between 1b-DDM and baryonic feedback parameter, which \Euclid is able to resolve much better than KiDS. We already explained that baryonic feedback should be more degenerate with 1b-DDM effects than with CWDM effects. Thus, in the CWDM case, there is no such factor and the \Euclid{} sensitivity remains closer to the KiDS one at least in the pessimistic case.

\subsection{Dark matter with two-body decay\label{sec:res_2bddm}}

{\it Main results.} For the 2b-DDM model, we perform forecasts using the parameters ($f_{\rm ddm}^{\rm ini}$, $\Gamma_{\rm ddm}$, $\varepsilon$) defined in Sect.\,\ref{sec:theo_2bddm}, with logarithmic priors defined in Table\,\ref{tab:priors_2bddm} for these parameters and Table\,\ref{tab:priors} for all other free parameters. As discussed in Sect.\,\ref{sec:theo_2bddm}, at the level of the linear spectrum, 2b-DDM induces a step-like suppression in the power spectrum with an amplitude controlled by ($f_{\rm ddm}^{\rm ini}$, $\Gamma_{\rm ddm}$) and a scale depending on $\varepsilon$. At the non-linear level the effects are more intricate and the suppression depends on all three parameters. The fiducial model of our forecast is chosen to be a pure $\Lambda$CDM model.  

\begin{table}[ht]
\centering
\caption{List of free parameters names, fiducial values, and top-hat prior ranges (in addition to those listed in Table\,\ref{tab:priors}) for the 2b-DDM model. The fiducial values correspond to the pure $\Lambda$CDM limit. The upper prior edges on $\log_{10}(\Gamma_{\rm ddm}/{\rm Gyr}^{-1})$ and $\log_{10}\varepsilon$ restrict the parameter space to the region in which the emulator was trained.
\label{tab:priors_2bddm}}
\begin{tabular}{ccc}
\hline
Parameter & Fiducial value & Range \\
\hline
\noalign{\vskip 2pt}
$\log_{10}f_{\rm ddm}^{\rm ini}$ & $-\infty$ & $[-1.3 ,\, 0]$ \\
$\log_{10}(\Gamma_{\rm ddm}/ {\rm Gyr}^{-1})$ & $-\infty$ & $ [-2.8,\, -1.13]$ \\
$\log_{10}\varepsilon$ & $-\infty$ & $[-3.5,\, -1.8]$ \\
\noalign{\vskip 2pt}
\hline
\end{tabular}
\end{table}

\begin{figure*}[t]
    \centering
    \includegraphics[width=0.49\textwidth]{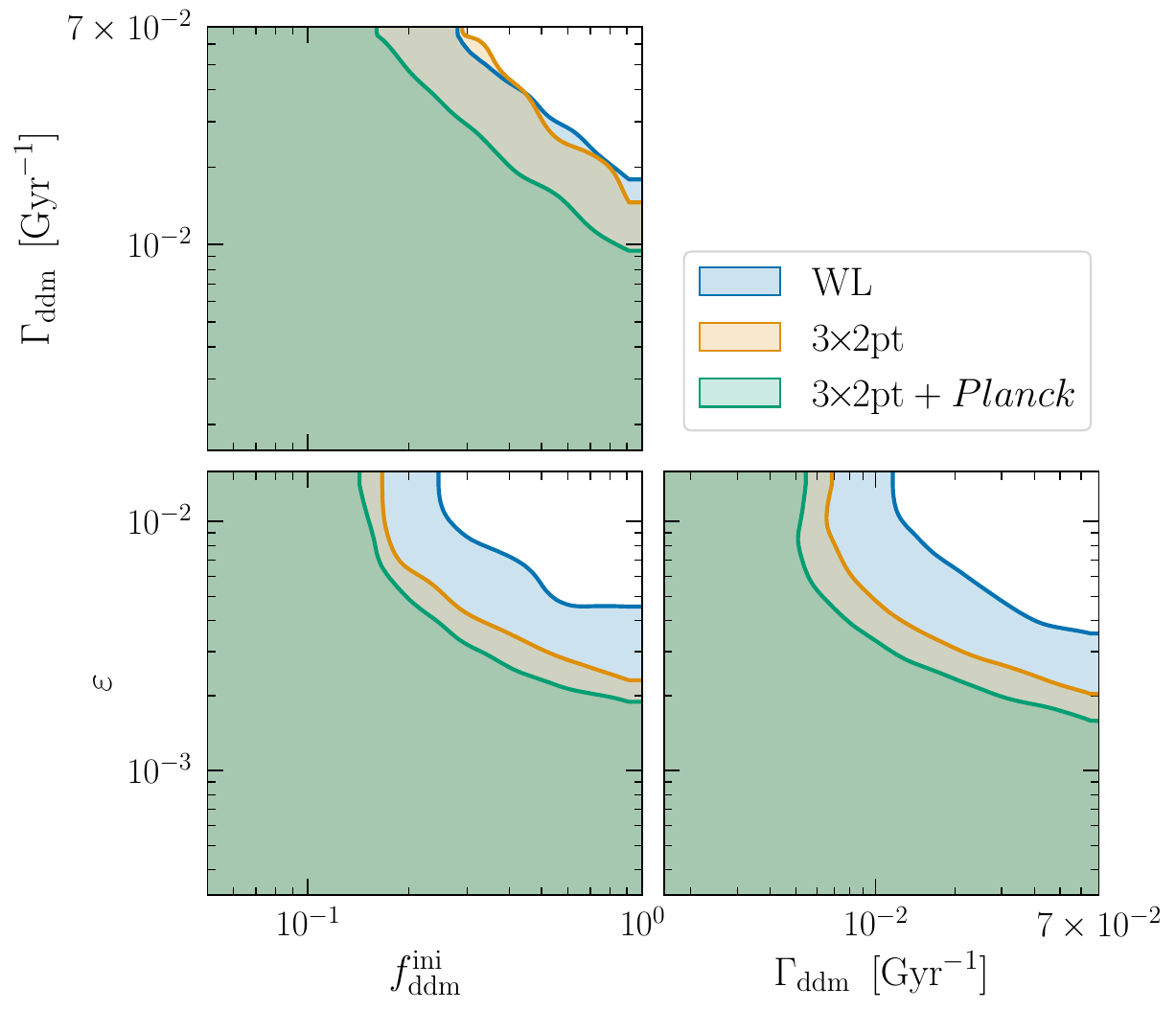}
    \includegraphics[width=0.49\textwidth]{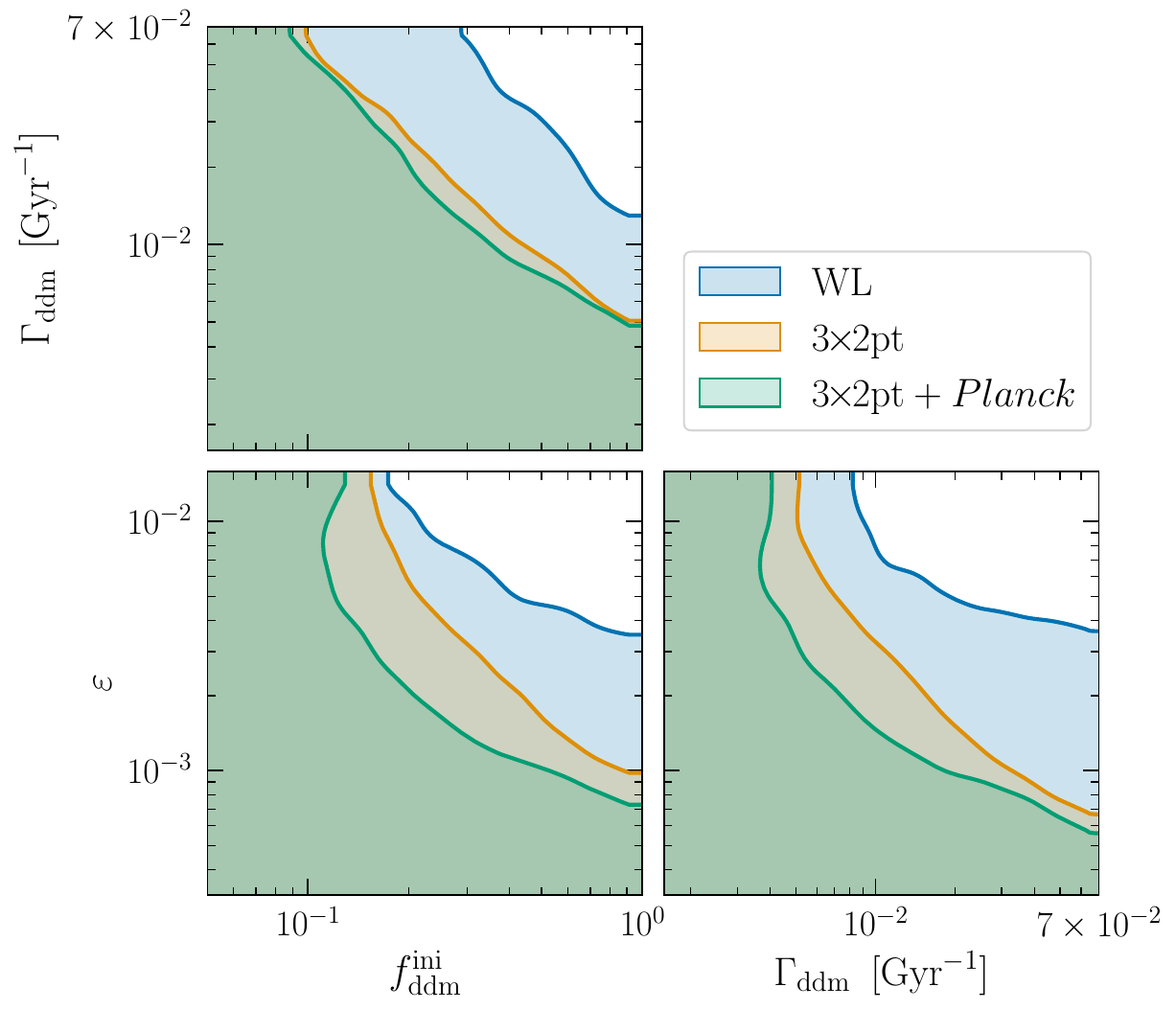}
    \caption{Same as Fig.\,\ref{fig:CWDM_pess_opt} for the parameters of the 2b-DDM model. The forecast assumes logarithmic priors on the DDM fraction $f_{\rm ddm}^{\rm ini}$, on the decay rate $\Gamma_{\rm ddm}$, and on the fraction of energy $\varepsilon$ going into the ultra-relativistic daughter at each decay. The model is equivalent to pure $\Lambda$CDM in the small $f_{\rm ddm}^{\rm ini}$ and/or small $\varepsilon$ and/or small  $\Gamma_{\rm ddm}$ limits.}
    \label{fig:2b_3x2_opt}
\end{figure*}

\begin{figure*}[t]
    \centering
    \includegraphics[width=0.99\textwidth]{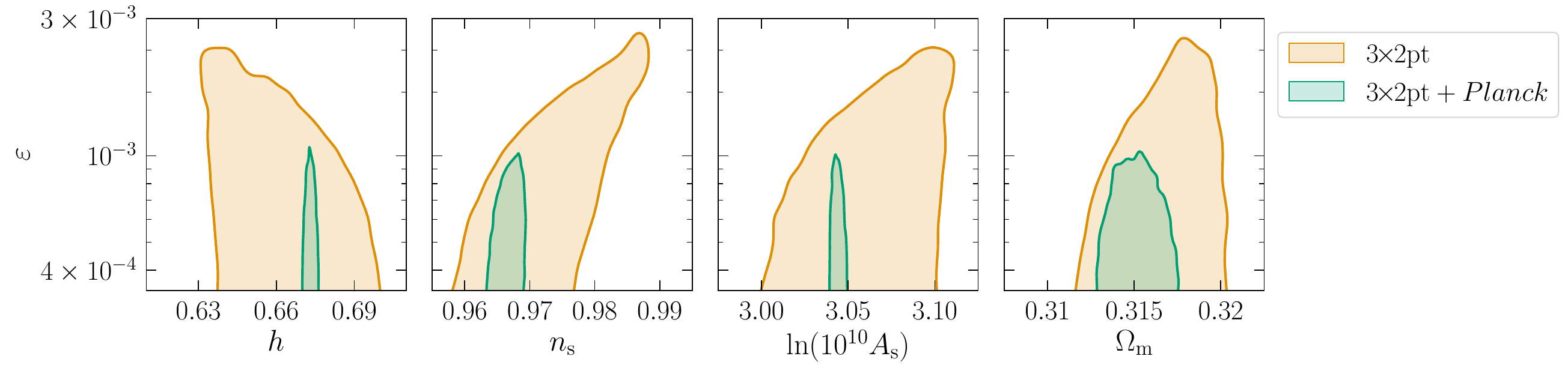}
    \caption{Degeneracies between the 2b-DDM parameter $\varepsilon$ and four other cosmological parameters for the optimistic case, in the particular case where $f_{\rm ddm}^{\rm ini} = 1$. The addition of \Planck{} data to 3\texttimes2pt data lifts these degeneracies.
    \label{fig:2bddm_planck}
    }
\end{figure*}

In Fig.\,\ref{fig:2b_3x2_opt} we show the 95\%CL contours on each pair of 2b-DDM parameters marginalised over cosmological and nuisance parameters for \Euclid{} WL only, \Euclid{} 3\texttimes2pt, and 3\texttimes2pt\,+\,\Planck{} data, under pessimistic (left panel) or optimistic (right panel) assumptions. We use the same colour scheme as in Figs.\,\ref{fig:CWDM_pess_opt} and \ref{fig:1b_pess_opt}. The fiducial $\Lambda$CDM model spans the left and lower axes of each panel. For each of the six cases shown in Fig.\,\ref{fig:2b_3x2_opt}, we ran 48 chains summing up to $\sim1$ MS in each optimistic case or $\sim1.3$ MS in each pessimistic case. The Gelman-Rubin convergence criterium reached about $|R-1| \sim 0.01$ for most parameters, with a worse value of 0.03 for a few parameters.

In ($f_{\rm ddm}^{\rm ini}$, $\Gamma_{\rm ddm}$) space, the contours follow lines of constant $\Gamma_{\rm ddm} \, f_{\rm ddm}^{\rm ini}$. This suggests that for both the 1b-DDM and 2b-DDM models the power spectrum suppression only depends on this product -- at least at the linear level. Thus, \Euclid{} can provide joint bounds on $\Gamma_{\rm ddm} \, f_{\rm ddm}^{\rm ini}$ and $\varepsilon$, while $f_{\rm ddm}^{\rm ini}$ or $\Gamma_{\rm ddm}$ will be left unconstrained.

We first comment the results of the pessimistic case. With WL only, our forecast returns the 95\%CL bound $\Gamma_{\rm ddm} \, f_{\rm ddm}^{\rm ini}<0.02\,{\rm Gyr}^{-1}$ (marginalised over $\varepsilon$). For $f_{\rm ddm}^{\rm ini}=1$ we find $\varepsilon<4\times10^{-3}$, while for $f_{\rm ddm}^{\rm ini}=0.3$  the 2b-DDM model is indistinguishable from $\Lambda$CDM at the 95\%CL, and $\varepsilon$ is unconstrained.
With the addition of 3\texttimes2pt data, the constraints remain stable. Finally, \Planck{} data is able to alleviate some degeneracies between cosmological parameters and shrink the bounds by about 25\%.

In the optimistic case, the WL-only bounds are identical, but the 3\texttimes2pt bounds shrink by a factor two compared to the 3\texttimes2pt pessimistic case, or a factor four compared to the WL optimistic case: $\Gamma_{\rm ddm} \, f_{\rm ddm}^{\rm ini}<0.005\,{\rm Gyr}^{-1}$ (with marginalisation over $\varepsilon$),
$\varepsilon<1\times10^{-3}$ for $f_{\rm ddm}^{\rm ini}=1$, and the 2b-DDM model is indistinguishable from $\Lambda$CDM below $f_{\rm ddm}^{\rm ini}=0.1$. In this case, the addition of \Planck{} data makes a difference for the bounds on $\varepsilon$, not because of \Planck{} data being directly sensitive to this parameter, but thanks to the better determination of other parameters. Figure\,\ref{fig:2bddm_planck} shows how \Planck{} data lift the degeneracy between, for instance, $\varepsilon$ and $n_{\rm s}$. In this case we obtain a bound $\varepsilon<0.7\times10^{-3}$ for $f_{\rm ddm}^{\rm ini}=1$.

\begin{figure*}[t]
    \centering
    \includegraphics[width=0.49\textwidth]{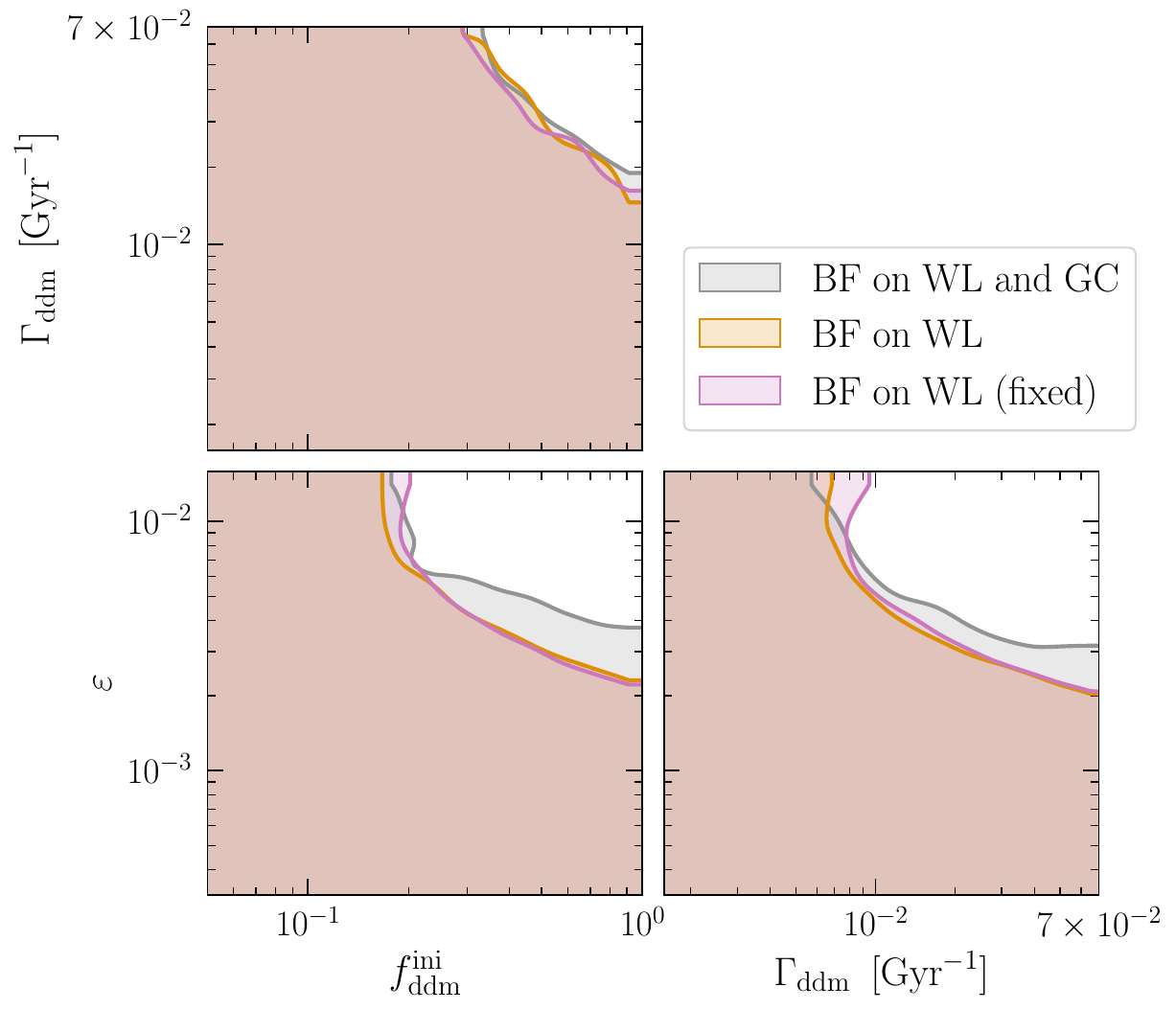}
    \includegraphics[width=0.49\textwidth]{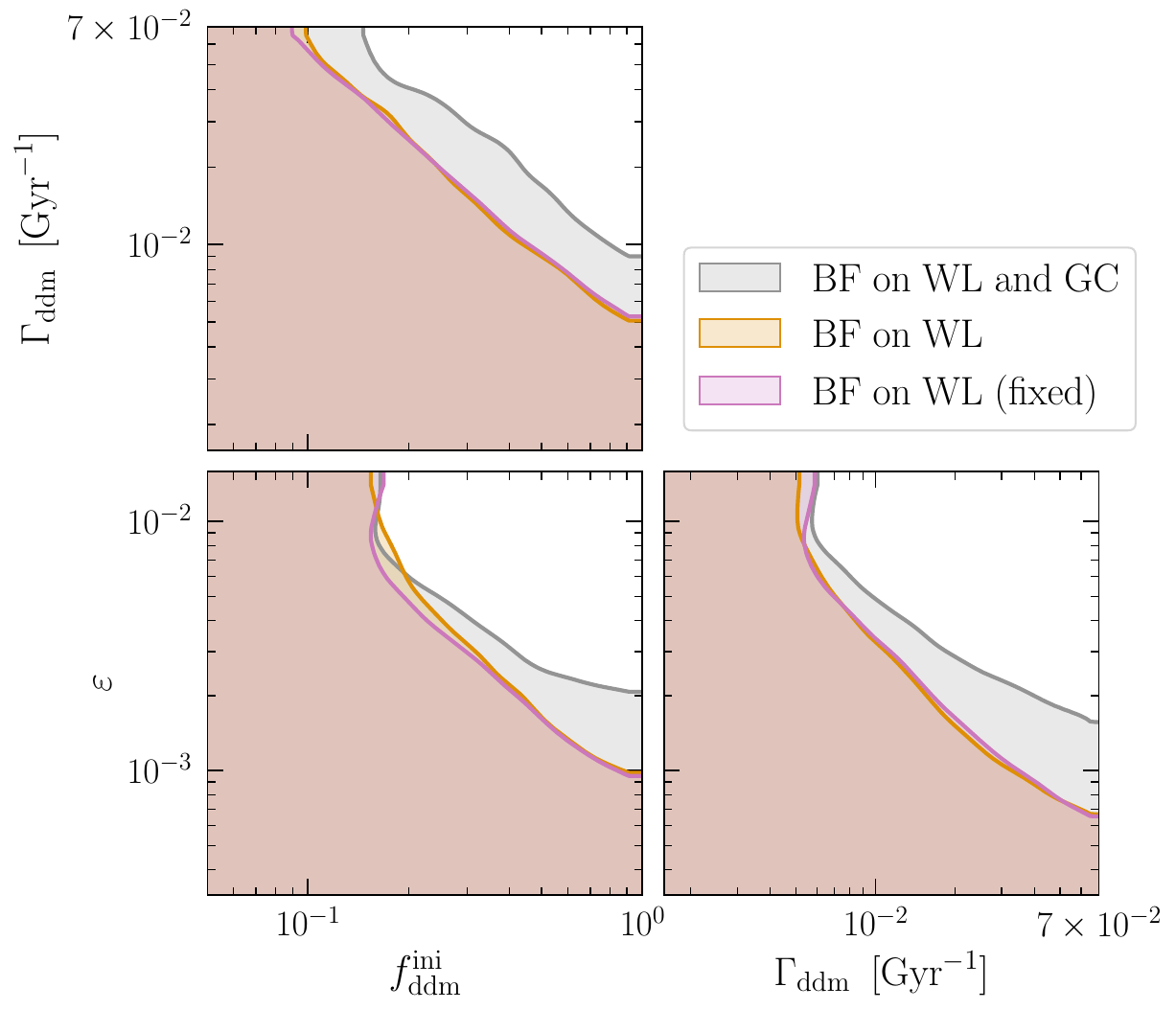}
    \caption{Same as Fig.\,\ref{fig:CWDM_BF_comparison} for the parameters of the 2b-DDM model.}
    \label{fig:2b_BF_comparison_opt}
\end{figure*}

{\it Importance of baryonic feedback.} Figure\,\ref{fig:2b_BF_comparison_opt} depicts how different baryonic feedback prescriptions influence the final posteriors, using the same colour and style as Figs.\,\ref{fig:CWDM_BF_comparison} and \ref{fig:1b_BF_comparison}. Like for 1b-DDM, there could be a degeneracy between 2b-DDM and baryonic feedback parameters since both effects tend to grow with time. Indeed, in the 2b-DDM model, the conversion of CDM into WDM particles appears dominantly at very late times and the non-linear matter power spectrum departs more and more from the $\Lambda$CDM limit.

However, we find that 2b-DDM effects and baryonic feedback are very weakly correlated. The constraints remain nearly stable when fixing the baryonic feedback parameters instead of marginalising over them, and become slightly weaker when baryonic feedback is applied also to the GC probe. The degradation is at most by a factor two. As a matter of fact, our forecasts predict 95\%CL bounds on the 2b-DDM parameter summarised by $\Gamma_{\rm ddm} \, f_{\rm ddm}^{\rm ini}<0.02\,{\rm Gyr}^{-1}$ (with marginalisation over $\varepsilon$) and
$\varepsilon<4\times10^{-3}$ for $f_{\rm ddm}^{\rm ini}=1$ in the pessimistic case; or 
$\Gamma_{\rm ddm} \, f_{\rm ddm}^{\rm ini}<0.008\,{\rm Gyr}^{-1}$ (with marginalisation over $\varepsilon$) and
$\varepsilon<2\times10^{-3}$ for $f_{\rm ddm}^{\rm ini}=1$ in the optimistic case.

\begin{figure}[ht]
    \centering
    \includegraphics[width=0.5\textwidth]{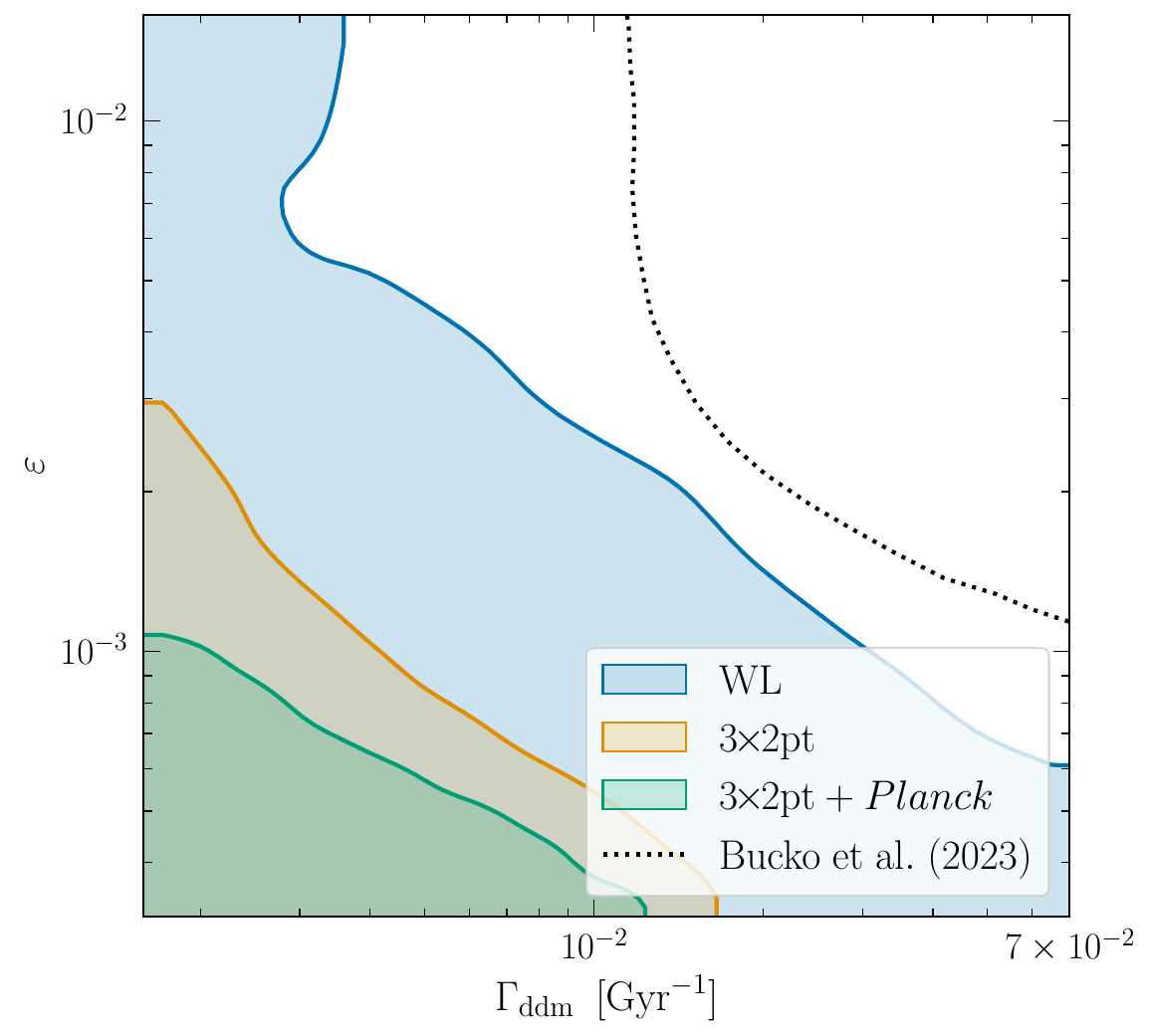}
    \caption{For the 2b-DDM model with $f_{\rm ddm}^{\rm ini}=1$, comparison of \Euclid{} WL-only, 3\texttimes2pt, and 3\texttimes2pt\,+\,\Planck{} bounds predicted by our sensitivity forecast with current constraints from KiDS \citep{Bucko:2023twobody}. The priors are identical in the four cases.
    \label{fig:2bddm_kids}}
\end{figure}

{\it Comparison to current bounds.} 
We believe that our forecast is the first one including as a free parameter the initial fraction $f_{\rm ddm}^{\rm ini}$ of DDM with two-body decay. Several studies in the past fixed $f^{\rm ini}_{\rm ddm} = 1$. Therefore, in order to compare the expected sensitivity of \Euclid{} to recent results obtained from real observations, we perform a few dedicated forecast with 100\% DDM. In Fig.\,\ref{fig:2bddm_kids}, we compare our results to the most recent limits on $\Gamma_{\rm ddm}$ and $\varepsilon$ from the KiDS survey \citep{Bucko:2023twobody}. In this case, we adopt precisely the same top-hat priors on the 2b-DDM parameters as \cite{Bucko:2023twobody}. We find that \Euclid{} with WL alone can improve over KiDS bounds roughly by a factor 3, while the full 3\texttimes2pt data would improve over KiDS by one order of magnitude. This improvement is closer to the one observed for 1b-DDM than for CWDM, since a precise measurement of the power spectrum on intermediate (linear and mildly non-linear) scales is crucial to constrain this model, while a better measurement on non-linear scales helps to discriminate decaying DM effects from baryonic feedback effects.

The 2b-DDM model with $f^{\rm ini}_{\rm ddm} = 1$ has also be confronted to current Lyman-$\alpha$ data in \cite{Fuss:2022zyt}. For very small values of $\varepsilon$, the suppression of the matter power spectrum could occur on such small scales that Lyman-$\alpha$ data would still probe the 2b-DDM effects while \Euclid{} data could not distinguish 2b-DDM from $\Lambda$CDM. However, for $\varepsilon \geq 10^{-3}$, \Euclid{} can probe the effects of 2b-DDM on linear and mildly non-linear scales, and thus can be expected to have more constraining power. This is confirmed by the results of \cite{Fuss:2022zyt}, who show that Lyman-$\alpha$ data from BOSS DR14 only constrain $\Gamma_{\rm ddm}$ to be smaller than ${\cal O}(0.1) \,{\rm Gyr}^{-1}$. This is about one order of magnitude weaker than the predicted \Euclid{} sensitivity. 

\subsection{ETHOS $n=0$ \label{sec:res_ethos}}

{\it Main results.} Our ETHOS $n=0$ forecasts use the parameters ($a_\mathrm{dark}$, $\xi_{\rm idr}$) defined in Sect.\,\ref{sec:theo_ethos}, with logarithmic priors defined in Table\,\ref{tab:priors_ethos} for these parameters and Table\,\ref{tab:priors} for all other free parameters. We have seen in Sects.\,\ref{sec:theo_ethos} and \ref{sec:nl_ethos} that the ETHOS $n=0$ model induces a  suppression in the power spectrum controlled at the linear level by $a_{\rm dark}\xi_{\rm idr}^4$ and at the non-linear level by the two ETHOS parameters.  The fiducial model of our forecasts is chosen to be a pure $\Lambda$CDM model.

\begin{table}[ht]
\centering
\caption{List of free parameters names, fiducial values, and top-hat prior ranges (in addition to those listed in Table\,\ref{tab:priors}) for the ETHOS $n=0$ model. In our runs, we additionally impose a prior $a_\mathrm{dark} \xi_{\rm idr}^4<0.05$ to exclude the region where the emulator should not be trusted (see Sect.\,\ref{sec:nl_ethos}).
The fiducial values correspond to the pure $\Lambda$CDM limit.
\label{tab:priors_ethos}}
\begin{tabular}{ccc}
\hline
Parameter & Fiducial value & Range\\
\hline
\noalign{\vskip 2pt}
$\loga$ & $-\infty$ & $[-6,\, 5]$ \\
$\log_{10} \xi_{\rm idr}$ & $-\infty$ & $ [-2,\, -0.4]$ \\
\noalign{\vskip 2pt}
\hline
\end{tabular}
\end{table}

\begin{figure*}[t]
    \centering
    \includegraphics[width=0.85\textwidth]{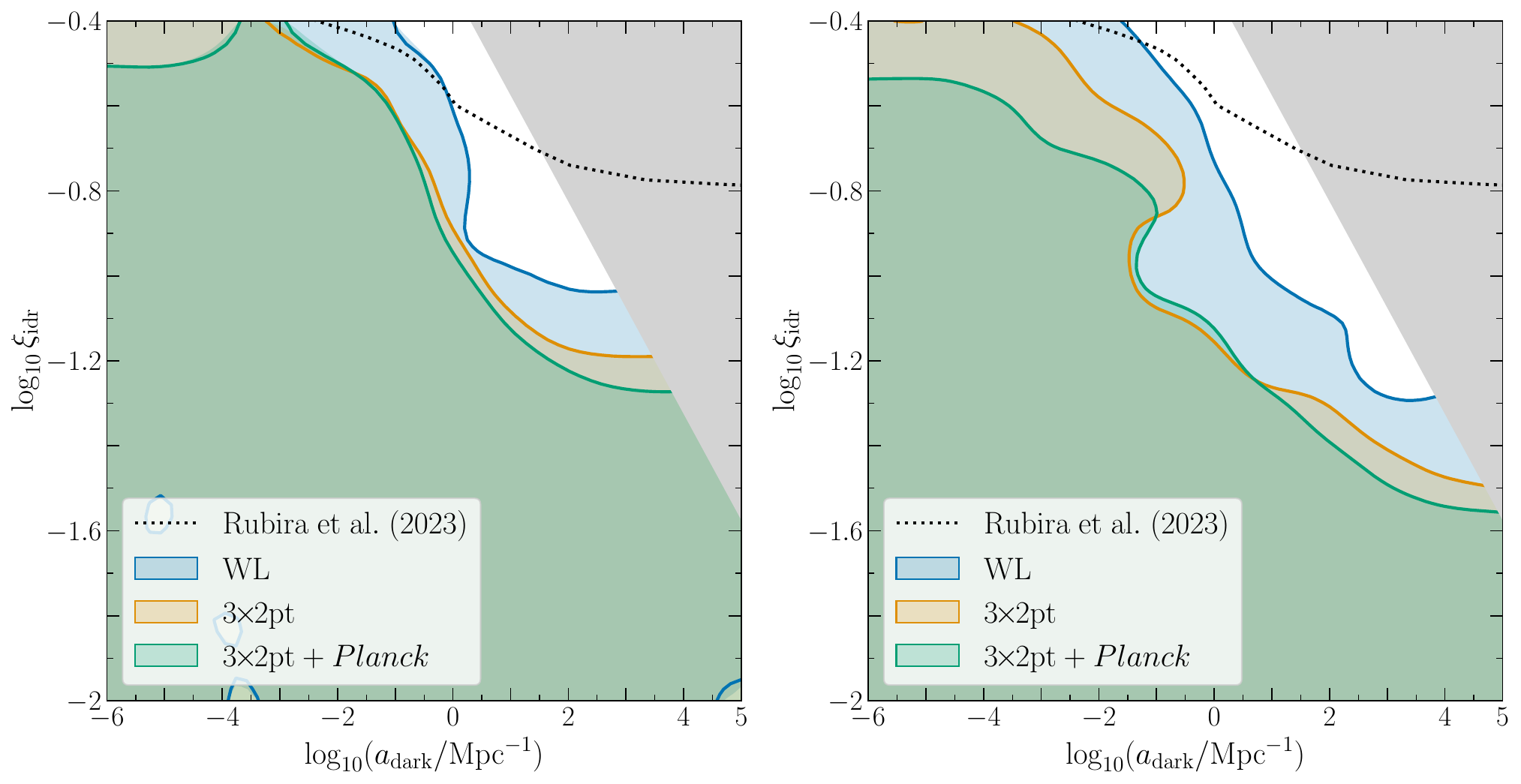}
    \caption{Same as Fig.\,\ref{fig:CWDM_pess_opt} for the parameters of the ETHOS $n=0$ model. Our forecast assumes logarithmic priors on the interaction strength $a_{\rm dark}$ and on the dark-radiation-to-photon temperature ratio $\xi_{\rm idr}$. The model is equivalent to pure $\Lambda$CDM in the small $a_{\rm dark}$ and/or small $\xi_{\rm idr}$ limits. The grey shade excludes the region $a_\mathrm{dark} \xi_{\rm idr}^4>0.05$ where the non-linear emulator cannot be trusted (see Sect.\,\ref{sec:nl_ethos}).
    We also show current constraints inferred from \Planck{}, BAO, and BOSS full-shape data by \cite{Rubira_2023} -- although these authors use different priors.
    \label{fig:ethos_WL_3x2_opt}
    }
\end{figure*}

In Fig.\,\ref{fig:ethos_WL_3x2_opt} we show the 95\% CL credible interval limits on the free parameters ($a_{\rm dark}$, $\xi_{\rm idr}$) coming from the \Euclid{} WL probe, from the full 3\texttimes2pt probe, and from the same in combination with \Planck{} data. For each of the six cases shown in Fig.\,\ref{fig:ethos_WL_3x2_opt}, we ran 48 chains summing up to $\sim1$ MS in each optimistic case or $\sim1.7$ MS in each pessimistic case. The Gelman-Rubin convergence criterium reached about $|R-1| \sim 0.005$ for most parameters, with a worse value of $\sim0.012$ for a few parameters.

In this case, the credible interval limits look slightly wobbly. However, the limits remain stable when pushing the MCMC chains to very high convergence criteria, or when performing multiple independent MCMC runs. Thus, the small oscillations of the contours in Fig.\,\ref{fig:ethos_WL_3x2_opt} are not caused by MCMC convergence issues but the emulator, which was trained on a slightly too coarse sample of models. In the future, the {\tt DMemu} emulator will be improved for this model. The oscillations are anyway sufficiently small to allow for a robust qualitative interpretation of our forecast results.

When $a_{\rm dark}$ is very small, this model is equivalent to a $\Lambda$CDM+$\Delta N_{\rm eff}$ model with $\Delta N_{\rm eff}=3.85\,\xi_{\rm idr}^4$. The radiation excess parameter $\Delta N_{\rm eff}$ can be constrained since it affects both the matter power spectrum and CMB anisotropy spectrum in a well-known way \citep{KP_nu}. In this limit, we expected bounds of the order of $\Delta N_{\rm eff} < {\cal O}(1)$ (95\%CL) from the 3\texttimes2pt optimistic probe and $\Delta N_{\rm eff} < {\cal O}(0.1)$ (95\%CL) from the combination 3\texttimes2pt\,+\,\Planck{} \citep{KP_nu}. This  translates respectively into $\log_{10}\xi_{\rm idr}<-0.1$ (95\%CL, 3\texttimes2pt) and $\log_{10}\xi_{\rm idr}<-0.4$ (95\%CL, 3\texttimes2pt\,+\,\Planck). Our choice of prior, $\xi_{\rm idr}\in[-2,-0.4]$, prevents us from seeing the upper bound in the 3\texttimes2pt case, but in the case of 3\texttimes2pt\,+\,\Planck{} we can see the upper limit on $\xi_{\rm idr}$ just below top axis of each panel in Fig.\,\ref{fig:ethos_WL_3x2_opt}.

For larger values of $a_{\rm dark}$, the model is further constrained by the impact of IDM-IDR interactions on the small-scale matter power spectrum. As already discussed, at the non-linear level, this effect depends on both $a_{\rm dark}$ and $\xi_{\rm idr}$ in a  non-trivial way. However, for $\loga < 1$, we find that the boundary of the preferred region can be approximately fitted by constant values of the combination $a_{\rm dark} \, \xi_{\rm idr}^4$ that controls the scattering rate of IDR off IDM.

In the pessimistic case, the most substantial part of the constraining power comes from the WL probe, since further addition of clustering and \Planck{} data do not improve the bounds significantly. In all cases, the bounds for $\loga < 1$ can be approximated as $a_{\rm dark} \,\xi_{\rm idr}^4<8\times 10^{-4}\,{\rm Mpc}^{-1}$ (95\%CL). With the 3\texttimes2pt probe, the data loses sensitivity to IDM-IDR interactions only for $\xi_{\rm idr}<0.06$ (i.e., $\Delta N_{\rm eff}< 5 \times 10^{-5}$). We stress that \Euclid{} would not detect such a tiny abundance of dark radiation through the effect of an enhanced radiation density, but through that of DM interactions.

In the optimistic case, the WL-only bound changes marginally, but the 3\texttimes2pt bound (with or without \Planck) shrinks by more than one order of magnitude. For $\loga < 1$ the bounds can be approximated by $a_{\rm dark} \,\xi_{\rm idr}^4<2\times 10^{-5}\,{\rm Mpc}^{-1}$. With 3\texttimes2pt information, the data loses sensitivity to the interaction rate only below $\xi_{\rm idr}<0.03$ (that is, $\Delta N_{\rm eff}< 3 \times 10^{-6}$).

\begin{figure*}[t]
    \centering
    \includegraphics[width=0.9\textwidth]{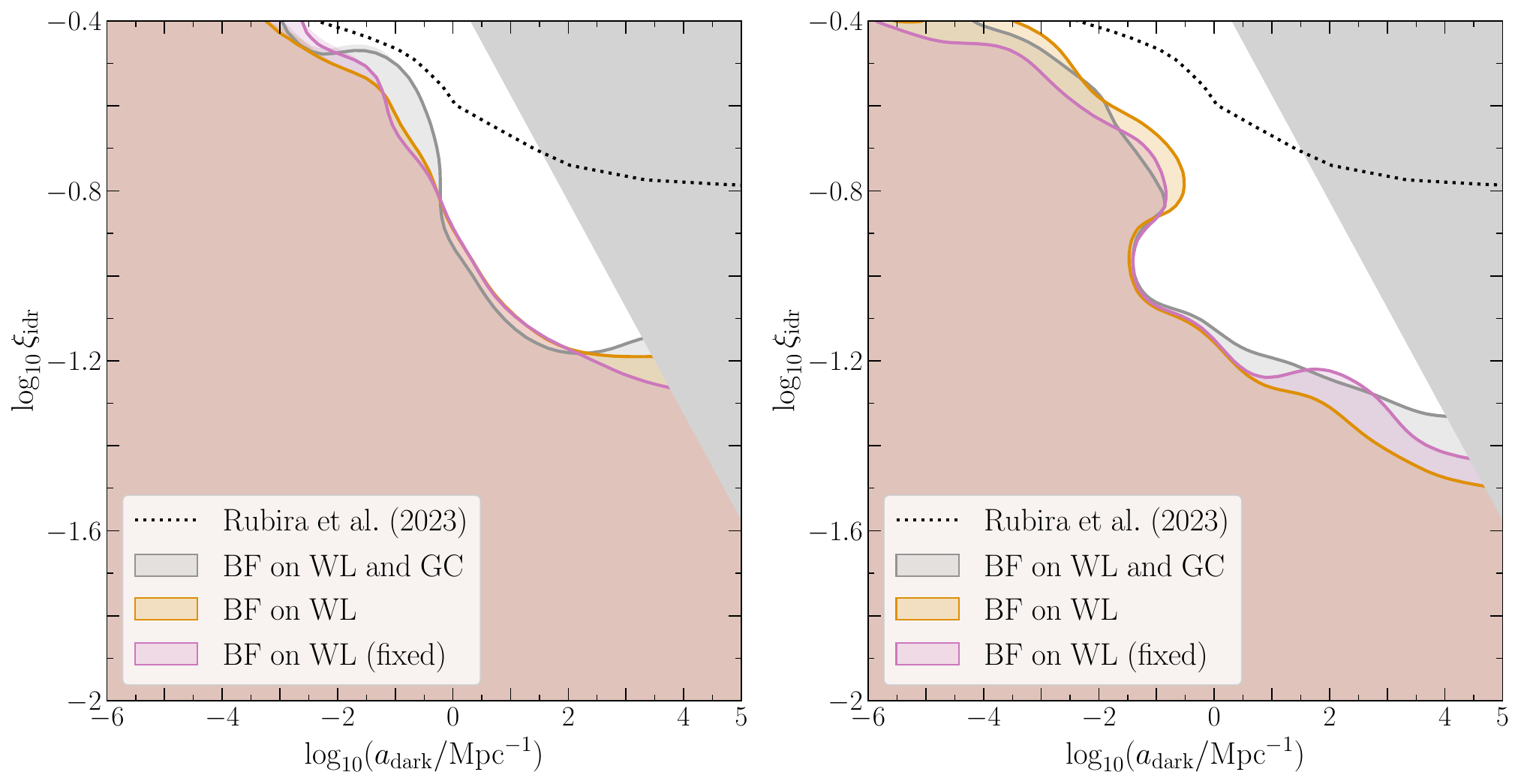}
    \caption{Same as Fig.\,\ref{fig:CWDM_BF_comparison} for the parameters of the ETHOS $n=0$ model. 
    \label{fig:ethos_BF_comparison_opt}
    }
\end{figure*}

{\it Importance of baryonic feedback.} In this case, the impact of baryonic feedback is illustrated in Fig.\,\ref{fig:ethos_BF_comparison_opt}. Interestingly, in the ETHOS $n=0$ case, we do not find any hint of degeneracies between the DM and baryonic feedback parameters. In the pessimistic and optimistic cases, the bounds remain roughly stable when the baryonic feedback parameters are fixed rather than marginalised, or when baryonic feedback is applied also to GC. A first explanation comes from the fact that the effect of the ETHOS $n=0$ model on the matter power spectrum always starts on linear scales ($k \sim 10^{-2}$--$10^{-1}\,h\,{\rm Mpc}^{-1}$) which are immune to baryonic feedback. If most of the information on this model resides in such scales, the bounds should indeed be independent from the modelling of baryonic feedback. In addition, like in the CWDM case, the redshift dependence of the DM-induced suppression is opposite to that of baryonic feedback. As a matter of fact, the effect of DM interactions is imprinted on the matter power spectrum at high redshift and subsequently smoothed out by non-linear clustering, while overall baryonic effects tend to grow with time, at least through most of the redshift range probed by \Euclid{}.  

{\it Comparison with current bounds.} For the same model, using flat priors on $\xi_{\rm idr}$ and logarithmic priors on $a_{\rm dark}$,
\cite{Archidiacono:2019wdp} found $a_{\rm dark} \xi_{\rm idr}^4<14\times 10^{-4}\,{\rm Mpc}^{-1}$ (95\%CL) using Planck, BAO, and high-resolution Lyman-$\alpha$ data from the HIRES/MIKE quasar sample. The comparison with our \Euclid{} bound is not straightforward due to the different prior shapes and edges, but indicates that \Euclid{} has a potential to improve over current CMB+Lyman-$\alpha$ bounds by a large factor (a factor 70 according to our predictions in the 3\texttimes2pt optimistic case). This is actually not so surprising since the ETHOS $n=0$ model leaves a signature already on linear scales, much larger than the scales probed by Lyman-$\alpha$ data.

\cite{Rubira_2023} also use flat priors on $\xi_{\rm idr}$ and logarithmic priors on $a_{\rm dark}$. Their results for IDM interacting with free-streaming IDR is reported in the left panel of their Fig.\,4. We extracted from this plot their joint bound on ($a_{\rm dark}$, $\xi_{\rm idr}$) inferred from \Planck, BOSS full-shape galaxy spectrum, and KiDS (the latter being implemented as a measurement of $S_8$). We display this bound in Figs.\,\ref{fig:ethos_WL_3x2_opt} and \ref{fig:ethos_BF_comparison_opt}. The comparison with our \Euclid{} forecast should be taken with a grain of salt since the priors are different in the two analyses. However, the main conclusion is that, on the one hand, their bound is similar to our \Euclid 3\texttimes2pt\,+\,\Planck{} bound in the limit of large $\xi_{\rm idr}$, which was expected since this bound only reflects the upper limit on $\Delta N_{\rm eff}$ from \Planck{}; but on the other hand, for smaller values of $\xi_{\rm idr}$, we find that \Euclid{} is much more sensitive to  $a_{\rm dark}$. According to \cite{Rubira_2023}, the BOSS data loses any sensitivity to $a_{\rm dark}$ when $\xi_{\rm idr}$ is equal to or smaller than $10^{-0.8}$ (that is, $\xi_{\rm idr}\leq 0.2$, or equivalently $\Delta N_{\rm eff}\leq 2\times 10^{-3}$), while for $\xi_{\rm idr}=0.2$ \Euclid{} 3\texttimes2pt data can still constrain the interaction rate to $a_{\rm dark}<1\,{\rm Mpc}^{-1}$ (95\%CL, pessimistic case) or $a_{\rm dark}<0.1\,{\rm Mpc}^{-1}$ (95\%CL, optimistic case). 

We conclude that \Euclid{} has a great potential to constrain the ETHOS $n=0$ model and to push the bounds well below those from any current experiment.

\section{Summary and conclusions \label{sec:discussion}}

In summary, we have estimated the sensitivity of the future \Euclid{} photometric probe (i.e., of 3\texttimes2pt statistics) to the parameters describing four non-minimal DM models. We have run several MCMC forecasts in which the fiducial model assumes plain CDM (with baryonic feedback) while the fitted model includes the effect of non-standard DM (with free baryonic freedback parameters). We have investigated the dependence of the results on various assumptions (cut-off multipole $\ell_{\rm max}$, modelling of baryonic feedback, combination with CMB data from \Planck{}). We have also compared the sensitivity predicted by our forecasts with current bounds derived from CMB data, Lyman-alpha data, WL data, and galaxy redshift survey data.

Each of the few non-minimal DM models considered here has a qualitatively different impact on the matter power spectrum. As a matter of fact, we reach significantly different conclusions for each of them in terms of degeneracy with baryonic feedback, constraining power of WL data compared to GC data, or sensitivity of \Euclid{} compared to current bounds. In this section, we put together a compact summary of the most striking conclusions.

For a mixture of cold and warm dark matter (CWDM), the key point is that the power spectrum looks exactly like that of $\Lambda$CDM up to a scale fixed by the WDM mass, beyond which a step-like suppression occurs. For large WDM fractions leading to a strong suppression, the mass bounds will always be dominated by data from Lyman-$\alpha$ forests probing smaller scales than \Euclid{}. However, the results of Sect.\,\ref{sec:res_cwdm} show that \Euclid{} can be very efficient at constraining small WDM masses when the WDM fraction is also small. The \Euclid 3\texttimes2pt analysis could even bound (or detect) masses of the order of just ${\cal O}(10)\,{\rm eV}$ (thermal WDM case) or ${\cal O}(100)\,{\rm eV}$ (Dodelson--Widrow scenario) even if WDM accounts for only 1\% of the total DM budget, while current observations are only sensitive to WDM contributing to at least 10\% of DM.

The situation is very different with models featuring a mixture of stable and unstable CDM, with the latter experiencing one-body decay into relativistic particles (1b-DDM). In this case, the DM parameters impact the evolution of perturbations up to very large scales, deep in the linear regime. Thus, CMB data is also highly sensitive to the decaying DM fraction $f_{\rm ddm}$ and decay rate $\Gamma_{\rm ddm}$ -- as a matter of fact, on their product $f_{\rm ddm} \Gamma_{\rm ddm}$.\footnote{In this summary, we use simplified notations. In the core of the paper, the fraction was denoted $f_{\rm ddm}^{\rm ini}$, see Sect.\,\ref{sec:theo_1bddm} for precise definitions.} However, the results of Sect.\,\ref{sec:res_1bddm} show that the \Euclid{} 3\texttimes2pt probe alone could provide twice stronger bounds on $f_{\rm ddm} \Gamma_{\rm ddm}$ than \Planck{}. In this case, there is a synergy between \Euclid{} and \Planck{}: the combined data sets can resolve parameter degeneracies and strengthen current bounds by a factor eight.

For a mixture of stable and unstable CDM such that the latter experiences two-body decay into one relativistic and one non-relativistic particle (2b-DDM), the power spectrum is step-like suppressed, a bit like the CWDM case but with a different suppression shape. As a matter of fact, this model can be understood as if a few CDM particles were gradually replaced by WDM particles at late times. In this case, WL and GC surveys are more sensitive to the parameters of the model than Lyman-$\alpha$ data in the limit of small decay rate and large $\varepsilon$, i.e., large velocity dispersion, since  in this limit the step-like suppression is small but occurs on relatively large scales. For such models, the results of Sect.\,\ref{sec:res_2bddm} show that \Euclid could improve the bound on the product $f_{\rm ddm} \Gamma_{\rm ddm}$ by one order of magnitude compared to current WL surveys like KiDS.

For DM interacting with dark radiation at a constant rate (ETHOS $n=0$), the power spectrum is suppressed on intermediate and small scales in a more progressive way than for CWDM and 2b-DDM. The suppression looks more like a broken power law than an exponential cut. The forecasts of Sect.\,\ref{sec:res_ethos} show that the constraining power of \Euclid{} is particularly strong in this case. \Euclid{} may improve current bounds from Lyman-$\alpha$ and \Planck by a factor 70. Through DM interaction effects, \Euclid{} can probe for the first time the limit in which the abundance of dark radiation is very small, namely in the range $\Delta N_{\rm eff} \sim [3\times 10^{-6}, 2\times 10^{-3}]$ that current galaxy surveys do not have the sensitivity to constrain. 

Note that our decision to choose $\Lambda$CDM as the fiducial model in each forecast was arbitrary: we could have chosen a fiducial model featuring non-standard DM and compatible with current bounds, and calculated the significance at which \Euclid{} could differentiate this model from $\Lambda$CDM. We did not perform such tests due to computational limits, but our forecasts suggest that \Euclid{} has a significant discovery potential. When we find that a $2\,\sigma$ bound from current data could shrink by a factor $n$ with future \Euclid{} data, we get a hint that a DM model described by some parameters saturating the current $2\,\sigma$ limit could be differentiated from $\Lambda$CDM roughly at the $2n\,\sigma$ level. Our forecasts show that, in some regions of parameter space, current bounds can improve by one to two orders of magnitude (e.g., a factor 70 in the ETHOS $n=0$ case). This shows that there are wide regions in parameter space where \Euclid{} could perform an actual discovery of non-minimal DM properties at a high level of significance.

If an experiment like \Euclid{} provides evidence in favour of a non-minimal DM model, the impact of such a discovery on cosmology will obviously be profound. Several future large-scale structure surveys -- e.g., from the Square Kilometer Array Observatory  
\citep{Santos:2015gra} or Rubin Observatory \citep{LSST:2008ijt} -- would have an opportunity to check this discovery independently, while our approach to understand and model the process of galaxy and star formation would be deeply impacted. Even if we conservatively assume that \Euclid{} will confirm plain CDM and strengthen all bounds on the parameters of non-minimal DM models, the results will be incredibly interesting, since the new limits will have implications for DM model building and cut into the parameter space of several possible dark sector models.

\begin{acknowledgements}
We acknowledge computing resources granted by RWTH Aachen University under project thes1340, rwth1411, rwth1437, and rwth1481. We also acknowledge funding from DFG project 456622116 and support from the IRAP and IN2P3 Lyon computing centers. \AckEC
\end{acknowledgements}

\bibliographystyle{aa}
\bibliography{biblio}

\begin{thebibliography}{106}
\expandafter\ifx\csname natexlab\endcsname\relax\def\natexlab#1{#1}\fi

\bibitem[{Abdalla {et~al.}(2022)Abdalla, Franco~Abellán, Aboubrahim,
  {et~al.}}]{Abdalla:2022yfr}
Abdalla, E., Franco~Abellán, G., Aboubrahim, A., {et~al.} 2022, JHEAp, 34, 49

\bibitem[{Anderhalden {et~al.}(2013)Anderhalden, Schneider, Maccio, Diemand, \&
  Bertone}]{Anderhalden:2012jc}
Anderhalden, D., Schneider, A., Maccio, A.~V., Diemand, J., \& Bertone, G.
  2013, JCAP, 03, 014

\bibitem[{Angulo {et~al.}(2021)Angulo, Zennaro, Contreras, Aric\`o,
  Pellejero-Iba\~nez, \& St\"ucker}]{Angulo:2020vky}
Angulo, R.~E., Zennaro, M., Contreras, S., {et~al.} 2021, Mon. Not. Roy.
  Astron. Soc., 507, 5869

\bibitem[{Aoyama {et~al.}(2014)Aoyama, Sekiguchi, Ichiki, \&
  Sugiyama}]{Aoyama:2014tga}
Aoyama, S., Sekiguchi, T., Ichiki, K., \& Sugiyama, N. 2014, JCAP, 07, 021

\bibitem[{Archidiacono {et~al.}(2019)Archidiacono, Hooper, Murgia, Bohr,
  Lesgourgues, \& Viel}]{Archidiacono:2019wdp}
Archidiacono, M., Hooper, D.~C., Murgia, R., {et~al.} 2019, JCAP, 10, 055

\bibitem[{Archidiacono {et~al.}(2024)}]{KP_nu}
Archidiacono, M. {et~al.} 2024 [\eprint[arXiv]{2405.06047}]

\bibitem[{Asgari {et~al.}(2021)Asgari, Lin, Joachimi, {et~al.}}]{KiDS:2020suj}
Asgari, M., Lin, C.-A., Joachimi, B., {et~al.} 2021, A\&A, 645, A104

\bibitem[{Audren {et~al.}(2013{\natexlab{a}})Audren, Lesgourgues, Benabed, \&
  Prunet}]{Audren:2012wb}
Audren, B., Lesgourgues, J., Benabed, K., \& Prunet, S. 2013{\natexlab{a}},
  JCAP, 02, 001

\bibitem[{Audren {et~al.}(2013{\natexlab{b}})Audren, Lesgourgues, Bird,
  Haehnelt, \& Viel}]{Audren:2012vy}
Audren, B., Lesgourgues, J., Bird, S., Haehnelt, M.~G., \& Viel, M.
  2013{\natexlab{b}}, JCAP, 01, 026

\bibitem[{Audren {et~al.}(2014)Audren, Lesgourgues, Mangano, Serpico, \&
  Tram}]{Audren:2014bca}
Audren, B., Lesgourgues, J., Mangano, G., Serpico, P.~D., \& Tram, T. 2014,
  JCAP, 12, 028

\bibitem[{Becker {et~al.}(2021)Becker, Hooper, Kahlhoefer, Lesgourgues, \&
  Sch\"oneberg}]{Becker:2020hzj}
Becker, N., Hooper, D.~C., Kahlhoefer, F., Lesgourgues, J., \& Sch\"oneberg, N.
  2021, JCAP, 02, 019

\bibitem[{Berezhiani {et~al.}(2015)Berezhiani, Dolgov, \&
  Tkachev}]{Berezhiani:2015yta}
Berezhiani, Z., Dolgov, A.~D., \& Tkachev, I.~I. 2015, Phys. Rev. D, 92, 061303

\bibitem[{Bertone {et~al.}(2005)Bertone, Hooper, \& Silk}]{Bertone:2004pz}
Bertone, G., Hooper, D., \& Silk, J. 2005, Phys. Rept., 405, 279

\bibitem[{Bird {et~al.}(2012)Bird, Viel, \& Haehnelt}]{Bird:2011rb}
Bird, S., Viel, M., \& Haehnelt, M.~G. 2012, MNRAS, 420, 2551

\bibitem[{Blas {et~al.}(2011)Blas, Lesgourgues, \& Tram}]{Blas:2011rf}
Blas, D., Lesgourgues, J., \& Tram, T. 2011, JCAP, 07, 034

\bibitem[{Bode {et~al.}(2001)Bode, Ostriker, \& Turok}]{Bode:2000gq}
Bode, P., Ostriker, J.~P., \& Turok, N. 2001, \apj, 556, 93

\bibitem[{Boehm {et~al.}(2001)Boehm, Fayet, \& Schaeffer}]{Boehm:2000gq}
Boehm, C., Fayet, P., \& Schaeffer, R. 2001, Phys. Lett. B, 518, 8

\bibitem[{{Bond} \& {Szalay}(1983)}]{Bond:1983aaa}
{Bond}, J.~R. \& {Szalay}, A.~S. 1983, \apj, 274, 443

\bibitem[{Bonvin \& Durrer(2011)}]{Bonvin:2011bg}
Bonvin, C. \& Durrer, R. 2011, Phys. Rev. D, 84, 063505

\bibitem[{Boyarsky {et~al.}(2009{\natexlab{a}})Boyarsky, Lesgourgues,
  Ruchayskiy, \& Viel}]{Boyarsky:2008xj}
Boyarsky, A., Lesgourgues, J., Ruchayskiy, O., \& Viel, M. 2009{\natexlab{a}},
  JCAP, 05, 012

\bibitem[{Boyarsky {et~al.}(2009{\natexlab{b}})Boyarsky, Lesgourgues,
  Ruchayskiy, \& Viel}]{Boyarsky:2008mt}
Boyarsky, A., Lesgourgues, J., Ruchayskiy, O., \& Viel, M. 2009{\natexlab{b}},
  Phys. Rev. Lett., 102, 201304

\bibitem[{Brinckmann \& Lesgourgues(2019)}]{Brinckmann:2018cvx}
Brinckmann, T. \& Lesgourgues, J. 2019, Phys. Dark Univ., 24, 100260

\bibitem[{Bucko {et~al.}(2023{\natexlab{a}})Bucko, Giri, Peters, \&
  Schneider}]{Bucko:2023twobody}
Bucko, J., Giri, S.~K., Peters, F.~H., \& Schneider, A. 2023{\natexlab{a}}
  [\eprint[arXiv]{2307.03222}]

\bibitem[{Bucko {et~al.}(2023{\natexlab{b}})Bucko, Giri, \&
  Schneider}]{bucko_2022_1bddm}
Bucko, J., Giri, S.~K., \& Schneider, A. 2023{\natexlab{b}}, A\&A, 672, A157

\bibitem[{Buen-Abad {et~al.}(2015)Buen-Abad, Marques-Tavares, \&
  Schmaltz}]{Buen-Abad:2015ova}
Buen-Abad, M.~A., Marques-Tavares, G., \& Schmaltz, M. 2015, Phys. Rev. D, 92,
  023531

\bibitem[{Buen-Abad {et~al.}(2018)Buen-Abad, Schmaltz, Lesgourgues, \&
  Brinckmann}]{Buen-Abad:2017gxg}
Buen-Abad, M.~A., Schmaltz, M., Lesgourgues, J., \& Brinckmann, T. 2018, JCAP,
  01, 008

\bibitem[{{Casas} {et~al.}(2023){Casas}, {Lesgourgues}, {Sch{\"o}neberg},
  {Sabarish V.}, {Rathmann}, {Doerenkamp}, {Archidiacono}, {Bellini}, {Clesse},
  {Frusciante}, {Martinelli}, {Pace}, {Sapone}, {Sakr}, {Blanchard},
  {Brinckmann}, {Camera}, {Carbone}, {Ili{\'c}}, {Markovic}, {Pettorino},
  {Tutusaus}, {Aghanim}, {Amara}, {Amendola}, {Auricchio}, {Baldi}, {Bonino},
  {Branchini}, {Brescia}, {Brinchmann}, {Capobianco}, {Cardone}, {Carretero},
  {Castellano}, {Cavuoti}, {Cimatti}, {Cledassou}, {Congedo}, {Conversi},
  {Copin}, {Corcione}, {Courbin}, {Cropper}, {Degaudenzi}, {Dinis}, {Douspis},
  {Dubath}, {Dupac}, {Dusini}, {Farrens}, {Frailis}, {Franceschi}, {Fumana},
  {Galeotta}, {Garilli}, {Gillis}, {Giocoli}, {Grazian}, {Grupp}, {Haugan},
  {Hormuth}, {Hornstrup}, {Jahnke}, {K{\"u}mmel}, {Kiessling}, {Kilbinger},
  {Kitching}, {Kunz}, {Kurki-Suonio}, {Ligori}, {Lilje}, {Lloro}, {Mansutti},
  {Marggraf}, {Marulli}, {Massey}, {Medinaceli}, {Mei}, {Meneghetti}, {Merlin},
  {Meylan}, {Moresco}, {Moscardini}, {Munari}, {Niemi}, {Padilla}, {Paltani},
  {Pasian}, {Pedersen}, {Percival}, {Pires}, {Polenta}, {Poncet}, {Popa},
  {Raison}, {Renzi}, {Rhodes}, {Riccio}, {Romelli}, {Roncarelli}, {Rossetti},
  {Saglia}, {Sartoris}, {Schneider}, {Secroun}, {Seidel}, {Serrano},
  {Sirignano}, {Sirri}, {Stanco}, {Starck}, {Surace}, {Tallada-Cresp{\'\i}},
  {Taylor}, {Tereno}, {Toledo-Moreo}, {Torradeflot}, {Valentijn}, {Valenziano},
  {Vassallo}, {Wang}, {Weller}, {Zamorani}, {Zoubian}, {Scottez}, \&
  {Veropalumbo}}]{Casas23}
{Casas}, S., {Lesgourgues}, J., {Sch{\"o}neberg}, N., {et~al.} 2023
  [\eprint[arXiv]{2303.09451}]

\bibitem[{Challinor \& Lewis(2011)}]{Challinor:2011bk}
Challinor, A. \& Lewis, A. 2011, Phys. Rev. D, 84, 043516

\bibitem[{Chisari {et~al.}(2018)Chisari, Richardson, Devriendt, Dubois,
  Schneider, Le Brun, Beckmann, Peirani, Slyz, \& Pichon}]{chisari_2018}
Chisari, N.~E., Richardson, M. L.~A., Devriendt, J., {et~al.} 2018, MNRAS, 480,
  3962

\bibitem[{Chudaykin {et~al.}(2016)Chudaykin, Gorbunov, \&
  Tkachev}]{Chudaykin:2016yfk}
Chudaykin, A., Gorbunov, D., \& Tkachev, I. 2016, Phys. Rev. D, 94, 023528

\bibitem[{Chudaykin {et~al.}(2018)Chudaykin, Gorbunov, \&
  Tkachev}]{Chudaykin:2017ptd}
Chudaykin, A., Gorbunov, D., \& Tkachev, I. 2018, Phys. Rev. D, 97, 083508

\bibitem[{Colombi {et~al.}(1996)Colombi, Dodelson, \& Widrow}]{Colombi:1995ze}
Colombi, S., Dodelson, S., \& Widrow, L.~M. 1996, \apj, 458, 1

\bibitem[{Cyr-Racine {et~al.}(2016)Cyr-Racine, Sigurdson, Zavala, Bringmann,
  Vogelsberger, \& Pfrommer}]{Cyr-Racine:2015ihg}
Cyr-Racine, F.-Y., Sigurdson, K., Zavala, J., {et~al.} 2016, Phys. Rev. D, 93,
  123527

\bibitem[{Dakin {et~al.}(2022)Dakin, Hannestad, \& Tram}]{Dakin_2022}
Dakin, J., Hannestad, S., \& Tram, T. 2022, MNRAS, 513, 991

\bibitem[{Diamanti {et~al.}(2017)Diamanti, Ando, Gariazzo, Mena, \&
  Weniger}]{Diamanti:2017xfo}
Diamanti, R., Ando, S., Gariazzo, S., Mena, O., \& Weniger, C. 2017, JCAP, 06,
  008

\bibitem[{Dodelson \& Widrow(1994)}]{Dodelson:1993je}
Dodelson, S. \& Widrow, L.~M. 1994, Phys. Rev. Lett., 72, 17

\bibitem[{Enqvist {et~al.}(2015)Enqvist, Nadathur, Sekiguchi, \&
  Takahashi}]{Enqvist:20151bddm}
Enqvist, K., Nadathur, S., Sekiguchi, T., \& Takahashi, T. 2015, JCAP, 09, 067

\bibitem[{{Euclid Collaboration: Blanchard} {et~al.}(2020){Euclid
  Collaboration: Blanchard}, Camera, Carbone, {et~al.}}]{Blanchard:2019oqi}
{Euclid Collaboration: Blanchard}, A., Camera, S., Carbone, C., {et~al.} 2020,
  A\&A, 642, A191

\bibitem[{{Euclid Collaboration: Cropper} {et~al.}(2024){Euclid Collaboration:
  Cropper}, {Al Bahlawan}, {Amiaux}, {et~al.}}]{EuclidSkyVIS}
{Euclid Collaboration: Cropper}, M., {Al Bahlawan}, A., {Amiaux}, J., {et~al.}
  2024, \aap, submitted, arXiv:2405.13492

\bibitem[{{Euclid Collaboration: Jahnke} {et~al.}(2024){Euclid Collaboration:
  Jahnke}, {Gillard}, {Schirmer}, {et~al.}}]{EuclidSkyNISP}
{Euclid Collaboration: Jahnke}, K., {Gillard}, W., {Schirmer}, M., {et~al.}
  2024, \aap, submitted, arXiv:2405.13493

\bibitem[{{Euclid Collaboration: Mellier} {et~al.}(2024){Euclid Collaboration:
  Mellier}, {Abdurro'uf}, {Acevedo~Barroso}, {Ach\'ucarro},
  {et~al.}}]{EuclidSkyOverview}
{Euclid Collaboration: Mellier}, Y., {Abdurro'uf}, {Acevedo~Barroso}, J.,
  {Ach\'ucarro}, A., {et~al.} 2024, \aap, submitted, arXiv:2405.13491

\bibitem[{Feng(2010)}]{Feng:2010gw}
Feng, J.~L. 2010, Ann. Rev. Astron. Astrophys., 48, 495

\bibitem[{Feng {et~al.}(2009)Feng, Kaplinghat, Tu, \& Yu}]{Feng:2009mn}
Feng, J.~L., Kaplinghat, M., Tu, H., \& Yu, H.-B. 2009, JCAP, 07, 004

\bibitem[{Franco~Abell\'an {et~al.}(2021)Franco~Abell\'an, Murgia, \&
  Poulin}]{FrancoAbellan:2021sxk}
Franco~Abell\'an, G., Murgia, R., \& Poulin, V. 2021, Phys. Rev. D, 104, 123533

\bibitem[{Franco~Abell\'an {et~al.}(2022)Franco~Abell\'an, Murgia, Poulin, \&
  Lavalle}]{FrancoAbellan:2020xnr}
Franco~Abell\'an, G., Murgia, R., Poulin, V., \& Lavalle, J. 2022, Phys. Rev.
  D, 105, 063525

\bibitem[{Fu\ss{} \& Garny(2023)}]{Fuss:2022zyt}
Fu\ss{}, L. \& Garny, M. 2023, JCAP, 10, 020

\bibitem[{{Gelman} \& {Rubin}(1992)}]{GelmanRubin}
{Gelman}, A. \& {Rubin}, D.~B. 1992, Statistical Science, 7, 457

\bibitem[{Giri \& Schneider(2021)}]{giri_emulation_2021}
Giri, S.~K. \& Schneider, A. 2021, JCAP, 12, 046

\bibitem[{{Giri} \& {Schneider}(2023)}]{Giri2023BCemu}
{Giri}, S.~K. \& {Schneider}, A. 2023, {BCemu: Model baryonic effects in
  cosmological simulations}, Astrophysics Source Code Library, record
  ascl:2308.010

\bibitem[{Gluscevic {et~al.}(2019)}]{Gluscevic:2019yal}
Gluscevic, V. {et~al.} 2019, Bull. Am. Astron. Soc., 51, 134

\bibitem[{Grandis {et~al.}(2023)Grandis, Aric\`o, Schneider, \&
  Linke}]{Grandis:2023qwx}
Grandis, S., Aric\`o, G., Schneider, A., \& Linke, L. 2023
  [\eprint[arXiv]{2309.02920}]

\bibitem[{Haridasu \& Viel(2020)}]{Haridasu:2020xaa}
Haridasu, B.~S. \& Viel, M. 2020, MNRAS, 497, 1757

\bibitem[{Hervas-Peters {et~al.}(2023)Hervas-Peters, Schneider, Bucko, Giri, \&
  Parimbelli}]{Peters_2023}
Hervas-Peters, F., Schneider, A., Bucko, J., Giri, S.~K., \& Parimbelli, G.
  2023 [\eprint[arXiv]{2309.03865}]

\bibitem[{Holm {et~al.}(2023)Holm, Herold, Hannestad, Nygaard, \&
  Tram}]{Holm:2022kkd}
Holm, E.~B., Herold, L., Hannestad, S., Nygaard, A., \& Tram, T. 2023, Phys.
  Rev. D, 107, L021303

\bibitem[{Hooper {et~al.}(2022)Hooper, Sch\"oneberg, Murgia, Archidiacono,
  Lesgourgues, \& Viel}]{Hooper:2022byl}
Hooper, D.~C., Sch\"oneberg, N., Murgia, R., {et~al.} 2022, JCAP, 10, 032

\bibitem[{Hubert {et~al.}(2021)Hubert, Schneider, Potter, Stadel, \&
  Giri}]{Hubert_2021}
Hubert, J., Schneider, A., Potter, D., Stadel, J., \& Giri, S.~K. 2021, JCAP,
  10, 040

\bibitem[{Ichiki {et~al.}(2004)Ichiki, Oguri, \& Takahashi}]{Ichiki:2004vi}
Ichiki, K., Oguri, M., \& Takahashi, K. 2004, Phys. Rev. Lett., 93, 071302

\bibitem[{Ivezi\'c {et~al.}(2019)Ivezi\'c, Kahn, Tyson,
  {et~al.}}]{LSST:2008ijt}
Ivezi\'c, Z., Kahn, S.~M., Tyson, J.~A., {et~al.} 2019, \apj, 873, 111

\bibitem[{Joachimi \& Bridle(2010)}]{Joachimi:2009ez}
Joachimi, B. \& Bridle, S.~L. 2010, Astron. Astrophys., 523, A1

\bibitem[{Joachimi \& Schneider(2009)}]{Joachimi:2009fr}
Joachimi, B. \& Schneider, P. 2009, Astron. Astrophys., 507, 105

\bibitem[{Kaiser(1992)}]{Kaiser:1991qi}
Kaiser, N. 1992, Astrophys. J., 388, 272

\bibitem[{Kilbinger {et~al.}(2017)}]{Kilbinger:2017lvu}
Kilbinger, M. {et~al.} 2017, Mon. Not. Roy. Astron. Soc., 472, 2126

\bibitem[{Knabenhans {et~al.}(2021)}]{Euclid:2020rfv}
Knabenhans, M. {et~al.} 2021, Mon. Not. Roy. Astron. Soc., 505, 2840

\bibitem[{Lepori {et~al.}(2022)}]{Euclid:2021rez}
Lepori, F. {et~al.} 2022, Astron. Astrophys., 662, A93

\bibitem[{Lesgourgues(2011)}]{Lesgourgues:2011re}
Lesgourgues, J. 2011 [\eprint[arXiv]{1104.2932}]

\bibitem[{Lesgourgues {et~al.}(2016)Lesgourgues, Marques-Tavares, \&
  Schmaltz}]{Lesgourgues:2015wza}
Lesgourgues, J., Marques-Tavares, G., \& Schmaltz, M. 2016, JCAP, 02, 037

\bibitem[{{Lesgourgues} \& {Pastor}(2006)}]{Lesgourgues_Pastor_2006}
{Lesgourgues}, J. \& {Pastor}, S. 2006, \physrep, 429, 307

\bibitem[{Lesgourgues \& Tram(2011)}]{Lesgourgues:2011rh}
Lesgourgues, J. \& Tram, T. 2011, JCAP, 09, 032

\bibitem[{Lesgourgues \& Verde(2022)}]{ParticleDataGroup:2022pth_JLLV}
Lesgourgues, J. \& Verde, L. 2022, Review on {\it Neutrinos in cosmology} of
  the RPP, PTEP, 2022, 083C01

\bibitem[{Ma {et~al.}(2005)Ma, Hu, \& Huterer}]{Ma:2005rc}
Ma, Z.-M., Hu, W., \& Huterer, D. 2005, Astrophys. J., 636, 21

\bibitem[{Maccio {et~al.}(2013)Maccio, Ruchayskiy, Boyarsky, \&
  Munoz-Cuartas}]{Maccio:2012rjx}
Maccio, A.~V., Ruchayskiy, O., Boyarsky, A., \& Munoz-Cuartas, J.~C. 2013,
  MNRAS, 428, 882

\bibitem[{{Mead} {et~al.}(2021){Mead}, {Brieden}, {Tr{\"o}ster}, \&
  {Heymans}}]{hmcode2020}
{Mead}, A.~J., {Brieden}, S., {Tr{\"o}ster}, T., \& {Heymans}, C. 2021, \mnras,
  502, 1401

\bibitem[{Murgia {et~al.}(2018)Murgia, Ir\v{s}i\v{c}, \& Viel}]{Murgia:2018now}
Murgia, R., Ir\v{s}i\v{c}, V., \& Viel, M. 2018, Phys. Rev. D, 98, 083540

\bibitem[{Murgia {et~al.}(2017)Murgia, Merle, Viel, Totzauer, \&
  Schneider}]{Murgia:2017lwo}
Murgia, R., Merle, A., Viel, M., Totzauer, M., \& Schneider, A. 2017, JCAP, 11,
  046

\bibitem[{Nygaard {et~al.}(2021)Nygaard, Tram, \& Hannestad}]{Nygaard:2020sow}
Nygaard, A., Tram, T., \& Hannestad, S. 2021, JCAP, 05, 017

\bibitem[{Oldengott {et~al.}(2016)Oldengott, Boriero, \&
  Schwarz}]{Oldengott:2016yjc}
Oldengott, I.~M., Boriero, D., \& Schwarz, D.~J. 2016, JCAP, 08, 054

\bibitem[{Pandey {et~al.}(2020)Pandey, Karwal, \& Das}]{Pandey:2019plg}
Pandey, K.~L., Karwal, T., \& Das, S. 2020, JCAP, 07, 026

\bibitem[{{Parimbelli} {et~al.}(2021){Parimbelli}, {Scelfo}, {Giri},
  {Schneider}, {Archidiacono}, {Camera}, \& {Viel}}]{Parimbelli_CWDM}
{Parimbelli}, G., {Scelfo}, G., {Giri}, S.~K., {et~al.} 2021, JCAP, 12, 044

\bibitem[{Potter {et~al.}(2017)Potter, Stadel, \&
  Teyssier}]{Potter:2017pkdgrav3}
Potter, D., Stadel, J., \& Teyssier, R. 2017, Computational Astrophysics and
  Cosmology, 4

\bibitem[{Poulin {et~al.}(2016)Poulin, Serpico, \&
  Lesgourgues}]{Poulin:2016nat}
Poulin, V., Serpico, P.~D., \& Lesgourgues, J. 2016, JCAP, 08, 036

\bibitem[{Rubira {et~al.}(2023)Rubira, Mazoun, \& Garny}]{Rubira_2023}
Rubira, H., Mazoun, A., \& Garny, M. 2023, JCAP, 01, 034

\bibitem[{{Santos} {et~al.}(2015){Santos}, {Bull}, {Alonso}, {Camera},
  {Ferreira}, {Bernardi}, {Maartens}, {Viel}, {Villaescusa-Navarro}, {Abdalla},
  {Jarvis}, {Metcalf}, {Pourtsidou}, \& {Wolz}}]{Santos:2015gra}
{Santos}, M., {Bull}, P., {Alonso}, D., {et~al.} 2015, in Advancing
  Astrophysics with the Square Kilometre Array (AASKA14), 19

\bibitem[{Schneider(2015)}]{Schneider:2014rda}
Schneider, A. 2015, MNRAS, 451, 3117

\bibitem[{Schneider {et~al.}(2022)Schneider, Giri, Amodeo, \&
  Refregier}]{schneider_constraining_2022}
Schneider, A., Giri, S.~K., Amodeo, S., \& Refregier, A. 2022, MNRAS, 514, 3802

\bibitem[{Schneider {et~al.}(2020)Schneider, Refregier, Grandis, Eckert,
  Stoira, Kacprzak, Knabenhans, Stadel, \& Teyssier}]{Schneider:2019xpf}
Schneider, A., Refregier, A., Grandis, S., {et~al.} 2020, JCAP, 04, 020

\bibitem[{Schneider \& Teyssier(2015)}]{schneider_new_2015}
Schneider, A. \& Teyssier, R. 2015, JCAP, 12, 049

\bibitem[{Schneider {et~al.}(2019)Schneider, Teyssier, Stadel, Chisari,
  Le~Brun, Amara, \& Refregier}]{Schneider:2018pfw}
Schneider, A., Teyssier, R., Stadel, J., {et~al.} 2019, JCAP, 03, 020

\bibitem[{Sch\"oneberg {et~al.}(2022)Sch\"oneberg, Franco~Abell\'an,
  P\'erez~S\'anchez, Witte, Poulin, \& Lesgourgues}]{Schoneberg:2021qvd}
Sch\"oneberg, N., Franco~Abell\'an, G., P\'erez~S\'anchez, A., {et~al.} 2022,
  Phys. Rept., 984, 1

\bibitem[{Simon {et~al.}(2022)Simon, Franco~Abell\'an, Du, Poulin, \&
  Tsai}]{Simon:2022ftd}
Simon, T., Franco~Abell\'an, G., Du, P., Poulin, V., \& Tsai, Y. 2022, Phys.
  Rev. D, 106, 023516

\bibitem[{{Sitzmann} {et~al.}(2020){Sitzmann}, {Martel}, {Bergman}, {Lindell},
  \& {Wetzstein}}]{sitzmann:2019siren}
{Sitzmann}, V., {Martel}, J. N.~P., {Bergman}, A.~W., {Lindell}, D.~B., \&
  {Wetzstein}, G. 2020, Advances in neural information processing systems
  [\eprint[arXiv]{2006.09661}]

\bibitem[{Smith {et~al.}(2003)Smith, Peacock, Jenkins, White, Frenk, Pearce,
  Thomas, Efstathiou, \& Couchmann}]{Smith:2002dz}
Smith, R.~E., Peacock, J.~A., Jenkins, A., {et~al.} 2003, Mon. Not. Roy.
  Astron. Soc., 341, 1311

\bibitem[{Spergel \& Steinhardt(2000)}]{Spergel:1999mh}
Spergel, D.~N. \& Steinhardt, P.~J. 2000, Phys. Rev. Lett., 84, 3760

\bibitem[{{Springel} {et~al.}(2005){Springel}, {White}, {Jenkins}, {Frenk},
  {Yoshida}, {Gao}, {Navarro}, {Thacker}, {Croton}, {Helly}, {Peacock}, {Cole},
  {Thomas}, {Couchman}, {Evrard}, {Colberg}, \& {Pearce}}]{Ngenic}
{Springel}, V., {White}, S. D.~M., {Jenkins}, A., {et~al.} 2005, \nat, 435, 629

\bibitem[{Takahashi {et~al.}(2020)Takahashi, Nishimichi, Namikawa, Taruya,
  Kayo, Osato, Kobayashi, \& Shirasaki}]{Takahashi:2019hth}
Takahashi, R., Nishimichi, T., Namikawa, T., {et~al.} 2020, \apj, 895, 113

\bibitem[{Takahashi {et~al.}(2012)Takahashi, Sato, Nishimichi, Taruya, \&
  Oguri}]{Takahashi:2012em}
Takahashi, R., Sato, M., Nishimichi, T., Taruya, A., \& Oguri, M. 2012, \apj,
  761, 152

\bibitem[{Tanidis \& Camera(2019)}]{Tanidis:2019teo}
Tanidis, K. \& Camera, S. 2019, Mon. Not. Roy. Astron. Soc., 489, 3385

\bibitem[{Tanidis {et~al.}(2024)}]{Euclid:2023pyq}
Tanidis, K. {et~al.} 2024, Astron. Astrophys., 683, A17

\bibitem[{Tram {et~al.}(2019)Tram, Brandbyge, Dakin, \& Hannestad}]{Tram_2019}
Tram, T., Brandbyge, J., Dakin, J., \& Hannestad, S. 2019, JCAP, 03, 022

\bibitem[{van Daalen {et~al.}(2020)van Daalen, McCarthy, \&
  Schaye}]{vanDaalen:2019pst}
van Daalen, M.~P., McCarthy, I.~G., \& Schaye, J. 2020, MNRAS, 491, 2424

\bibitem[{Vattis {et~al.}(2019)Vattis, Koushiappas, \& Loeb}]{Vattis:2019efj}
Vattis, K., Koushiappas, S.~M., \& Loeb, A. 2019, Phys. Rev. D, 99, 121302

\bibitem[{Verde {et~al.}(2019)Verde, Treu, \& Riess}]{Verde:2019ivm}
Verde, L., Treu, T., \& Riess, A.~G. 2019, Nature Astron., 3, 891

\bibitem[{Voruz {et~al.}(2014)Voruz, Lesgourgues, \& Tram}]{Voruz:2013vqa}
Voruz, L., Lesgourgues, J., \& Tram, T. 2014, JCAP, 03, 004

\bibitem[{Workman {et~al.}(2022)}]{ParticleDataGroup:2022pth}
Workman, R.~L. {et~al.} 2022, PTEP, 2022, 083C01

\bibitem[{Xiao {et~al.}(2020)Xiao, Zhang, An, Feng, \& Wang}]{Xiao:2019ccl}
Xiao, L., Zhang, L., An, R., Feng, C., \& Wang, B. 2020, JCAP, 01, 045

\bibitem[{Yoo {et~al.}(2009)Yoo, Fitzpatrick, \& Zaldarriaga}]{Yoo:2009au}
Yoo, J., Fitzpatrick, A.~L., \& Zaldarriaga, M. 2009, Phys. Rev. D, 80, 083514

\bibitem[{Yoo \& Zaldarriaga(2014)}]{Yoo:2014sfa}
Yoo, J. \& Zaldarriaga, M. 2014, Phys. Rev. D, 90, 023513

\end{thebibliography}

\end{document}